\documentclass[pra, twocolumn, superscriptaddress]{revtex4}
\usepackage{enumitem}
\usepackage{graphicx}
\usepackage{bm}
\usepackage{amsmath}
\usepackage{amssymb}
\usepackage{underscore}
\usepackage{color}
\usepackage{float}
\usepackage[caption = false]{subfig}
\usepackage{amsmath,amsthm,amsfonts,amssymb,amscd}

\definecolor{brown}{rgb}{0.59, 0.29, 0.0}

\def\squareforqed{\hbox{\rlap{$\sqcap$}$\sqcup$}}
\def\qed{\ifmmode\squareforqed\else{\unskip\nobreak\hfil
\penalty50\hskip1em\null\nobreak\hfil\squareforqed
\parfillskip=0pt\finalhyphendemerits=0\endgraf}\fi}

\def\duzomniejsze{<\kern-.7mm<}
\def\duzowieksze{>\kern-.7mm>}

\def\textbf#1{{\bf #1}}
\def\beq{\begin{equation}}
\def\eeq{\end{equation}}
\def\be{\begin{equation}}
\def\ee{\end{equation}}
\def\ben{\begin{eqnarray}}
\def\een{\end{eqnarray}}
\def\beqa{\begin{eqnarray}}
\def\eeqa{\end{eqnarray}}
\def\eea{\end{array}}
\def\bea{\begin{array}}

\newcommand{\bei}{\begin{itemize}}
\newcommand{\eei}{\end{itemize}}
\newcommand{\bee}{\begin{enumerate}}
\newcommand{\eee}{\end{enumerate}}
\newcommand{\nc}{\newcommand}
\nc{\1}{{\openone}}

\def\>{\rangle}
\def\<{\langle}

\newtheorem{lemma}{Lemma}

\newtheorem{prop}{Proposition}
\newtheorem{theorem}{Theorem}

\newtheorem*{definition}{Definition}

\def\bed{\begin{definition}}
\def\eed{\end{definition}}
\def\bel{\begin{lemma}}
\def\eel{\end{lemma}}
\def\bet{\begin{theorem}}
\def\eet{\end{theorem}}
\def\be{\begin{equation}}
\def\ee{\end{equation}}

\def\pmin{p_{\min}^{\text{SV}}}
\def\pmax{p_{\max}^{\text{SV}}}
\def\phardy{p_H}
\def\bhardy{B_H}
\def\zhardy{z_H}

\def\emp{\cdot}
\def\bsv{\bar{B}^{\text{SV}}_{\epsilon}}

\usepackage{mathtools}
\newcommand{\defeq}{\vcentcolon=}
\newenvironment{protocol*}[1]
  {
    \begin{center}
      \hrulefill\\
      \textbf{#1}
  }
  {
    \vspace{-1\baselineskip}
    \hrulefill
    \end{center}
  }
\begin{document}

%
%
%

\author{Ravishankar Ramanathan}
\affiliation{Laboratoire d'Information Quantique, Universit\'{e} Libre de Bruxelles, Belgium}

\author{Micha\l{} Horodecki}
\affiliation{Institute of Theoretical Physics and Astrophysics and the National
	Quantum Information Centre, Faculty of Mathematics, Physics and
	Informatics, University of Gdansk, 80-309 Gdansk, Poland.}

\author{Hammad Anwer}
\affiliation{Department of Physics, Stockholm University, S-10691 Stockholm, Sweden}

\author{Stefano Pironio}
\affiliation{Laboratoire d'Information Quantique, Universit\'{e} Libre de Bruxelles, Belgium}

\author{Karol Horodecki}
\affiliation{Institute of Informatics and the National
	Quantum Information Centre, Faculty of Mathematics, Physics and
	Informatics, University of Gdansk, 80-309 Gdansk, Poland.}

\author{Marcus Gr\"{u}nfeld}
\affiliation{Department of Physics, Stockholm University, S-10691 Stockholm, Sweden}

\author{Sadiq Muhammad}
\affiliation{Department of Physics, Stockholm University, S-10691 Stockholm, Sweden}

\author{Mohamed Bourennane}
\affiliation{Department of Physics, Stockholm University, S-10691 Stockholm, Sweden}

\author{Pawe{\l} Horodecki}
\affiliation{Faculty of Applied Physics and Mathematics
	and the National Quantum Information Centre, Gdansk University of
	Technology, 80-233 Gdansk, Poland.}
\affiliation{International Centre for Theory of Quantum Technologies, University of Gdansk, 80–952 Gdansk, Poland}

\title{Practical No-Signalling proof Randomness Amplification using Hardy paradoxes and its experimental implementation}

\begin{abstract}
Device-Independent (DI) security is the gold standard of quantum cryptography, providing information-theoretic security based on the very laws of nature. In its highest form, security is guaranteed against adversaries limited only by the no-superluminal signalling rule of relativity. The task of randomness amplification, to generate secure fully uniform bits starting from weakly random seeds, is of both cryptographic and foundational interest, being important for the generation of cryptographically secure random numbers as well as bringing deep connections to the existence of free-will. DI no-signalling proof protocols for this fundamental task have thus far relied on esoteric proofs of non-locality termed pseudo-telepathy games, complicated multi-party setups or high-dimensional quantum systems, and have remained out of reach of experimental implementation. In this paper, we construct the first practically relevant no-signalling proof DI protocols for randomness amplification based on the simplest proofs of Bell non-locality and illustrate them with an experimental implementation in a quantum optical setup using polarised photons. 
	Technically, we relate the problem to the vast field of Hardy paradoxes, without which it would be impossible to achieve amplification of arbitrarily weak sources in the simplest Bell non-locality scenario consisting of two parties choosing between two binary inputs. Furthermore, we identify a deep connection between proofs of the celebrated Kochen-Specker theorem and Hardy paradoxes that enables us to construct Hardy paradoxes with the non-zero probability taking any value in $(0,1]$. Our methods enable us, under the fair-sampling assumption of the experiment, to realize up to $25$ bits of randomness in $20$ hours of experimental data collection from an initial private source of randomness $0.1$ away from uniform. 

\end{abstract}

\maketitle

\date{\today}

\textit{Introduction.-} 
Device-independent (DI) cryptography \cite{Pironio} achieves the highest form of security in quantum cryptography, namely one where the users do not need to trust the very devices executing the cryptographic protocol. Instead, DI cryptography guarantees information-theoretic security based on the very laws of nature. Furthermore, the honest users can verify the correct execution of the cryptographic protocol and the security of their output by simple statistical tests on the devices, in the form of Bell tests for quantum non-local correlations. In its highest form, such a cryptographic protocol guarantees security based on the simplest and most fundamental law of nature, namely the rule of no-superluminal signalling of relativity. Such a high form of cryptographic security is desirable for many tasks of critical importance, but naturally comes with concomitant stringent requirements on its experimental implementation. 

A paradigmatic cryptographic task is that of randomness amplification \cite{CR12}, namely the task of converting a source of partially random bits into one of fully uniform bits. Besides its cryptographic importance in the secure generation of private random bits (which have applications in secure encrypted communications and numerical simulations to gambling), this problem is also of foundational interest with implications for the philosophical problem of the existence of free-will \cite{Myrvold}. The amplification of arbitrarily weak random bits into fully random ones is there thought of as the statement that the existence of any weak random physical process implying the existence of a fully random one, in other words, the existence of a complete freedom of choice. Thus, for both fundamental and practical reasons, it is of great interest to have experimentally feasible DI protocols for randomness amplification, and show their security based on the no-signalling principle. 

However, all no-signalling proof DI protocols proposed so far for this fundamental task have remained out of reach of experimental implementation. They have relied on esoteric proofs of non-locality termed pseudo-telepathy games, complicated multi-party setups or high-dimensional quantum systems, placing stringent requirements on practical implementations. Therefore, the question of practical feasibility of such protocols, and the experimental realizability of any no-signaling DI protocol (for any cryptographic task even going beyond randomness amplification) has remained open.


In this paper, we provide the first practically feasible no-signalling proof DI protocol for randomness amplification. We do this by relating the problem of finding experimentally friendly randomness amplification schemes to the vast field of Hardy paradoxes (see Fig. \ref{DIRA-explanation-part1}) and, as a consequence, present a device-independent randomness amplification protocol secure against no-signaling adversaries in the simplest experimentally feasible Bell scenario of two parties with two binary inputs. Furthermore, we show that just as proofs of the Kochen-Specker theorem give rise to pseudo-telepathy games, substructures within these proofs termed $01$-gadgets give rise to Hardy paradoxes and we use them to construct Hardy paradoxes with the non-zero probability taking any value in $(0,1]$. The inter-relationship between Hardy paradoxes and Kochen-Specker proofs, also enables us to construct customized Hardy paradoxes with interesting properties. Finally, we provide a partial characterization of the Bell scenarios in which Hardy paradoxes can be used to certify randomness against a no-signaling adversary. We illustrate the realizability of our protocol with state-of-art experimental setups by performing an experimental implementation in a quantum optical setup using polarised photons. Up to the fair-sampling assumption, this thus constitutes the first experimental realization of a DI protocol that enables a weakening of the fundamental freedom-of-choice assumption. 

\begin{figure}
	\includegraphics[width=1.0\columnwidth]{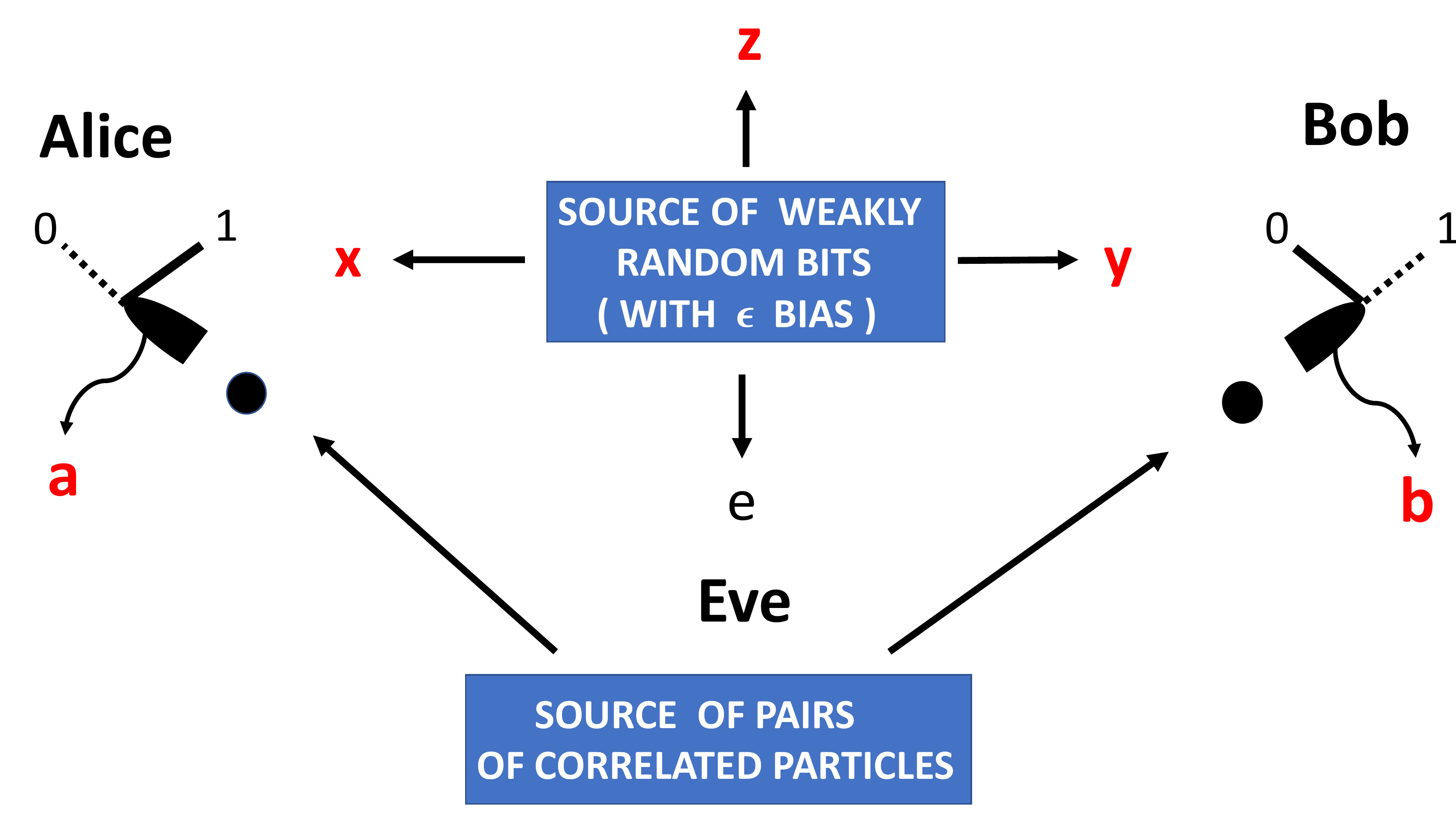}
\caption{\label{DIRA-explanation} \textit{Pictorial depiction of the Device-Independent Randomness Amplification (DIRA) protocol. First step: Bell experiment with weakly random settings. An eavesdropper Eve has a source that distributes quantum particles to two parties Alice and Bob. These parties perform measurements on the received particles using apparata with two settings each, constituting the simplest possible $(2,2,2)$ Bell scenario. The choice of the settings (in each run of the experiment) is made according to bits $(x, y)$ obtained from a Santha-Vazirani (SV) source of weakly random bits, which also produces further bits $z$ to be fed into a randomness extractor in a subsequent step of the protocol. Eve is taken to hold some classical side information $e$ about the weakly random source, as well as some no-signalling side information about the devices held by Alice and Bob. The outputs $(a, b)$ produced by these devices are described by a family of probabilities $\{P_{AB|XY}(ab|xy) \}$. Note that in the figure, a single run of the experiment is depicted while in practice the scheme is repeated many times producing sequences of bits. The assumption here is that the SV source is private, i.e., that the bits held by Alice and Bob are unknown to Eve. The goal is to produce, out of the outcomes $(a, b)$ and some further bits $z$ from the SV source, a sequence of final output bits that are secure and fully random from the perspective of Eve. This is achieved by further processing of the outputs in the second step of the protocol (Fig. 2).}}
\end{figure}

\begin{figure}
	\includegraphics[width=1.0\columnwidth]{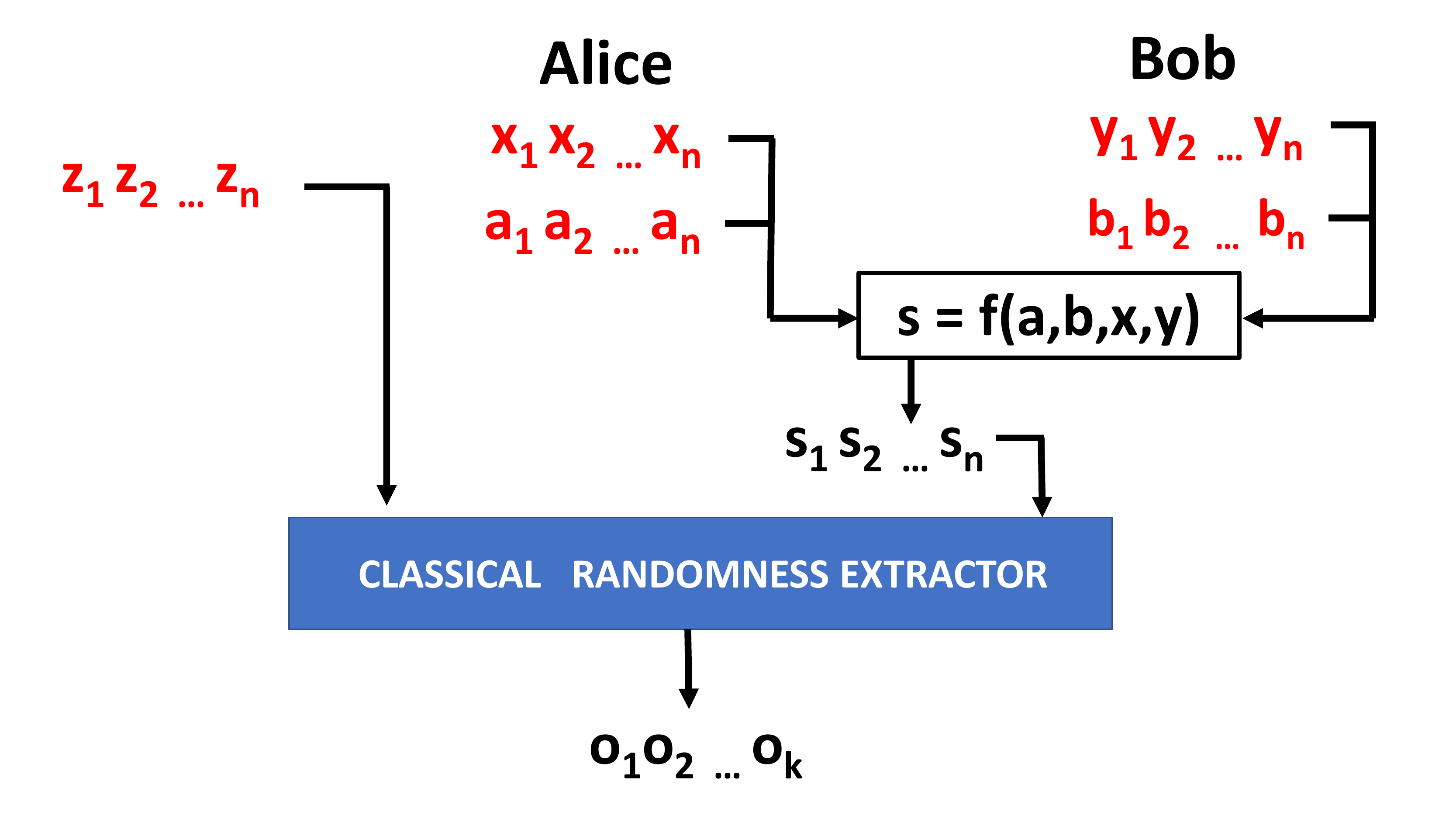}
\caption{\label{DIRA-explanation-p2} \textit{Pictorial depiction of the Device-Independent Randomness Amplification (DIRA) protocol. Second step: classical processing of the outputs of the Bell experiment. When the Bell experiment is performed as in Fig. 1, a sequence of outputs in $n$ runs of the experiment, denoted as $\textbf{a} = a_1, \dots, a_n$, $\textbf{b} = b_1, \dots, b_n$ is obtained for inputs $\textbf{x} = x_1, \dots, x_n$, $\textbf{y} = y_1, \dots, y_n$. Alice and Bob first calculate the Bell parameter $L_n^{\epsilon}(\textbf{a}, \textbf{b}, \textbf{x}, \textbf{y})$ and verify that this is above some threshold value $\delta > 0$. They apply a hash function $f$ to the bits $(\textbf{a}, \textbf{b}, \textbf{x}, \textbf{y})$ to obtain a bit sequence $\textbf{s} = s_1, \dots, s_n$ that is partially random and secure from Eve. They feed this bit sequence together with a further bit sequence $z_1, \dots, z_n$ from the weakly random SV source to a classical randomness extractor to obtain a final output sequence $o_1, \dots, o_k$ that is fully random and secure from Eve. The magic of the quantumness is manifested in the fact that the bits $\textbf{s}$ happen to be decorrelated from $z$, which could never  happen if the particles in the experiment from Fig. 1 were classical. This - quantumly originated - decorrelation is the crucial fact that makes the classical independent-source extractor work. The surprising power of the whole scenario is that the final bits stay strongly random even from the perspective of an eavesdropper who is only limited by the fundamental no-superluminal signalling principle of special relativity.}}
\end{figure}

For clarity, all the proofs of the Propositions and the security proof of the randomness amplification protocol are deferred to the Appendices.

\textit{Background and Statement of the problem.-}
The problem of randomness amplification has gained interest since the initial breakthrough work by Colbeck and Renner who showed that quantum non-local correlations enable the amplification of weak sources of specific type, a task which was shown to be impossible with purely classical resources. The model of a source of randomness is the Santha-Vazirani (SV) source \cite{SV84}, a model of a biased coin where the individual coin tosses are not independent but rather the bits $Y_i$ produced by the source obey
\begin{eqnarray}
\frac{1}{2} - \epsilon \leq P(Y_i = 0 | Y_{i-1}, \dots, Y_1) \leq \frac{1}{2} + \epsilon.
\end{eqnarray}
Here $0 \leq \epsilon < \frac{1}{2}$ is a parameter describing the reliability
of the source, the task being to convert a source with $\epsilon < \frac{1}{2}$ into one with $\epsilon \rightarrow 0$. Since then, several works have proposed DI protocols for the task, and proven security against both quantum \cite{KF17, ChSW14} and non-signalling adversaries \cite{GMTD+13, BRGH+16, RBHH+15}. The latter paradigm, constrained by the impossibility of sending messages instantaneously (i.e., of signaling) has an appealing advantage over the paradigm with quantum 
adversaries. 
Namely, while in order to build the devices implementing the protocol, 
the knowledge of quantum mechanics is necessary, yet verifying whether the device provides 
true randomness  does not require any such knowledge. Security of the randomness produced by the device 
can be thus verified just by experts in statistics. It is therefore desirable to develop these protocols, and to make them as feasible 
for implementation as possible. This practical issue goes in parallel with a fundamental one: which quantum correlations constitute the strongest and simplest source of randomness certified by impossibility of signalling. 



Usually, the scheme for randomness amplification consists of two ingredients:
(i) quantum correlations  - whose interaction with the initial weak source of randomness generate additional quantum randomness, and (ii) a classical protocol for amplifying the obtained randomness. Regarding implementation of the scheme in practice, the first ingredient describes the quantum hardware to be built, 
while the second ingredient describes the needed traditional software, i.e., an algorithm to be run on a standard computer.
The main technological challenge is therefore to implement the quantum part, hence it is mandatory to make it as simple as possible.

The level of technological challenge is to a large extent related to 
(a) the property of each single device, and here it is desired 
to have the minimal number of settings and outputs, 
(b) the number of devices  (= the number of parties needed to have quantum correlations),
(c) the feasibility of implementation of the quantum states and measurements required (here, two qubit entangled states are preferred in order to achieve the high fidelities required in DI applications)
(d) the quality of "raw" quantum randomness - i.e. the probability 
distribution of a chosen outcome (to be later amplified by the software part) should be as close as possible to the fair coin distribution 
($\frac12,\frac12)$.
The ultimate bound for (a) and (b) is the Bell scenario $(2,2,2)$, meaning, one needs at least  two devices, with each device having at least
binary inputs and binary outputs (note that throughout this paper we will denote by $(n,k,m)$ the Bell scenario with $n$ parties, each with $k$ inputs and $m$ outputs per input). For typical photonic implementations, the cheapest here seems to be the number of settings, which may be increased  if it could lead to better quality of raw randomness. 

As yet, none of the protocols based on the simple no-signalling paradigm, has entered the regime of experimental realizability, so that the question of practical feasibility and experimental implementation of any no-signalling proof DI protocol has remained open. There are two  basic problems with the present schemes of randomness amplification against a no-signaling adversary \cite{CR12, GMTD+13, BRGH+16, RBHH+15, WBGH+16}.
The first problem is that they have a pretty complicated quantum part:  either the number of devices,  
the number of settings/outputs per device, or the dimensionality of quantum states and measurement required is not minimal. This implies a large level of noise, which restricts the range of parameters 
of the input weak source to be amplified.
The simplest existing schemes are in the Bell scenarios (2,9,4) \cite{RBHH+15}, 
and (4,2,2) \cite{BRGH+16}. 
These are still at the edge of the capabilities of present-day technology. And even if the technology were to reach the required level, there will always be a  demand for simpler, and cheaper schemes. 

The second problem is more conceptual, namely, there are no general easy methods for finding such new schemes. 
For instance, the scheme of \cite{RBHH+15} was obtained through extensive symbolic search.
In this paper, we resolve both problems, by providing a general method of finding new schemes, as well as achieving the simplest possible 
scheme involving two devices each having binary inputs and outputs, which was beyond reach thus far.

\begin{figure}
	\includegraphics[width=1.0\columnwidth]{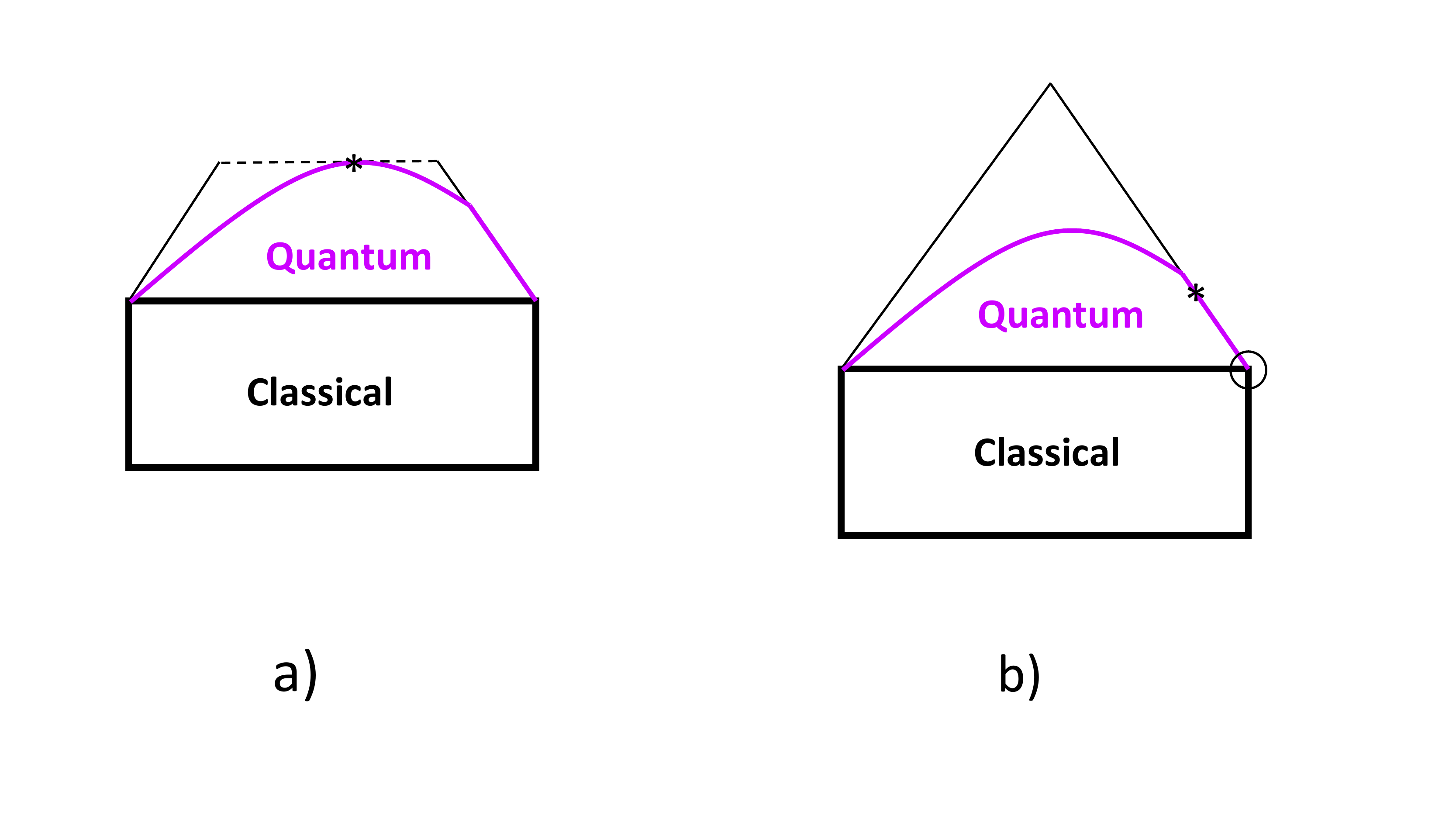}
\caption{\label{Hardy-explanation} \textit{A novelty of the present approach that enables practical implementation. The convex sets of statistical behaviors $\{P_{A,B|X,Y}(a,b|x,y)\}$ obtainable in Bell experiments using classical, quantum and general no-signalling correlations are depicted here. The quantum correlations outside the classical set enable violation of a Bell inequality. Thus far, it was believed that randomness amplification schemes against no-signalling adversaries required quantum correlations that reached a ``pure" no-signalling boundary as in the figure (a) above. These ``pseudo-telepathic" correlations (depicted by the dashed line in (a)) met the stringent requirement that their convex decomposition admitted no classical fraction. However, such a stringent requirement was only possible in Bell scenarios with many inputs and outputs and high-dimensional quantum systems making them completely infeasible for practical experimental implementation. A novelty in this paper is the identification that much simpler quantum correlations that violate Hardy paradoxes, i.e., which only reach ``partial" no-signalling boundaries as in figure (b) above, can still enable DIRA against a general no-signalling adversary. Crucially, our protocol enables the application of these correlations to the task, despite the fact that these correlations admit a significant classical fraction, denoted by the circle $\circ$,  when considered as a convex mixture of general no-signalling behaviors. Furthermore, the special Hardy behaviors such as the point denoted by (*) in the figure (b), are already realizable in the simplest possible $(2,2,2)$ Bell scenario of two parties choosing two binary inputs, which corresponds to simple polarization-entangled photon experiments that are realized everyday in photonic laboratories with high fidelities. It is noteworthy that the pseudo-telepathic part of the boundary is well-known to not exist in this simple case, as illustrated in (b). }}
\end{figure}

\textit{Methodology.-}
To this end we combine three ingredients. One of them is the application of the simplest Hardy paradox, following Ref. \cite{LPR+2015} where it was shown 
that the Hardy paradox is a natural tool for generating randomness - namely, if the correlations exhibit Hardy paradox, then 
for a quantum adversary the probability of the so-called Hardy output is often bounded both from both below and from above 
(however it must be noted that the important problem of randomness amplification was not considered in this context). The power of using Hardy paradoxes as opposed to the pseudo-telepathy games considered so far, is illustrated in Fig. \ref{Hardy-explanation}.

Second, we employ a novel form of Bell inequalities - ones testing so called "measurement dependent locality" 
(MDL) of \cite{PRBL+14, KF17}. Finally, we employ a protocol of randomness amplification of \cite{RBHH+15} which turns out to be
ideal to amplify randomness just by use of Hardy paradoxes. 

We show that the three ingredients combined together 
result in a qualitative advance in the case of a no-signaling adversary - the possibility of randomness amplification in $(2,2,2)$ (the protocol is depicted in Figures \ref{DIRA-explanation} and \ref{DIRA-explanation-p2}). 
We further analyse which Hardy paradoxes can be  used for randomness amplification, and prove that  any 
Hardy paradox with $2$ settings for one party and arbitrary $n\geq 2$ settings  for the other party gives rise to a randomness amplification scheme.  

Remarkably, in the simplest $(2,2,2)$ case, we show that for every input, half of the outputs has probability bounded 
from above and from below. This allows for huge simplification of the original protocol of Ref. \cite{RBHH+15}, 
resulting in dramatic improvement of noise tolerance.  

These findings provide additional motivation  
for the community developing Hardy type paradoxes \cite{Hardy92,Hardy93,MV14,CCXS+13}, 
to search for  new ones, tailored to randomness amplification scheme. Indeed, as a side product of our investigation, we also relate the generation of Hardy paradoxes to proofs of the Kochen-Specker theorem. In particular, we show that substructures within these proofs termed as $01$-gadgets in \cite{RRHH+17} directly give rise to Hardy paradoxes. We then use this connection to solve an open problem in the design of such paradoxes, by generating Hardy paradoxes where the non-zero probability takes all values in the range 
$(0,1]$. 

Our results once more show that 
quantum information is "applied philosophy" \cite{Zukowski}: 
the field of Hardy-type paradoxes, developed initially as a fancy way of proving that Nature violates local-realism,
can be directly related to practical issues, providing a qualitative jump in the 
randomness amplification technology. For other applications of Hardy paradoxes for cryptographic tasks see \cite{ZukowskiHardy}.

\textit{Bell inequalities for the task of randomness amplification.-}
In the task of device-independent randomness amplification (DIRA) against no-signaling adversaries, inequalities where quantum theory allows to reach the no-signaling limit have been considered to be of prime importance. Beginning with Colbeck and Renner's use of the Braunstein-Caves chained Bell inequality in \cite{CR12}, successive works have used versions of the GHZ paradox \cite{GMTD+13, BRGH+16} or two-player pseudo-telepathy games \cite{RBHH+15}. A failure to reach the no-signaling limit implies that there exists some bias $\epsilon$ of the source for which the observed Bell violation can be simulated with classical boxes, leading to an adversarial attack strategy \cite{GHH+14}. A central aim of this work is to bring DIRA closer to practical realization, by reducing the number of inputs. To this end, we present a DIRA protocol based on Hardy paradoxes, which are proofs of non-locality with a similar flavour to pseudo-telepathy games, in that they also impose the probability of a particular subset of events to be zero. 

Hardy's original paradox \cite{Hardy92, Hardy93} was regarded by Mermin to be the ``simplest form of Bell's theorem" \cite{Mermin}. In the $(2,2,2)$ Bell scenario, the two parties Alice and Bob choose between two inputs $X, Y$ taking values $x,y \in \{0,1\}$ respectively, and obtain outcomes $A, B$ taking values $a, b \in \{0,1\}$ respectively. The observed probability distribution of their outputs conditioned upon the inputs is then denoted by $P_{\text{A,B}|\text{X,Y}}(a,b|x,y)$. In this scenario, the paradox is formulated by the following four constraints:
\begin{eqnarray}
\label{eq:Hardy-paradox-2}
\text{(i)} \; P_{\text{A,B}|\text{X,Y}}(0, 1|0, 1) &=& 0, \nonumber \\
\text{(ii)} \; P_{\text{A,B}|\text{X,Y}}(1, 0|1, 0) &=& 0, \nonumber \\
\text{(iii)} \; P_{\text{A,B}|\text{X,Y}}(0, 0|1, 1) &=& 0, \nonumber \\
\text{(iv)}\; P_{\text{A,B}|\text{X,Y}}(0, 0|0, 0) &>& 0. 
\end{eqnarray} 
While classically, it is simple to verify that conditions (i)-(iii) impose the probability of the "Hardy output" to be zero, i.e., $P_{\text{A,B}|\text{X,Y}}(0, 0|0, 0) = 0$, there exist a suitable two-qubit non-maximally entangled state and dichotomic measurements such that all four conditions are obeyed. 
Explicitly, Alice and Bob perform on the shared state
\begin{eqnarray}
\label{eq:hardy-st}
|\psi_{\theta} \rangle = \frac{1}{\sqrt{1+\cos{(\theta)}^2}} \left[\cos{(\theta)} \left( |01 \rangle + |10 \rangle \right) + \sin{(\theta)} | 11 \rangle \right], 
\end{eqnarray}
measurements in the bases
\begin{eqnarray}
\label{eq:hardy-meas}
&&\{|0 \rangle, |1 \rangle\} \; \; \;  \; \text{for} \; x,y=1, \nonumber \\
&&\{\sin{\theta} | 0 \rangle - \cos{\theta} |1 \rangle, \cos{\theta} | 0 \rangle + \sin{\theta} | 1 \rangle\} \; \; \text{for} \; x,y = 0. \nonumber \\
\end{eqnarray}
The constraints (i)-(iv) are satisfied for any value $0 < \theta < \pi/2$, and the optimal value of $P_{\text{A,B}|\text{X,Y}}(0, 0|0, 0)$ ($= \frac{5 \sqrt{5}-11}{2} \approx 0.09$) is achieved at $\theta = \arccos{\left(\sqrt{\frac{\sqrt{5}-1}{2}}\right)}$. 
Hardy's paradox is thus a proof of ``non-locality without inequalities". Yet it is also a probabilistic proof, in the sense that the difference between classical and
quantum worlds in the paradox lies in the possibility of occurrence of some type of events. 

\textit{Hardy paradoxes and randomness certification.-}
Since the discovery of the original Hardy paradox in the $(2,2,2)$ Bell scenario, the paradoxes have been intensively studied and various extensions have been proposed \cite{CCXS+13, BBMH97, ACY16}, especially with a view to boost the probability of the Hardy output. In \cite{LPR+2015} it was noted that the Hardy paradox  is especially suited to reveal intrinsic randomness of quantum statistics. 
Namely, the authors show (in the case of quantum adversary) that the Hardy output is a source of so-called min-entropy, identifying thereby
a natural source of Bell inequalities dedicated for randomness generation. Note here, that while the Hardy paradox is termed a paradox without inequalities, in the presence of noise it becomes a Bell inequality (similar to how Kochen-Specker paradoxes in contextuality are tested in labs by means of 
Cabello-type inequalities \cite{Cab08}). 
 
On the other hand, one can notice that the Bell inequality used in a DIRA protocol of  
Ref. \cite{RBHH+15} -- which allowed for the first time to amplify arbitrarily weak randomness by use of  two devices -- 
can be seen as a variant of Hardy paradox, and the Hardy's output probability is used as a source of min-entropy needed for obtaining randomness. 

Last but not least, recently a concept of measurement dependent locality inequalities appeared, which is suitable for 
randomness amplification problems. 

Combining the above concepts, we will show in this paper that a DIRA protocol can be constructed starting from any Hardy paradox that certifies randomness against a no-signaling adversary for arbitrary initial $\epsilon$. Before present this main result, let us discuss  a couple of important questions that arise in this regard. 
Firstly, it is interesting to see if all Hardy paradoxes certify randomness against a no-signaling adversary. Secondly, it is important to understand how far the probability of the Hardy output can be boosted. 

In answering the first question, we observe that any no-signaling box in the (2,2,2) Bell scenario that satisfies conditions (i) - (iv) of the original Hardy paradox also satisfies $P_{A,B|X,Y}(0,0|0,0) \leq \frac{1}{2}$. This is due to the fact that the only non-local no-signaling extremal box in this scenario is the Popescu-Rohrlich (PR) box (up to relabeling of parties, inputs and outputs) whose entries are in $\{0, \frac{1}{2} \}$, \cite{PR94} and any box that satisfies conditions (i) - (iii) is a convex mixture of a PR box and classical boxes which have $P_{A,B|X,Y}(0,0|0,0) =0$. 
This therefore implies partial randomness in the outputs for the input $(0,0)$ in any no-signaling box. Specifically, assigning bit value $0$ to the output pair $(0,0)$ and bit value $1$ to the output pairs $(0,1),(1,0),(1,1)$, one obtains a partially random bit $S$ of the form $0 < P_{S|X,Y}(s = 0|0,0) \leq \frac{1}{2}$. We will have more to say about the randomness in this particular scenario in Section \ref{sec:NS-(2,2,2)-box} of the Appendix. We now consider whether such a phenomenon is generic to all Hardy paradoxes, i.e., is it the case that every two-party Hardy paradox which certifies that $P_{A,B|X,Y}(a^*,b^*|x^*,y^*) > 0$ also guarantees that $P_{A,B|X,Y}(a^*,b^*|x^*,y^*) < 1$? 

In Section \ref{subsec:rand-2outs} of the Appendix, we show (in Proposition \ref{prop:bin-out-Hardy}) that this is the case when both parties have binary outputs $|\mathcal{A}| = |\mathcal{B}| = 2$.
We then move to the case when Alice measures 2 observables and Bob measures $n > 2$ observables, each with an arbitrary number of outputs, and show in Prop. \ref{prop:upperbound1xn} of Section \ref{subsec:rand-2xn} that in this case the probability of the Hardy output is bounded as $P_{A,B|X,Y}(a^*,b^*|x^*,y^*) \leq \frac{n-1}{n}$, again giving rise to partial randomness. Finally, we consider the case where both parties measure more than two observables with more than two outputs each, i.e., $|\mathcal{A}| = |\mathcal{B}| = m > 2$. In this case, we show in Section \ref{subsec:rand-beyond2xn} that for all $m > 2$, there exist Hardy paradoxes such that $0< P^q_{A,B|X,Y}(a^*,b^*|x^*,y^*) < 1$ and yet $P^{ns}_{A,B|X,Y}(a^*,b^*|x^*,y^*) = 1$. In other words, in this case, even though the probability of the Hardy output is strictly bounded below 1 in quantum theory, there exist no-signaling boxes that achieve the value $1$ while satisfying the same Hardy constraints. Therefore, in using paradoxes with more than two outputs for device-independent randomness certification, a linear programming check is needed in each specific instance to ensure that the Hardy probability is strictly bounded below 1. We now proceed to answer the second interesting question as to how far the Hardy output probability can be boosted, and relate this problem to a class of local contextuality sets recently studied in \cite{RRHH+17, CPSS18}. 

\textit{Generating new Hardy paradoxes using contextuality.-}
Several improvements on the probability of the Hardy output have been proposed in the literature \cite{BBMH97, MV14, CCXS+13}. A "ladder paradox" was introduced in \cite{BBMH97} in the $(2,k,2)$ scenario of two parties, each choosing from $k$ binary inputs. In the limit of a large number of inputs ($k \rightarrow \infty$), the Hardy probability was shown to approach $0.5$, with the corresponding state getting close to the maximally entangled state.  An extension to two qudit systems for $d > 2$ considered in \cite{CCXS+13} makes use of the following conditions in the $(2,2,d)$ scenario: $P_{A,B|X,Y}(a < b|1,0) = 0$, $P(b < a|0,0) = 0$, $P(a < b|0,1) = 0$ and $P(a < b|1,1) > 0$. While classically it is easy to see that the above conditions cannot be satisfied simultaneously, it was numerically verified that in quantum theory the maximum value of the non-zero probability in this case is $\approx 0.417$ for large $d$. The occurrence of the original Hardy paradox as a universal sub-structure within the Hardy paradoxes in $(2,k,2)$ and $(2,2,d)$ Bell scenarios was studied by Mansfield and Fritz in \cite{MF12}. In \cite{MV14}, Man\v{c}inska and Vidick proposed a generalization of Hardy's paradox, increasing the probability of the Hardy output to the entire interval $(0,1]$, but at the expense of a large number of outputs and increasing dimensionality of the shared quantum state. More precisely they showed that in the $(2,2,2^d)$ scenario, the probability of the Hardy output can be boosted to $1 - (1-p_Q^*)^d$ where $p_Q^* = \frac{5 \sqrt{5} - 11}{2}$, so that infinite dimensional states are required to achieve the entire interval $(0,1]$. In this paper, we relate the field of Hardy paradoxes to so-called $01$-gadgets - basic ingredients of the Kochen-Specker paradox \cite{RRHH+17, CPSS18}. In particular, we show how to construct Hardy paradoxes using these local contextuality proofs to achieve the entire interval $(0,1]$ using a finite number of inputs, and as few as four outputs, with a shared two-ququart maximally entangled state. The details of the construction are shown in Section \ref{subsec:Hardy-gadgets} in the Appendix. It is well-known that proofs of the Kochen-Specker theorem give rise to pseudo-telepathy games \cite{BBT05, RW04}, and we now find that in a foundational analogy, the so-called gadget substructures within these proofs similarly lead to Hardy paradoxes neatly connecting these domains of study. We now outline a generic procedure to design DIRA protocols against no-signaling adversaries using Hardy paradoxes.


\textit{A generic procedure to obtain DIRA protocols using Hardy paradoxes.-}
\bei
\item[(a)] Given a Hardy paradox, identify the pair of Hardy settings $(x^*,y^*)$, and Hardy output $(a^*,b^*)$. 
\item[(b)] obtain the maximal quantum value probability $P_{\text{A,B}|\text{X,Y}}(a^*, b^*|x^*,y^*)$ denoted by $p^*_Q$.
\item[(c)] Use linear programming to find the maximal no-signaling value of the  probability $P_{\text{A,B}|\text{X,Y}}(a^*, b^*|x^*,y^*)$ denoted by $p^*_{NS}$. 
\item[(d)] Check for each input if at least one of the output probabilities is bounded strictly below unity. This can be verified using linear programming.
\item[(e)] If item (d) applies, use the software part Protocol I (Fig. \ref{protocolsingle}).
If not, apply the software part Protocol II (Fig. \ref{protocolsingle-2}) stated in Section \ref{sec:DIRA-security} of the Appendix.   
\eei
{\it Remark.} The linear programming part is replacing the semidefinite programming used in works such as \cite{LPR+2015} for a quantum adversary. 

Note that as we show in Section \ref{sec:NS-(2,2,2)-box} of the Appendix, the condition in item (d) directly applies in the (2,2,2) scenario without resorting to linear programming, i.e., we establish for each input in this scenario, strict and achievable bounds on the output probabilities. 

\begin{figure}
	\begin{protocol*}{Protocol I}
		\begin{enumerate}
			\item The $\epsilon$-SV source is used to choose the measurement settings $(x_i, y_i)$ for $n$ runs on a single device consisting of two components. The device produces output bits $x = (a_i, b_i)$ with $i \in \{1,\dots, n\}$.
			\item The parties perform an estimation of the violation of a measurement-dependent locality inequality from the Hardy paradox by computing the empirical average $\textit{L}^{\epsilon}_n(\textbf{a},\textbf{b}, \textbf{x}, \textbf{y}) \defeq  \frac{1}{n} \sum_{i=1}^{n} w_i(\epsilon) B_{\text{H}}(a_i,b_i,x_i,y_i)$. The protocol is aborted unless $\textit{L}_n^{\epsilon}(\textbf{a},\textbf{b}, \textbf{x}, \textbf{y}) \geq \delta$ for fixed constant $\delta > 0$. 
			\item Conditioned on not aborting in the previous step, the parties apply an independent source extractor \cite{CG, one-bit-extr} to the sequence of outputs from the device and further $n$ bits from the SV source.
		\end{enumerate}
	\end{protocol*}
	\caption{\label{protocolsingle} Protocol for device-independent randomness amplification using the two-party Hardy paradox.}
\end{figure}

\textit{Randomness amplification in the simplest (2,2,2) Bell scenario.-}
We exemplify the procedure by means of the simplest Hardy paradox in the experimentally feasible (2,2,2) Bell scenario, with the security proof provided in Appendix \ref{sec:DIRA-security}. 
The Protocol 
I of Fig. \ref{protocolsingle} 
for extracting randomness from Hardy paradoxes is a modification of the protocol we designed in \cite{RBHH+15}. In that protocol, the honest parties were required to perform two tests: one for the Hardy constraints such as (i)-(iii) from Eq.(\ref{eq:Hardy-paradox-2}), and a second test corresponding to constraint (iv). This second test served to lower bound the value of the Hardy output $P_{\text{A,B}|\text{X,Y}}(a^*,b^*|x^*,y^*)$ in a linear number of runs. In the modified Protocol I 
proposed above, we combine these two tests into a single test estimating the violation of a \textit{measurement-dependent locality inequality} (see Section \ref{subsec:MDL-ineq} of the Appendix) first introduced in \cite{PRBL+14}. 
The proof of security is appropriately modified \cite{RBHH+15, BRGH+16} and is sketched with the particular parameters of the simplest (2,2,2) Hardy paradox in Section \ref{sec:DIRA-security} and Section IV of the Appendix. In particular, when the test is passed (i.e., the measurement-dependent locality quantity is observed to satisfy $L_n^{\epsilon}(\textbf{a},\textbf{b},\textbf{x}, \textbf{y}) \geq \delta$ for some $\delta > 0$), we certify that the outputs constitute a min-entropy source of linear min-entropy $h_{box}$ given by (see Appendix IV A for the derivation of this expression)
\begin{eqnarray}
&& h_{box} \geq \nonumber \\
&& \max_{\delta_{Az}, \kappa} \bigg[ \frac{n}{4} \min \bigg\{-\left(\frac{\delta - \delta_{Az}- \kappa}{\frac{1}{16}-\kappa} \right)\log_2\left( 1 - \frac{\kappa}{2(\frac{1}{4}-\epsilon^2)^2} \right), \nonumber \\
&&  \qquad \qquad \log_2(e) \frac{\delta_{Az}^2}{4}  \bigg\}-\frac{3}{4} \bigg]
\end{eqnarray}
for parameters $0 < \delta_{Az} < \delta$ and $0 < \kappa < \delta - \delta_{Az}$.
Feeding these bits together with some further bits from SV source into a randomness extractor \cite{Raz2005}, and following an analogous proof to \cite{BRGH+16, RBHH+15}, we obtain $\Omega(n^{1/4})$ bits that are guaranteed to be secure under the strong universal composability criterion. The universally composable security parameter $d_c$ given by \cite{PR14} 
\begin{eqnarray}
\label{eq:dc-def}
d_c := &&\sum_{o,e} \max_{w} \sum_z \big| P_{O,Z,E|W,ACC}(o,z,e|w,ACC) - \nonumber \\ && \qquad \qquad \quad \frac{1}{|O|} P_{Z,E|W,ACC}(z,e|w,ACC) \big|,
\end{eqnarray}
is shown to satisfy
\begin{eqnarray}
d_c \times P(\text{ACC}) \leq 2^{-\Omega(n^{1/4})},
\end{eqnarray}
where $P(\text{ACC})$ denotes the probability with which the test in the protocol is passed. 

\begin{figure}
	\includegraphics[width=1.0\columnwidth]{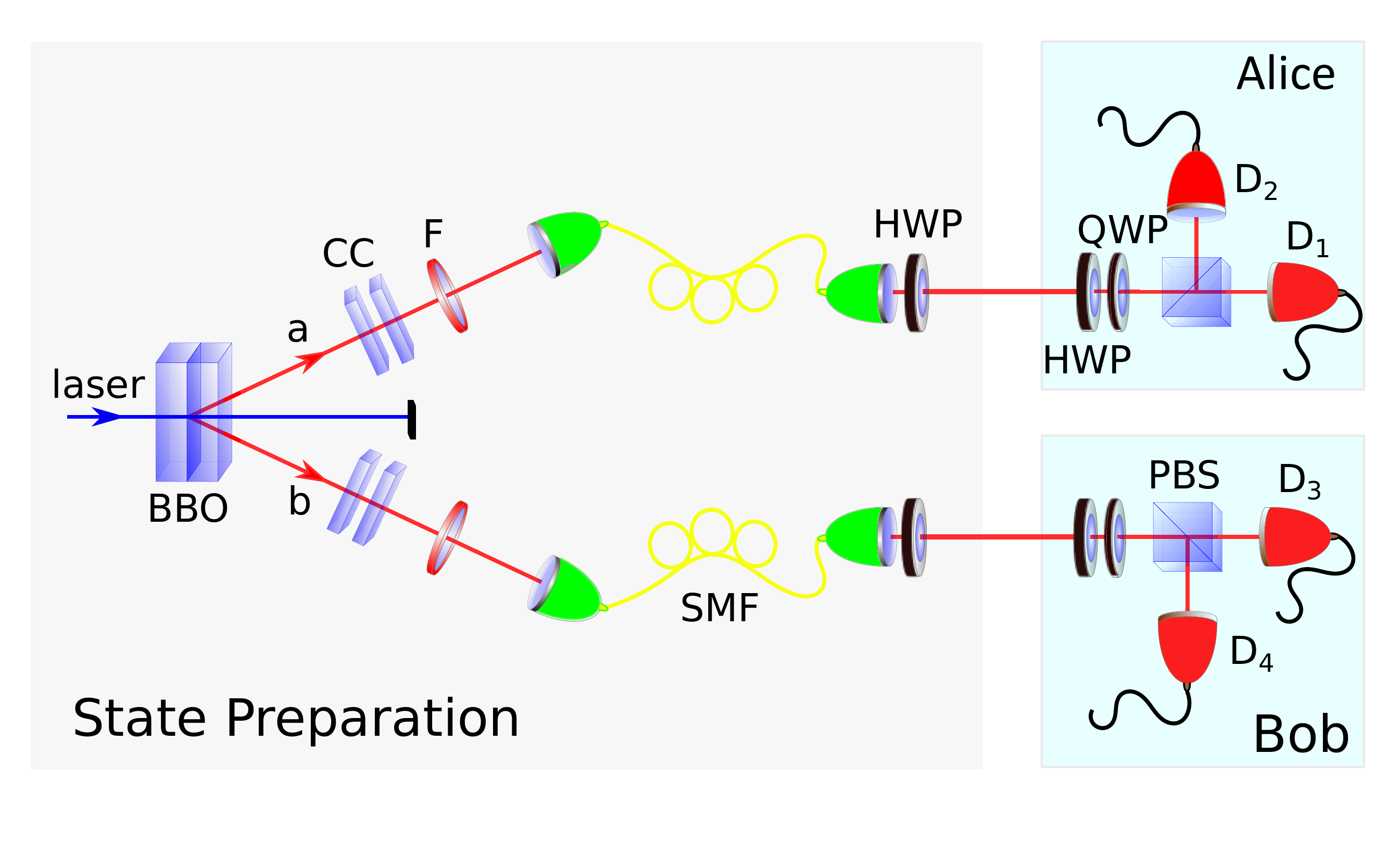}
	\caption{\label{setup} Preparation and measurement setup. A UV pump laser at $390 nm$ was focused onto two beta barium borate (BBO) crystals places in cross-configuration to produce photon pairs emitted into two spatial modes "a" and "b" through type-I SPDC process. To remove any spatial, temporal or spectral distinguishability between the photons we use a pair of $YVO_4$ crystals (CC), narrow-bandwidth filters (F) ($\lambda=1$ nm), and coupling into single-mode fibers (SMF). To prepare the state \ref{eq:hardy-st}, the photon polarizations in each mode are rotated through a half wave plate (HWP). Alice’s and Bob’s measurements were performed using a HWP, a quarter wave plate (QWP), a polarizing beam splitter (PBS) and single-photon avalanche photodiodes $D_i (i =\{1, 2, 3, 4\})$.}
\end{figure}

\textit{Experiment.-}
In our experiment, the physical qubits are single-photon polarization states and the computational basis corresponds to the horizontal (H) and vertical (V) polarization, i.e., $|H \rangle \equiv | 0 \rangle$  and $|V \rangle \equiv| 1 \rangle$. To produce the state in Eq.(\ref{eq:hardy-st}), an ultraviolet pump laser at $390 nm$ was focused onto two beta barium borate (BBO) crystals places placed in cross-configuration to produce photon pairs emitted into two spatial modes "a" and "b" through a type-I Spontaneous Parametric Down-Conversion (SPDC) process. Two crystals were used for compensation of longitudinal and transversal walk-offs. The emitted photons were coupled into single-mode optical fibers and passed through a narrow-bandwidth interference filter  to secure well-defined spatial and spectral emission modes (see  Fig. \ref{setup}). To observe the desired state, we have used a half wave plate (HWP) oriented at $27.86$ degrees and placed after the output fiber coupler in each of the two modes. The polarization measurement was performed using HWPs and polarizing beam splitters  followed by single photon detectors. We have performed the full state tomography and we have obtained the desired state with very high  fidelity of  $F = 0.997 \pm 0.001$
where the fidelity is given by $F = \langle	\psi_{\theta}| 	\rho_{\exp}|\psi_{\theta} \rangle$, of the experimentally prepared state $\rho_{\exp}$ with respect to the optimal $\psi_{\theta}$.
For the Hardy test, the settings of Alice  and Bob (for $x,y=1$) and (for $x,y=0$) \ref{eq:hardy-meas} were implemented with a help of a HWP oriented at the angle for $64.09$ and $0$ degrees respectively. The average  rate of the two photon counting coincidence  events was $10^5$ per second and the measurement time for each setting was $5$ hours (giving a total of $20$ hours of experimental data collection).  

During each measurement, the total number of coincidence counts for Alice and Bob's settings  $(x = 0, y = 0)$, $(x = 0, y = 1)$, $(x = 1, y = 0)$, and $(x = 1, y = 1)$ were $1.90\times 10^9$, $1.91\times 10^9$, $1.91 \times 10^9$, and $1.93 \times 10^9$ respectively. The obtained joint probability $P_{AB,XY}(ab,xy)$  and corresponding errors (standard deviations in the estimated probabilities) for $a, b = \{0,1\}$ the outputs of Alice and Bob's measurements respectively and $x, y = \{0,1\}$, Alice and Bob's settings respectively are listed in the table \ref{Table1-exp1} . In particular the probabilities $P_{\text{AB},\text{XY}}(00,00) = 0.022668 \pm 0.000035$,  
$P_{\text{AB},\text{XY}}(01,01)  =0.000384 \pm 0.000004$, $P_{\text{AB},\text{XY}}(10,10) = 0.000363\pm 0.000004$,and $P_{\text{AB},\text{XY}}(00,11)  = 0.000203\pm 0.000003$. For further details on the experimental setup and results  see Appendix III. 

\begin{table}[t]
\centering
\caption{\label{Table1-exp1}{ The  obtained joint number of coincidence counts $n_{AB,XY}(ab,xy)$ (in multiples of $10^6$), joint probabilities $P_{AB,XY}(ab, xy)$ and corresponding errors (standard deviations in the estimated probabilities) for $a, b = \{0,1\}$  the outputs of Alice and Bob's measurements respectively and or $x, y = \{0,1\}$, Alice and Bob's settings respectively}}
{\begin{tabular}{ |l|l|l|l| }
	\hline
	{\bf  $(ab,xy)$} & {\bf Counts $10^6$}& {\bf Probabilities}& {\bf Error } \\ \hline \hline
	{\bf $(00,00)$ }  &173.54  &0.022668  &0.000035 \\
	{\bf $(01,00)$ }  &284.78  &0.037198  &0.000045 \\
	{\bf $(10,00)$ }  &301.86  &0.039430  &0.000046  \\
	{\bf $(11,00)$ }  &1143.84 &0.149410  &0.000095  \\
	
	{\bf $(00,01)$ }  &459.53  &0.060024  &0.000058  \\
	{\bf $(01,01)$ }  &2.94    &0.000384  &0.000004  \\
	{\bf $(10,01)$ }  &257.02  &0.033572  &0.000043  \\
	{\bf $(11,01)$ }  &1190.90 &0.155556  &0.000097  \\
	
	{\bf $(00,10)$ }  &479.39  &0.062619  &0.000059  \\
	{\bf $(01,10)$ }  &270.59  &0.035345  &0.000044  \\
	{\bf $(10,10)$ }  &2.78    &0.000363  &0.000004  \\
	{\bf $(11,10)$ }  &1162.20 &0.151788  &0.000096 \\
	
	{\bf $(00,11)$ }  &1.56    &0.000204  &0.000003  \\
	{\bf $(01,11)$ }  &715.59  &0.093471  &0.000073  \\
	{\bf $(10,11)$ }  &754.50  &0.098554  &0.000075 \\
	{\bf $(11,11)$ }  &454.88  &0.059417  &0.000057  \\
	\hline
	\end{tabular} }
\end{table}

\textit{Final output randomness from the experimental data.-} 
The detailed analysis of the final output randomness, obtained by applying the randomness extractor to the min-entropy from the device and further bits from the SV source, is performed in Appendix IV.  With respect to the specific randomness extractor from \cite{Raz2005} that we apply here, our experimental data allows for the production of $k(\epsilon, t)$ bits of final output randomness, where  
\begin{equation}
k(\epsilon, t)=\frac{g(\delta_{exp}, \epsilon, n ) n - (t+3)}{2},
\end{equation}
for experimentally obtained MDL value $\delta_{exp}$, experimental number of runs $n$ and a security parameter $t > 0$.
The distribution $\{q_i\}, i = 1, \dots, 2^k$ of the final $k(\epsilon, t)$ output bits satisfies
\begin{equation}
q_i \leq \frac{1}{2^k}\left(1+2^{-t}\right).
\end{equation}
The explicit form of the function $g(\delta_{exp}, \epsilon, n)$, encapsulating the details of the protocol including also the experimental data, 
is provided in the Appendix IV. Exemplary data showing the number of output bits $k(\epsilon, t)$ as a function of the initial $\epsilon$ from the SV source are shown in the Fig. \eqref{fig:krate} for particular values of the security parameter $t$ ($t = 5, 10, 100$, with the final bits deviating from uniform by $2^{-t-1}$). One can see that the state-of-art experimental setups are able to achieve the parameters required for reasonable randomness production by our protocol. It must be stressed that, up to the fair-sampling assumption, this is the first time ever that any device-independent protocol secure against a no-signalling adversary has been implemented in the lab, since all previous schemes had stringent requirements far beyond the scope of any experimental technology known to date. 

\begin{figure}[bth]
	\includegraphics[width=0.5\textwidth]
	{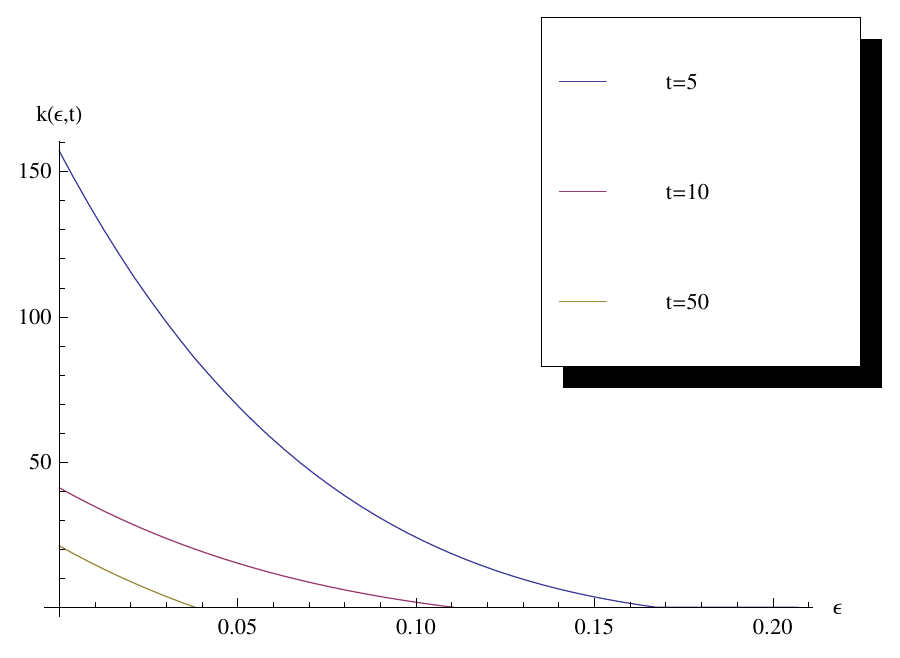}
	\caption{\label{fig:krate} The number of output bits $k(\epsilon, t)$ versus $\epsilon$ for three values of security parameter $t$ (recall that the final $k$ bits deviate from uniform by $2^{-t-1}$).}
\end{figure}


\textit{Conclusions.-}
In this paper, we have shown a generic application of Hardy paradoxes to a scheme of device-independent randomness amplification secure against general no-signaling adversaries. Remarkably, the Hardy paradox allows for experimentally friendly parameters in the simplest (2,2,2) Bell scenario, providing the first practically feasible device-independent application against a no-signaling adversary. 
We have illustrated this feasibility with help of the routine two-photon experiment, achieving the satisfactory rates. This is the first time, when 
such an illustration is provided, since the previous protocols were out of reach of the state-of-art technology.
 
Furthermore, we answered interesting questions arising with regard to Hardy paradoxes and randomness certification. We have shown that Hardy paradoxes with binary outputs, and those with two observables for one party allow for generic randomness certification against no-signaling adversaries, while more general Bell scenarios require specific linear programming checks. Moreover, we  have shown that just as Kochen-Specker sets give rise to two-player pseudo-telepathy games, subsets within the KS proofs provide a systematic method to construct Hardy paradoxes. We use this to solve an open problem showing the existence of Hardy paradoxes with finite numbers of inputs and outputs for which the non-zero Hardy probability takes any value in the interval $(0,1]$.    

An interesting open question is to optimize the construction of Hardy paradoxes to obtain the DIRA scheme with highest min-entropy rate, under constant noise-tolerance. A similar question arises to find the minimal Hardy paradox (with minimum input, output parameters) that allows the probability of the Hardy output to take all values in $(0,1]$. An important question in general DIRA schemes is to relax the assumption of independence between source and device which has so far been employed in all finite-device randomness amplification schemes against no-signaling adversaries \cite{WBGH+16}.   

%

{\it Acknowledgements.-}
R.R. acknowledges support from the research project  ``Causality in quantum theory: foundations and applications'' of the Fondation Wiener-Anspach and from the Interuniversity Attraction Poles 5 program of the Belgian Science Policy Office under the grant IAP P7-35 photonics@be. This work is supported by the Start-up Fund 'Device-Independent Quantum Communication Networks' from The University of Hong Kong. This work was supported by the National Natural Science Foundation of China through grant 11675136, the Hong Kong Research Grant Council through grant 17300918, and the John Templeton Foundation through grants 60609, Quantum Causal Structures, and 61466, The Quantum Information Structure of Spacetime (qiss.fr). MH and PH are supported by the John Templeton Foundation. The opinions expressed in this publication are those of the authors and do not necessarily reflect the views of the John Templeton Foundation. M.H. is supported by the National Science Centre, Poland, grant OPUS 9. 2015/17/B/ST2/01945. P.H and MH. are also supported by the Foundation for Polish Science (IRAP project, ICTQT, contract no.2018/MAB/5, co-financed by EU within the Smart Growth Operational Programme). K.H. acknowledges support from the grant Sonata Bis 5 (grant number: 2015/18/E/ST2/00327) from the National Science Centre. S.P. is a Research Associate of the Fonds de la Recherche Scientifique (F.R.S.-FNRS). We acknowledge support from the EU Quantum Flagship project QRANGE.


\section{Appendix: Security of the DIRA protocol using the simplest Hardy paradox}
\label{sec:DIRA-security}
\subsection{Bell inequality and bounds on output probabilities}
Suppose in a two-party Bell experiment in the $(2,2,2)$ Bell scenario that Alice and Bob choose the (binary) inputs to their respective devices using an $\epsilon$-SV source \cite{SV84}. Let $\bar{B}^{\text{SV}}_{\epsilon}$ denote the value achieved by the box $P_{\text{A,B}|\text{X,Y}}$ used in a single run, for the Measurement-dependent locality (MDL) \cite{PRBL+14} quantity $\bar{B}^{\text{SV}}_{\epsilon}$ obtained from the Hardy paradox in this scenario, i.e., 
\begin{eqnarray}
\label{eq:B-SV-Hardy}
\bar{B}^{\text{SV}}_{\epsilon} := \sum_{a,b,x,y} \nu_{\text{SV}}(x,y) B_{\text{H}}^{\epsilon}(a,b,x,y) P_{\text{A,B}|\text{X,Y}}(a,b|x,y),
\end{eqnarray}
where $B_{\text{H}}^{\epsilon}$ is an indicator taking values as
\begin{eqnarray}
\label{eq:BH-SV-Hardy}
 B_{\text{H}}^{\epsilon}(a,b,x,y) := \left\{
\begin{array}{lr}
\cr-\left(\frac{1}{2} + \epsilon \right)^2  : (a,b,x,y) \in \{(0,1,0,1),  \nonumber \\ \qquad \qquad \qquad \qquad (1,0,1,0), (0,0,1,1)\} \\
\cr\left(\frac{1}{2} - \epsilon \right)^2  : (a,b,x,y) = (0,0,0,0) \\
\cr 0 : \text{otherwise} 
\end{array}
\right.
\end{eqnarray}
and $\nu_{\text{SV}}$ is the distribution from the unknown SV source. The maximal value that a classical box can achieve for the MDL quantity (\ref{eq:B-SV-Hardy}) is $0$ \cite{PRBL+14}, which violates the Hardy constraints (i) - (iv) detailed in the main text. Since here each party chooses a single bit input, we have that 
\begin{eqnarray}
\label{eq:l-h-1bit}
\left(\frac{1}{2} - \epsilon \right)^2 \leq \nu_{\text{SV}}(x,y) \leq \left(\frac{1}{2} + \epsilon \right)^2.
\end{eqnarray} 
In an ideal experiment, the parties would measure exactly the state given by Eq.(\ref{eq:hardy-st}) with the measurements given by Eq.(\ref{eq:hardy-meas}) and consequently the box would achieve the optimal quantum value of 
$\bar{B}^{\text{SV}}_{\epsilon} = \left(\frac{1}{2} - \epsilon \right)^4 p_Q^*$. However, in a real experiment with noise, suppose that the value achieved is $\bar{B}^{\text{SV}}_{\epsilon} \geq \delta$ for some constant $\delta > 0$. The following Lemma then shows a bound on the maximum value of the Hardy probability $P_{\text{A,B}|\text{X,Y}}(0,0|0,0)$ for any no-signaling box. 
\begin{lemma}
	\label{lem:lin-prog-bound}
	Consider a two-party no-signaling box $\{P_{\text{A,B}|\text{X,Y}}\}$ satisfying $\bar{B}^{\text{SV}}_{\epsilon} \geq \delta$ for some constant $\delta > 0$ where $\bar{B}^{\text{SV}}_{\epsilon}$ is given by Eq.(\ref{eq:B-SV-Hardy}).  
	Then for all inputs and outputs we have
	\begin{eqnarray}
	\label{eq:bound-lin-prog}
	P_{\text{A,B}|\text{X,Y}}(a,b|x,y) \leq 1 - \frac{\delta}{\left( \frac{1}{4} - \epsilon^2 \right)^2}. 
	\end{eqnarray}
	while for input-output combinations satisfying the CHSH condition  $a\oplus b =x \cdot y$ we have 
	\begin{eqnarray}
	\label{eq:bound-lin-prog2}
	\frac{\delta}{\left( \frac{1}{4} - \epsilon^2 \right)^2} \leq 
	P_{\text{A,B}|\text{X,Y}}(a,b|x,y) \leq 1 - \frac{\delta}{\left( \frac{1}{4} - \epsilon^2 \right)^2}. 
	\end{eqnarray}

\end{lemma}   

\begin{proof}
	Let $\{P_{\text{A,B}|\text{X,Y}}\}$ denote a two-party no-signaling box satisfying $\bar{B}^{\text{SV}}_{\epsilon} \geq \delta$. 
	Then, by Eqs.(\ref{eq:B-SV-Hardy}) and (\ref{eq:l-h-1bit}) we have 
	\begin{eqnarray}
	\label{eq:B-U-Hardy}
	\sum_{a,b,x,y} 4 B_{\text{H}}^{0}(a,b,x,y) P_{\text{A,B}|\text{X,Y}}(a,b|x,y) \geq \frac{\delta}{\left(\frac{1}{4} - \epsilon^2 \right)^2},
	\end{eqnarray} 
	with $B_{\text{H}}^{\epsilon}(a,b,x,y)$ given by Eq.(\ref{eq:B-SV-Hardy}). 
	The upper and lower bounds on the value of any  $P_{\text{A,B}|\text{X,Y}}(a,b| x,y)$ given the constraints \eqref{eq:B-U-Hardy} are computed by means of a linear program, with the analytical bounds obtained by means of a feasible solution to the dual program.  
\end{proof}

To make the claim more transparent, we present in Section \ref{sec:NS-(2,2,2)-box} a more explicit proof, through  Proposition \ref{prop:svphbound} which directly implies the above 
Lemma making it more convenient to generalize to other Hardy paradoxes. 

\subsection{Estimation}
Let us state the following Lemma \ref{lemmaazuma} shown in \cite{BRGH+16}(based on the Azuma-Hoeffding inequality) which we will use to estimate the arithmetic average of the values of the MDL quantity $\bar{B}^{\text{SV}}_{\epsilon}$ for the conditional boxes, over all the runs of the protocol.
\begin{lemma}[\cite{BRGH+16}] 
	\label{lemmaazuma} 
	Consider  arbitrary random variables $W_i$ for $i=0,1,\ldots,n$, and binary random variables $B_i$ for $i=1,\ldots n$ that are functions of 
	$W_i$, i.e. $B_i=f_i(W_i)$ 
	for some functions $f_i$. Let us denote $\overline{B}_i=\mathbb{E}(B_i|W_{i-1},\ldots,W_1,W_0)$ for $i=1,\ldots,n$ 
	(i.e. $\overline{B}_i$ are conditional means).
	Define for $k = 1, \ldots, n$, the empirical average
	\begin{eqnarray}
	L_k=\frac{1}{k}\sum_{i=1}^k B_i
    \label{eq:means}
	\end{eqnarray}
	and the arithmetic average of conditional means 
	\begin{eqnarray}
	\label{eq:cond-means}
	\overline{L}_k=\frac{1}{k}\sum_{i=1}^k \overline{B}_i.
	\end{eqnarray}
	Then we have 
	\begin{eqnarray}
	\text{Pr}(|L_n-\overline{L}_n|\geq  \delta_{Az})\leq 2 e^{-n\frac{\delta_{Az}^2}{2}}
	\label{eq:PrAzumaL-Ln}
	\end{eqnarray}
\end{lemma}
We also state the following fact
\begin{lemma}
	\label{lem:num-good-runs}
	If the arithmetic average $\overline{L}_n$ of $n$ conditional means in Eq.(\ref{eq:cond-means}) satisfies $\overline{L}_n \geq \frac{\delta}{2}$ for some parameter $\delta > 0$, with $\overline{B}_i \leq \frac{1}{16}$ for every $i$, then in at least $\frac{\delta/2 - \kappa}{1/16 - \kappa}n$ positions $i$ we have that $\overline{B}_i \geq \kappa$ for $0 < \kappa < \delta/2$.
\end{lemma}
\begin{proof}
	Let $\overline{L}_n = \frac{1}{n} \sum_{i=1}^{n} \overline{B}_i \geq \delta/2$ with $\overline{B}_i \leq \frac{1}{16}$ for all $i$. Consider the set $I := \{ i | \overline{B}_i \geq \kappa\}$. Then 
	\begin{eqnarray}
	&&\sum_{i \in I} \overline{B}_i + \sum_{i \notin I} \overline{B}_i \geq \frac{\delta}{2} n \nonumber \\
	&&\Rightarrow \frac{1}{16} |I| + \kappa (n - |I|) \geq \frac{\delta}{2} n \nonumber \\
	&&\Rightarrow |I| \geq \frac{\delta/2 - \kappa}{1/16 - \kappa}n, 
	\end{eqnarray}
	where we have used $\overline{B}_i \leq \frac{1}{16}$ for $i \in I$ and $\overline{B}_i < \kappa$ for $i \notin I$.
\end{proof}
Lemma \ref{lemmaazuma} states that with high probability, the arithmetic average of the mean values $\overline{L}_n$ for the conditional boxes is close to the 
observed value $L_n$. This implies that when the test is passed so that the observed MDL value satisfies $L_n \geq \delta$, with high probability (specifically $1 - 2 e^{-n\frac{\delta^2}{8}}$ from Lemma \ref{lemmaazuma}) we have $\overline{L}_n \geq \frac{\delta}{2}$. 
Lemma \ref{lem:num-good-runs} then states that in this case, choosing $\kappa = \frac{\delta}{4}$, in at least $\frac{4\delta}{1-4\delta} n$
runs, the true Bell value of the box when the settings are chosen with an $\epsilon$ SV source is at least $\frac{\delta}{4}$, i.e., $\overline{B}_i^{\epsilon} \geq \frac{\delta}{4}$ in at least $\frac{4\delta}{1-4\delta} n$ runs. Lemma \ref{lem:lin-prog-bound} then assures that for these runs, all outputs have probabilities (conditioned on past inputs and outputs)  bounded above by $\gamma = 1 - \frac{\delta}{4\left( \frac{1}{4} - \epsilon^2 \right)^2}$ as in Eq. \ref{eq:bound-lin-prog} (with $\frac{\delta}{4}$ in place of $\delta$). Also, for input-output combinations satisfying the CHSH condition, with $\overline{B}_i^{\epsilon} \geq \frac{\delta}{4}$ we obtain that 
\begin{eqnarray}
\label{eq:bound-lin-prog3}
\frac{\delta}{4\left( \frac{1}{4} - \epsilon^2 \right)^2} \leq 
P_{\text{A,B}|\text{X,Y}}(a,b|x,y) \leq 1 - \frac{\delta}{4\left( \frac{1}{4} - \epsilon^2 \right)^2}.
\label{eq:box-upper-bound} 
\end{eqnarray}

We may then hash these outputs to obtain a single partially random bit $S$ in each of these runs as follows. For each $(x,y)$ assign bit value $S = 0$ to one output pair $(a,b)$ satisfying the CHSH condition $a \oplus b = x \cdot y$ and $S=1$ otherwise (i.e., for input $(x,y) = (0,0)$ assign $S=0$ if the output is $(a,b) = (0,0)$ and $S=1$ otherwise, while for input $(x,y) = (1,1)$ assign $S=0$ if the output is $(a,b) = (0,1)$ and $S=1$ otherwise, etc.). We thus obtain in the $i$-th run, with $i \in I$ and  $|I| \geq \frac{4 \delta}{1 - 4 \delta} n$, partially random bit $S_i$ satisfying
\begin{eqnarray}
 \label{eq:bound-lin-prog4}
  \frac{\delta}{4\left( \frac{1}{4} - \epsilon^2 \right)^2} \leq 
P_{\text{S}_i|X_{\leq i},Y_{\leq i},A_{<i},B_{<i},Z,E}(s_i|x_{\leq i},y_{ \leq i},a_{<i},b_{<i},z,e) \nonumber \\
\leq 1 - \frac{\delta}{4\left( \frac{1}{4} - \epsilon^2 \right)^2}. \nonumber \\
\end{eqnarray}

The following Lemma proved in \cite{BRGH+16, RBHH+15} then shows that these outputs constitute a min-entropy source of linear min-entropy. Recall that the min-entropy of a random variable $X$ is given by 
\begin{eqnarray}
H_{\text{min}}(X) = \min_{x \in \text{supp}(X)} - \log_2 P(X=x),
\end{eqnarray}
where supp(X) denotes the support of $X$.  

\begin{lemma}[\cite{BRGH+16,RBHH+15}]
	\label{lem:min-entropy}
	Fix any measure $P$ on the space of sequences $(z,e,x_1,y_1,a_1,b_1,\dots,x_n,y_n,a_n,b_n)$. Suppose that for a sequence $(z,e,x_1,y_1,a_1,b_1,\dots,x_n,y_n,a_n,b_n)$, there exists $K \subseteq [n]$ of size larger than $\mu n$ such that for all $i \in K$ the boxes 
	$P_{A_i,B_i|X_{\leq i},Y_{\leq i},A_{<i},B_{<i},Z,E}(a_i,b_i|x_{\leq i},y_{ \leq i},a_{<i},b_{<i},z,e)$ satisfy 
	\begin{eqnarray}
	P_{A_i,B_i|X_{\leq i},Y_{\leq i},A_{<i},B_{<i},Z,E}(a_i,b_i|x_{\leq i},y_{\leq i},a_{<i},b_{<i},z,e) \leq \gamma. \nonumber
	\end{eqnarray}
	Then, 
	\begin{eqnarray}
	\label{eq:nrun-bound}
	P_{\textbf{A},\textbf{B}|\textbf{X},\textbf{Y}, Z,E}(\textbf{a},\textbf{b}|\textbf{x}, \textbf{y}, z, e) \leq \gamma^{\mu n}.
	\end{eqnarray}
\end{lemma}
In our case $\mu\geq \frac{4\delta}{1-4\delta}$, and $\gamma = 1 - \frac{\delta}{4\left( \frac{1}{4} - \epsilon^2 \right)^2}$ as mentioned above, so that we obtain a min-entropy source of linear min-entropy $\mu n \log_2 \left( \frac{1}{\gamma} \right)$ with these parameters depending on the observed MDL value $\delta$ and the source quality $\epsilon$.

\subsection{Noise tolerance}

With the use of the state and measurements from Eqs.(\ref{eq:hardy-st}) and (\ref{eq:hardy-meas}), we obtain a quantum box $P^q_{A,B|X,Y}(a,b|x,y)$ that achieves the value at least as high as $\left(\frac{1}{2} - \epsilon \right)^4 p_Q^*$ for the MDL quantity $\overline{B}_{\epsilon}^{\text{SV}}$ in Eq.(\ref{eq:B-SV-Hardy}). In a real experiment with noise, suppose the state is of the form 
\begin{eqnarray}
\rho_{\theta} = (1 - \eta) | \psi_{\theta} \rangle \langle \psi_{\theta} | + \eta \frac{\textbf{1}}{4},
\end{eqnarray}
with a noise parameter $0 \leq \eta < 1$. Such a state achieves the value of $\overline{B}_{\epsilon}^{\text{SV}}$ given by
\begin{eqnarray}
\overline{B}_{\epsilon}^{\text{SV}} \geq \left(\frac{1}{2} - \epsilon \right)^4 \left[p_Q^*(1- \eta) - \eta \frac{3}{4} \frac{\left( \frac{1}{2} + \epsilon \right)^4}{\left(\frac{1}{2} - \epsilon \right)^4} + \frac{1}{4} \eta  \right]
\end{eqnarray}
In order to ensure that the test is passed with $\overline{B}_{\epsilon}^{\text{SV}} > 0$ with such noisy quantum boxes, we obtain the condition on the noise parameter $\eta$ of the state to be
\begin{eqnarray}
\eta < \frac{p_Q^*}{\frac{3}{4} \left[ \frac{\left( \frac{1}{2} + \epsilon \right)^4}{\left(\frac{1}{2} - \epsilon \right)^4} \right]- (1/4 - p_Q^*)}.
\end{eqnarray} 
%
\subsection{Security of the protocol}
To complete this section, we state that when the test in Protocol \ref{protocolsingle} is passed, we obtain $\Omega(n^{1/4})$ bits that are guaranteed to be secure under the strong universal composability criterion, the proof follows along analogous lines to those in \cite{BRGH+16, RBHH+15}. In other words, with $P(\text{ACC})$ being the probability that the test in the protocol is passed, and with the security parameter $d_c$ defined as in \cite{PR14} by
\begin{eqnarray}
\label{eq:dc-def-Appendix}
d_c := &&\sum_{o,e} \max_{w} \sum_z \big| P_{O,Z,E|W,ACC}(o,z,e|w,ACC) - \nonumber \\ && \qquad \qquad \quad \frac{1}{|O|} P_{Z,E|W,ACC}(z,e|w,ACC) \big|,
\end{eqnarray}
following analogous arguments to those in \cite{BRGH+16, RBHH+15} we obtain the following result.
\begin{theorem}[\cite{BRGH+16,RBHH+15}]
	Let $n$ denote the number of runs in the Protocol \ref{protocolsingle} and suppose we are given $\epsilon > 0$ and the test is passed with observed MDL value $\delta > 0$.
	Then there exists a non-explicit extractor $\text{Ext}(x,t)$ with $|S| = 2^{\Omega(n^{1/4})}$ values, such that 
	\begin{eqnarray}
	d_c \times P(\text{ACC}) \leq 2^{- \Omega(n^{1/4})},
	\end{eqnarray} 
	with $d_c$ given by Eq.(\ref{eq:dc-def-Appendix}). 
	Alternatively, there exists an explicit extractor $\text{Ext}'(x,t)$ producing a single bit of randomness with
	\begin{eqnarray}
	d_c \times P(\text{ACC}) \leq 2^{- \Omega(n^{1/(2C)})},
	\end{eqnarray}  
	for a fixed constant $C$.
\end{theorem}

\subsection{Properties of a no-signaling box exhibiting the simplest Hardy paradox}
\label{sec:NS-(2,2,2)-box}
\begin{figure}[bth]
	\includegraphics[width=0.5\textwidth]{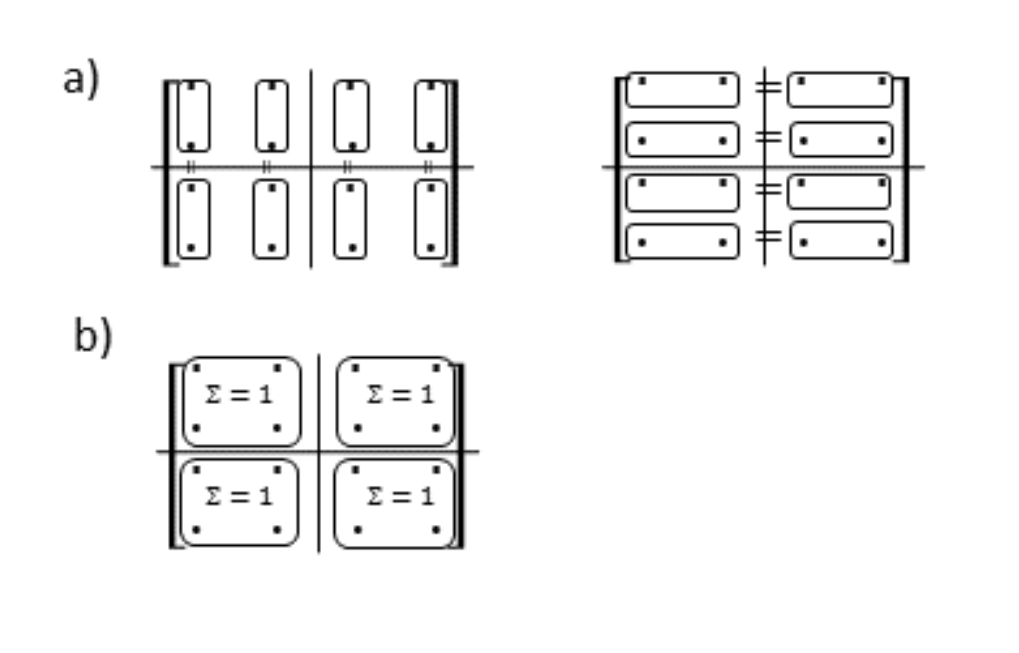}
	\caption{\label{fig:ns-nor}No-signaling box: (a) no-signaling constraints (b) normalization constraints}
\end{figure}

In this section, we study in detail the randomness obtained in the simplest $(2,2,2)$ Bell scenario against a no-signaling adversary using the MDL version of the Hardy paradox.  In particular  we give an explicit proof of the Lemma	\ref{lem:lin-prog-bound}. 	
We denote 
\be
\phardy :=P_{A,B|X,Y}(a_H, b_H|x_H,y_H)
\ee with $(a_H,b_H,x_H,y_H)=(0,0,0,0)$ 
and 
\be
\zhardy :=\sum_{(a,b,x,y)\in S_0} P_{A,B|X,Y}(a,b|x,y)
\ee  
where $S_0=\{(0,1,0,1),(1,0,1,0),(0,0,1,1)\}$. 
In the noiseless scenario, the Hardy paradox means that we have $\zhardy=0$ and $\phardy>0$. 
In the noisy case, we consider the difference $\bhardy$ defined by
\be
\bhardy :=\phardy - \zhardy = \sum_{a,b,x,y} 4 B_{\text{H}}^{0}(a,b,x,y) P_{\text{A,B}|\text{X,Y}}(a,b|x,y).
\ee
The table of conditional probabilities $P_{A,B|X,Y}(a,b|x,y)$ constituting a no-signaling box exhibiting the Hardy paradox 
is then the following:
\ben
\label{eq:hardy-simple}
\left[
\begin{array}{c|c}
	\bea{cc}  p_H & \emp  \\ \emp   & \emp  \\ \eea   
	& 
	\bea{cc} \emp & \delta_1  \\ \emp   & \emp  \\ \eea  
	\\
	\hline
	\bea{cc}  \emp  & \emp  \\ \delta_3   & \emp \\ \eea     
	&
	\bea{cc}  \delta_2 & \emp  \\ \emp   & \emp  \\ \eea  \\
\end{array}
\right],
\een
where $\delta_1+\delta_2+\delta_2=\zhardy$, and dots denote other entries. Note that in the table, each square denotes a pair of inputs with the top left square indicating the setting $(0,0)$ for Alice and Bob respectively, the top right $(0,1)$, the bottom left $(1,0)$ and the bottom right $(1,1)$. Similarly, within each square, the four entries denote the four possible output combinations, with the top left being the output $(0,0)$ for Alice and Bob respectively, and so on to the bottom right denoting the output pair $(1,1)$. The entries of the table satisfy the 
constraints of no-signaling, and normalization. 
The no-signaling constraints imply that (i) for every row of the table, its left and right parts sum up to the same number, and  
(ii) the same for every column. The normalization constraints imply that the entries in each square sum up to 1 (see Fig. \ref{fig:ns-nor})
We shall now see, how $\bhardy>0$ certifies randomness in the Hardy output (ergo $\bhardy\leq 0$ is a Bell inequality). Namely we have

\begin{lemma}
	\label{lem:boundpHsimple}
	All no-signaling boxes with 2 binary inputs for Alice and Bob satisfy
	\be
	\label{eq:boundpH-1}
	\bhardy \leq \phardy \leq 1- \bhardy.
	\ee
	In addition 
	\be
	\label{eq:boundzH-1}
	\zhardy\leq 1 - 2\bhardy.
	\ee
\end{lemma}
\begin{proof}
	To prove the lemma, we need the following proposition
	
	\begin{prop}
		\label{prop:upperboundnoisy2}
		No-signaling boxes with 2 binary inputs for Alice and Bob satisfy 
		\be
		\phardy\leq \frac12+ \frac12 z_H. 
		\ee
		Moreover, the bound is tight, i.e. there exists an explicit no-signaling box, which saturates the bound.
	\end{prop}

	\begin{proof} 
		We shall consider the chain of inequalities implied by no-signaling,  which will "propagate" $\phardy$ towards the bottom right square, and then use the normalization constraint for this square. The inequalities are depicted by means of arrows in Fig. \ref{fig:phns}.
		\begin{figure}[bth]
			\centering
			\includegraphics[width=0.2\textwidth]{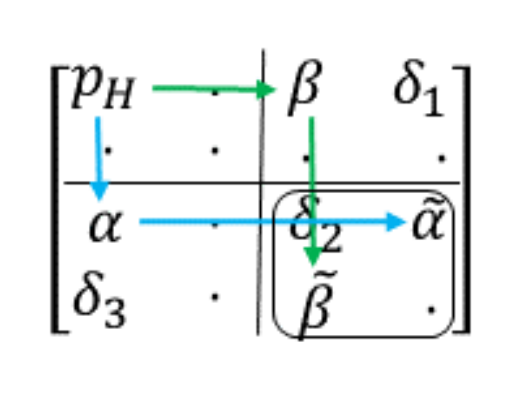}
			\caption{\label{fig:phns} The "propagation" in no-signaling box: arrows go from smaller to larger quantities (up to terms like $\delta_i$).}
		\end{figure}
		The arrows all point from a smaller entry to a larger entry (with the difference being an appropriate $\delta_i > 0$). 
		For instance, the horizontal green arrow denotes that $\phardy\leq \beta +\delta_1$, and the vertical green arrow denotes $\beta\leq \tilde\beta+\delta_2$ (from the no-signaling constraints). Similarly, the vertical blue arrow indicates that $\phardy \leq \alpha + \delta_3$ and the horizontal blue arrow indicates that $\alpha \leq \tilde\alpha + \delta_2$.
		Altogether, we obtain:
		\be
		\phardy \leq \tilde\beta +\delta_1+\delta_2,\quad  \phardy \leq \tilde\alpha +\delta_2+\delta_3
		\ee
		The normalization constraint gives 
		\be
		\tilde\alpha +\tilde\beta+\delta_2\leq 1
		\ee
		Adding the former two inequalities and then substituting the latter, we obtain 
		\be
		\phardy\leq \frac12 + \frac12 \zhardy,
		\ee
		where recall that $\zhardy = \delta_1 + \delta_2 + \delta_3$.
	\end{proof}
	
	Now, in Eq. (\ref{eq:boundpH-1}), the lower bound for $\phardy$ ($\phardy \geq \bhardy$)  is obvious, since $\zhardy \geq 0$ being the sum of probabilities. The upper bound ($\phardy \leq 1 - \bhardy$) we obtain by inserting 
	$\zhardy = \phardy -\bhardy$ into the inequality  from Proposition \ref{prop:upperboundnoisy2}:
	\be
	\phardy\leq \frac12+ \frac12 \zhardy,
	\ee
	giving 
	\be 
	\label{eq:ph-uppb}
	\phardy \leq 1 - \bhardy.
	\ee
	Finally, we obtain the third inequality (\ref{eq:boundzH-1}) of the Lemma as
	\be
	\zhardy=\phardy - \bhardy \leq 1 - 2\bhardy,
	\ee
	using Eq.(\ref{eq:ph-uppb}).
	The following explicit no-signaling box shows that the upper bounds are tight:
	\ben
	\left[
	\begin{array}{c|c}
		\bea{cc}  \frac{1+z_H}{2} & 0  \\  0  & \frac{1-	z_H}{2}  \\ \eea   
		& 
		\bea{cc} \frac12 & \frac{z_H}{2}  \\ 0   & \frac{1-	z_H}{2}  \\ \eea  
		\\
		\hline
		\bea{cc}  \frac12  & 0 \\ \frac{z_H}{2}    & \frac{1-	z_H}{2} \\ \eea     
		&
		\bea{cc}  0 & \frac12 \\ \frac12   & 0  \\ \eea  \\
	\end{array}
	\right]
	\een
	It is easy to verify by inspection that the box is no-signaling and satisfies the constraint $\zhardy=\sum_{(a,b,x,y)\in S_0} P_{A,B|X,Y}(a,b|x,y)$. Moreover, the box satisfies $\phardy = \frac{1+z_H}{2}$, and $\bhardy = \frac{1-z_H}{2} = 1 - \phardy$, so that $z_H = 1 - 2 \bhardy$.
\end{proof}

Let us remark here that the  inequality \eqref{eq:boundpH-1} would already be enough to obtain true random bits, 
using  the protocol of \cite{RBHH+15}. To elaborate, the parties would perform an estimation procedure and infer the value of the Bell parameter $\bhardy$ and use \eqref{eq:boundpH-1} to bound the Hardy probability $P_{A,B|X,Y}(a_H, b_H|x_H,y_H)$ from above and below (the bounds being valid in all no-signaling theories). Yet, since the bounds are only proven for $\phardy$, we would obtain randomness from only a single setting pair, meaning that we would require the boxes to have good Bell value in a sufficiently large number of runs where this setting appears, leading to low noise-tolerance.  (We will return to this point also at the end of section \ref{subsec:MDL-ineq}).

To overcome this problem and obtain a good level of noise-tolerance, we shall now prove something much stronger in
 Prop. \ref{prop:bounds-all}. Namely for every setting pair, there is an output whose probability 
is bounded in the same way as $\phardy$  (we will actually show more, by providing bounds for all outputs).

\begin{prop}
	\label{prop:bounds-all}
	Consider the $(2,2,2)$ Bell scenario with $\phardy=P_{A,B|X,Y}(a_H,b_H|x_H,y_H)$, and $\zhardy=\sum_{(x,y,a,b)\in S_0} P_{A,B|X,Y}(a,b|x,y)$ where $S_0 = \{(0,1,0,1),(1,0,1,0),(0,0,1,1)\}$ and $\bhardy=\phardy-\zhardy$. 
	Then, assuming the no-signaling conditions hold, 
	\bei
	\item For any $(a,b,x,y)$ such that $a\oplus b=xy$ we have that
	\be
	\label{eq:boundPR}
	\bhardy \leq P_{A,B|X,Y}(a,b|x,y)\leq 1-\bhardy
	\ee
	\item For all other $(a,b,x,y)$ (i.e. such that  $a\oplus b\not=xy$ we have 
	\be
	P_{A,B|X,Y}(a,b|xy)\leq 1 - 2\bhardy
	\ee	
	\eei
\end{prop}

\begin{proof} 
	Consider a no-signaling box with the values as given in the following table:
	\ben
	\left[
	\begin{array}{c|c}
		\bea{cc}  p_H & x  \\ x'   & f  \\ \eea   
		& 
		\bea{cc} c & \delta_1  \\ s   & c'  \\ \eea  
		\\
		\hline
		\bea{cc}  e  & s'  \\ \delta_3   & e' \\ \eea     
		&
		\bea{cc}  \delta_2 & d'  \\ d   & t  \\ \eea  \\
	\end{array}
	\right]
	\een
	We now want to get bounds on all the elements in terms of $\bhardy$.
	We shall  consider a chain of inequalities, obtained by  "propagating" within the table of 
	conditional probabilities,using the no-signaling and normalization constraints shown in Fig. \ref{fig:ns-nor}.
	The arrows in Fig. \ref{fig:bluegreens}a
	\begin{figure}[bth]
		\centering
		\includegraphics[width=0.5\textwidth]{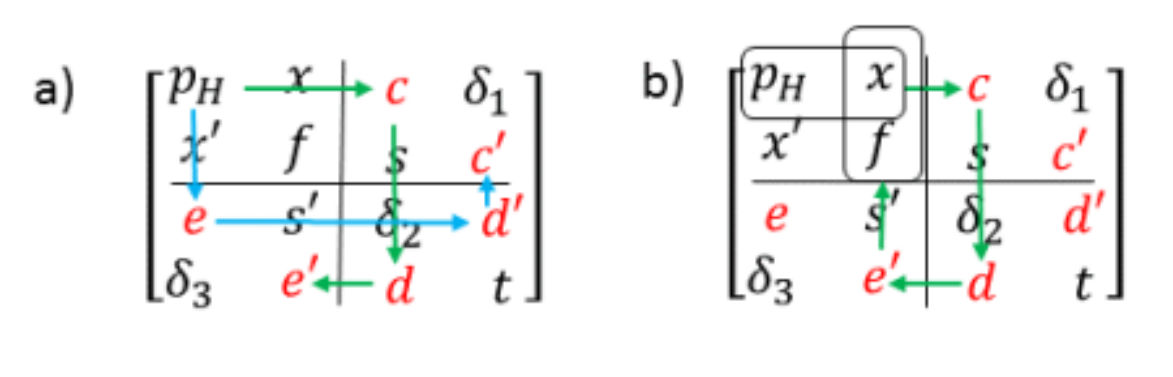}
		\caption{\label{fig:bluegreens} Chains of inequalities for upper bound for entries satisfying PR-box condition. }
	\end{figure}
	all point from a smaller entry to a larger entry (with the difference between the entries given by the appropriate $\delta_i$). For example, the top horizontal green arrow and the left vertical blue arrow denote the following two inequalities (obtained simply from no-signaling)
	\be
	c \geq \phardy-\delta_1,\quad e\geq \phardy- \delta_3
	\ee
	Then, using $d \geq c - \delta_2$ and $d' \geq e - \delta_2$,  we obtain
	\be
	d \geq \phardy  - \delta_1 - \delta_2, \quad d' \geq \phardy - \delta_2 - \delta_3
	\ee
	Finally, using $e' \geq d - \delta_3$ and $c' \geq d' - \delta_1$, we obtain 
	\be
	e' \geq \phardy - \zhardy, \quad c'\geq \phardy 	- \zhardy.
	\ee	
	where we recall that  $\zhardy=\delta_1+\delta_2+\delta_3$.
	Then since by definition $\bhardy = \phardy - z_H$, we have that the six entries \textit{(c,c',d,d',e,e')} are all bounded from below by $\bhardy$.
	Having already shown that $\phardy \geq \bhardy$ in Lemma \ref{lem:boundpHsimple}, the last quantity from \eqref{eq:boundPR} which we want to bound from below is $f$. The needed chain of inequalities is given in Fig. \ref{fig:bluegreens}b,
	where again the arrows point from smaller (up to appropriate $\delta$'s) to larger values. Explicitly, we have that $\phardy + x = c + \delta_1$, $d \geq c - \delta_2$, $e' \geq d - \delta_3$ and $f \geq e' - x$, all together giving $f \geq \phardy - z_H = \bhardy$.

	Let us now bound the entries satisfying $a \oplus b = x \cdot y$ from above. First we have from normalization
	\be
	c+c'+\delta_1\leq 1
	\ee
	This combined with the lower bounds for $c$ and $c'$ derived above gives 
	\be
	c\leq 1 - \phardy +\delta_2+\delta_3,\quad c'\leq 1-\phardy
	\ee
	From this we get $c,c'\leq 1-\bhardy$. Similarly we proceed with the pair $e,e'$ and obtain 
	\be
	e \leq 1 -\phardy + \delta_1 + \delta_2, \quad e'\leq 1 - \phardy.
	\ee
	Finally, for the pair $d,d'$ we obtain
	\be
	d\leq 1- \phardy - \delta_3 \quad, d'\leq 1- \phardy + \delta_1
	\ee
	Thus all these six elements are bounded from above by $1-\bhardy$.
	Regarding $f$, by the normalization constraint $f+x+x'+\phardy=1$ we get that $f\leq 1-\phardy$ so that also $f\leq 1 -\bhardy$. We are thus done with all entries satisfying $a\oplus b=xy$, i.e., we have shown \eqref{eq:boundPR}.
	
	Let us now consider the entries with $a\oplus b \not=x \cdot y$.
	For the entries $\delta_i$ we have simply that $\delta_i\leq \zhardy$, and the bound for $\zhardy$ is given by Lemma \ref{lem:boundpHsimple}.
	For the other entries, we obtain 
	the needed estimate directly (i.e. without chains of inequalities), using either no-signaling or normalization,
	see Fig. \ref{fig:nsnor} for details.
	\begin{figure}[bth]
		\centering
		\includegraphics[width=0.5\textwidth]{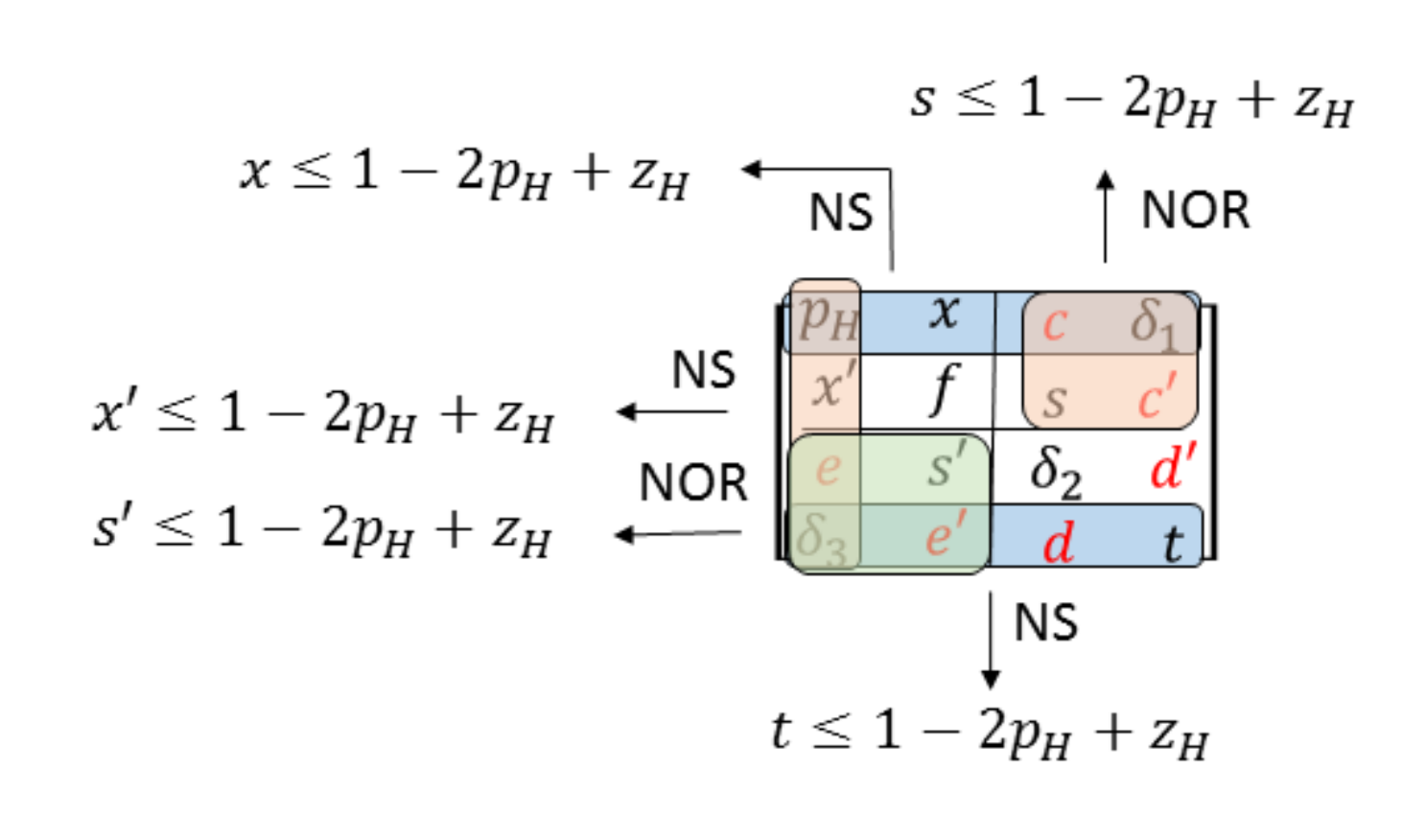}
		\caption{\label{fig:nsnor}Inequalities to get upper bound for entries satisfying PR-box condition.}
	\end{figure}	
\end{proof}

{\it Measured value versus true value of $\bhardy$.}
As we have seen, each of the entries $P_{A,B|X,Y}$ in the $(2,2,2)$ Bell scenario can be bounded (both from above and from below) in terms of the Hardy Bell parameter $\bhardy$. 
The quantity $\bhardy$ is not directly accessible in our protocol however, as our inputs $(x,y)$ are not uniformly distributed, but are rather drawn from an SV-source. We recall that the latter satisfies
\be
\pmin\leq  \nu(x,y)\leq \pmax
\ee
where $\nu$ is distribution of SV source, and 
\be
\pmax=\left(\frac12+\epsilon\right)^{2}, \quad   \pmin=\left(\frac12-\epsilon\right)^{2}
\ee
where the exponent $2$  is due to the fact that two bits are needed to  choose the two inputs $x,y \in \{0,1\}$ for Alice and Bob. Therefore the quantity that will be directly measured will be a Bell parameter 
with tilted weights giving rise to the {\it measurement dependent 
	locality} inequality introduced by P\"utz et al \cite{PRBL+14}
\ben
\bar{B}^{\text{SV}}_{\epsilon} &=& \pmin \nu_{SV}(x_H,y_H)P_{A,B|X,Y}(a_H,b_H|x_H,y_H) -\nonumber \\
&&\pmax  \sum_{(a,b,x,y)_\in S_0} \nu_{SV}(x,y) P_{A,B|X,Y}(a,b|x,y). \nonumber \\
\een 
Let us now show, that this quantity is in a simple way related to $\bhardy$.  
The following holds
\begin{lemma}
	\label{lem:pzBsv}
	We have 
	\ben
	\label{eq:pzBsv}
	\bhardy \geq \frac{\bar{B}^{\text{SV}}_{\epsilon}}{\pmin \pmax},
	\een
\end{lemma} 
\begin{proof}
	We have 
	\ben
	\label{eq:boundBSV}
	\bar{B}^{\text{SV}}_{\epsilon} &&= \pmin \nu_{SV}(x_H,y_H) P_{A,B|X,Y}(a_H,b_H|x_H,y_H)   \nonumber \\
	&&  \; \; \;- \pmax \sum_{(x,y,a,b)\in S_0}\nu_{SV}(x,y) P_{A,B|X,Y}(a,b|x,y) \nonumber\\
	&&  \leq \pmin\pmax \phardy - \pmax \pmin \zhardy = \pmin \pmax \bhardy. \nonumber \\
	\een
\end{proof}
We have thus arrived at the main result of this section
\begin{prop}
	\label{prop:svphbound}
	For an arbitrary no-signaling box $P_{A,B|X,Y}(a,b|x,y)$  with $2$ binary inputs for Alice and Bob we have 
	\be
	\frac{\bar{B}^{\text{SV}}_{\epsilon}}{\left(\frac14 - \epsilon^2\right)^2}\leq 
	P_{A,B|X,Y}(a,b|x,y)\leq 1-\frac{\bar{B}^{\text{SV}}_{\epsilon}}{\left(\frac14 - \epsilon^2\right)^2}.
	\ee
	for (a,b,x,y) satisfying  $a\oplus b=xy$ and 
	\be
	P_{A,B|X,Y}(a,b|x,y)\leq 1-\frac{\bar{B}^{\text{SV}}_{\epsilon}}{\left(\frac14 - \epsilon^2\right)^2}.
	\ee
	for the other entries. 
\end{prop}
The above proposition directly implies the lemma \ref{lem:lin-prog-bound}. We have thus shown in explicit fashion the bound on the probabilities in this simple $(2,2,2)$ scenario. Note that similar bounds were obtained using a simple, although not as explicit, linear programming duality relation in previous works \cite{BRGH+16, RBHH+15, GMTD+13}.

\subsection{Measurement-dependent locality inequalities.}
\label{subsec:MDL-ineq}
The Bell inequality tested in a device-independent randomness amplification protocol against no-signaling adversaries is required to obey specific properties: (i) it should not be possible to simulate the quantum value of the Bell expression using classical strategies, under any bias $\epsilon$ of the source; 
(ii) the Bell test must certify randomness against a no-signaling adversary, i.e., there must exist a hashing function $f: |\textbf{A}| \times |\textbf{B}| \times |\textbf{X}| \times |\textbf{Y}| \rightarrow \{0,1\}$ which when applied to any no-signaling box $P(\textbf{a},\textbf{b}| \textbf{x}, \textbf{y})$ that achieves the quantum value, obeys $P(f(\textbf{a}, \textbf{b}, \textbf{x}, \textbf{y}) | \textbf{x}, \textbf{y}) < 1$. 

In a previous work \cite{RBHH+15}, we showed how the randomness certified by the min-entropy condition (ii) above can be extracted using a device-independent protocol. One of the key techniques in that protocol was the introduction of a second test which helped to certify that $0< P(f(\textbf{a}, \textbf{b}, \textbf{x}, \textbf{y}) | \textbf{x}, \textbf{y})$ in the boxes used in a linear fraction of the runs. In the overlapping fraction of runs in which both certification procedures enable the guarantee that $0 < P(f(\textbf{a}, \textbf{b}, \textbf{x}, \textbf{y}) | \textbf{x}, \textbf{y}) < 1$, the bit $\{f(\textbf{a}, \textbf{b}, \textbf{x}, \textbf{y}), f(\textbf{a}, \textbf{b}, \textbf{x}, \textbf{y})  \oplus 1\}$ is partially random. While this procedure is effective in achieving its aim, the noise-tolerance in the resulting protocol was found to be too low to be implementable in current experiments. To remedy this defect, in this paper we employ an alternative approach to extract the randomness guaranteed by Bell tests satisfying (i) and (ii), through the measurement-dependent locality (MDL) inequalities introduced by P\"utz et al in \cite{PRBL+14}.


The MDL inequalities were designed to certify quantum non-locality in the situation of limited measurement-independence, i.e., to satisfy condition (i) above. 
In the (2,2,2) scenario, each party chooses their input $x, y$ with a single bit from the Santha-Vazirani source so that
\begin{eqnarray}
\left(\frac{1}{2} - \epsilon \right)^2 \leq P_{X,Y}(x,y) \leq \left(\frac{1}{2} + \epsilon \right)^2.
\end{eqnarray}  
The simplest MDL inequality then states that all classical correlations satisfy \cite{PRBL+14} 
\begin{widetext}
	\begin{eqnarray}
	\label{eq:MDL-2}
	\left( \frac{1}{2} - \epsilon \right)^2 P_{A,B,X,Y} (0,0,0,0) -  \left( \frac{1}{2} + \epsilon \right)^2 \left[P_{A,B,X,Y} (0,1,0,1) + \nonumber  P_{A,B,X,Y} (1,0,1,0) + P_{A,B,X,Y}(0,0,1,1) \right] \leq 0.
	\end{eqnarray}
\end{widetext}
Recall that by the Hardy constraints in this scenario, any classical box for which $P_{A,B|X,Y}(0,0|0,0) > 0$ satisfies that at least one of the probabilities $P_{A,B|X,Y}(0,1|0,1)$, $P_{A,B|X,Y}(1,0|1,0)$ and $P_{A,B|X,Y}(0,0|1,1)$ is also non-zero. This implies that even when the source generates input $(0,0)$ with probability $P_{X,Y}(0,0) = \left(\frac{1}{2} + \epsilon \right)^2$, the inequality cannot be violated by a classical box. An analogous inequality can be derived from a Hardy paradox for any number of inputs and outputs, simply as
\begin{widetext}
	\begin{eqnarray}
	\label{eq:MDL-gen}
	\left( \frac{1}{2} - \epsilon \right)^{|X||Y|} P_{A,B,X,Y}(a^*,b^*,x^*,y^*) - \left( \frac{1}{2} + \epsilon \right)^{|X||Y|} \sum_{(a,b,x,y) \in \text{Hardy-constraints}} P_{A,B,X,Y}(a,b,x,y) \leq 0,
	\end{eqnarray}
\end{widetext}
where $(a^*,b^*,x^*,y^*)$ denotes as usual the Hardy input-output and the sum is over all $(a,b,x,y)$ for which the Hardy constraints impose $P_{A,B|X,Y}(a,b|x,y) = 0$. 
On the other hand, the corresponding quantum strategies satisfy the Hardy constraints and achieve $P_{A,B|X,Y}(a^*,b^*|x^*,y^*) > 0$, thereby violating the inequality. 

The inequality (\ref{eq:MDL-gen}) from any Hardy paradox satisfies both conditions (i) and (ii) with $f(a,b,x,y)$ defined by
\[ f(a,b,x,y) = \begin{cases*}
0 & if  $(a,b,x,y) = (a^*,b^*,x^*,y^*)$  \\
1 & otherwise
\end{cases*} \]
The advantage in testing the MDL inequality (\ref{eq:MDL-gen}) is that it is more noise-tolerant compared to the approach of testing multiple Bell estimators in \cite{RBHH+15}. The latter approach required multiple tests to be passed in a significant proportion of the runs each, in order to ensure a non-trivial overlapping set of ''good" runs where all tests succeed and the corresponding outputs are random.  

However, we can only use this advantage, if for each setting there is at least one output,
	whose probability is bounded from below and from above. We have shown that this is the case in the $(2,2,2)$ scenario. 
	For more general Hardy paradoxes, we would still need to use a multi-testing procedure, more specifically we employ the protocol of \cite{RBHH+15} re-stated for convenience in Fig. \ref{protocolsingle-2}. In this protocol, as stated there is an additional step 3 where the parties perform an additional test which ensures when passed that a sufficient number of runs were performed with 
	boxes that have randomness in their outputs. Specifically, they check that $\textit{S}_n(\textbf{a},\textbf{b}, \textbf{x}, \textbf{y}) := \frac{1}{n} \sum_{i=1}^{n} D(a_i, b_i, x_i, y_i) \geq \mu_1$ for fixed constant $\mu_1 > 0$. Here $D$ is an indicator with entries $D(a_i, b_i, x_i, y_i) = 1$ if $(x_i, y_i)$ correspond to an input pair with randomness in the outputs and $(a_i, b_i)$ is an output pair corresponding to bit value $0$ of this random bit, and $D(a_i, b_i, x_i, y_i) = 0$ otherwise. For instance, in the situation where the randomness is only present in the Hardy input, $D(a_i, b_i, x_i, y_i) = 1$ if $(a_i,b_i, x_i, y_i) = (a^*, b^*, x^*, y^*)$ (the Hardy input-output combination) and $D(a_i, b_i, x_i, y_i) = 0$ otherwise. Note that for generic $2 \times n$ Hardy paradoxes, we will show in Section \ref{subsec:rand-2xn} that randomness is obtainable in half the number of settings, so that the requirement of overlap between the set of runs with "good" Bell value and the runs where these settings appear is not very stringent and the procedure of \cite{RBHH+15} can be applied with larger noise tolerance.

	\begin{figure}
		\begin{protocol*}{Protocol II}
			\begin{enumerate}
				\item The $\epsilon$-SV source is used to choose the measurement settings $(x_i, y_i)$ for $n$ runs on the single device consisting of two components. The device produces output bits $(a_i, b_i)$ with $i \in \{1, \dots, n\}$.
				\item The parties perform an estimation of the violation of the Bell inequality in the  device by computing the empirical average $L_n^{\epsilon}(\textbf{a}, \textbf{b}, \textbf{x}, \textbf{y}) \defeq  \frac{1}{n} \sum_{i=1}^{n} w_i(\epsilon) B(a_i, b_i, x_i, y_i)$. The protocol is aborted unless $L_n^{\epsilon}(\textbf{a}, \textbf{b}, \textbf{x}, \textbf{y})  \geq \delta$ for fixed constant $\delta > 0$. 
				\item Conditioned on not aborting in the previous step, the parties subsequently check if $\textit{S}_n(\textbf{a},\textbf{b}, \textbf{x}, \textbf{y}) := \frac{1}{n} \sum_{i=1}^{n} D(a_i, b_i, x_i, y_i) \geq \mu_1$. The protocol is aborted if this condition is not met for fixed $\mu_1 > 0$.  
				\item Conditioned on not aborting in the previous steps, the parties apply an independent source extractor \cite{CG, one-bit-extr} to the sequence of outputs from the device and further $n$ bits from the SV source.
			\end{enumerate}
		\end{protocol*}
		\caption{Protocol from \cite{RBHH+15} for device-independent randomness amplification from a single device with two no-signaling components.}
		\label{protocolsingle-2}
	\end{figure}


\subsection{Remarks}
Let us note, that we have actually not used apriori the fact that $\bhardy\leq 0$ is a Bell inequality. 
Rather we showed that it implies randomness, which is impossible for classical boxes indicating that it must be a Bell inequality. 
Also, one may be puzzled as to why showing that the probabilities are strictly bounded from above and below implies true randomness.
Suppose that we do not just estimate $\bhardy$ but know all the entries, and they are all $1/4$, then do we have true randomness? Clearly not necessarily, because this may just be a mixture of deterministic boxes. 
So why does the bound of probabilities from above we have obtained give rise to certified randomness? 
The answer is that we obtained the bounds for probabilities by means of a {\it linear quantity}. 
We have shown that on one side of the hyperplane given by $\bhardy=0$ (the side determined by violation of the inequality) 
there are no deterministic boxes. And therefore, a box violating the inequality cannot be a mixture thereof,
hence the randomness must be of non-local origin. 
Of course this comment applies  also to previous works on randomness amplification. 

Finally, let us mention that  there is relation between $\bhardy$ and the CHSH inequality. The latter is given by
\be
B_{CHSH}=\sum_{a,b,x,y: a\oplus b = xy} P_{A,B|X,Y}(ab|xy)\leq 3
\ee 
Using linear programming one can find that 
\be
B_{CHSH}\geq 3 + 2 \bhardy,
\ee
so that $\bhardy > 0$ implies $B_{CHSH} > 3$. 
The reason why we have not used CHSH inequality, is that with the latter it is not possible
to certify randomness for all $\epsilon$, simply it is possible to simulate the violation of the CHSH inequality using classical boxes when inputs are chosen from an SV source, above some threshold value of $\epsilon$. 
%
In contrast, we have seen in lemma \ref{lem:pzBsv}, that the inequality $\bhardy\leq 0 $ is violated if and only if 
$\bsv\leq 0$ is violated.

\section{Appendix: Hardy paradoxes in randomness certification}
\label{sec:Hardy-vs-randomness}

\subsection{Preliminaria}
Here we treat more general Hardy paradoxes. As we have seen in the example with the simplest Hardy paradox, 
we had to bound the output probabilities from above and below. We shall now consider more general Hardy paradoxes
and show that they are also useful for randomness amplification.

Let us first revisit the original Hardy paradox. One distinguishes there 
a set of input-output combinations corresponding to the Hardy zero constraints $\{(0,1,0,1), (1,0,1,0), (0,0,1,1)\}$. We called this set $S_0$. 
Moreover one also considers a special Hardy input-output combination $(a_H,b_H,x_H,y_H)$ which in the (2,2,2) case was $(0,0,0,0)$. 
Recall that we introduced the notation 
\ben
\phardy &=& P_{A,B|X,Y}(a_H,b_H|x_H,y_H), \nonumber \\
\zhardy &=& \sum_{(a,b,x,y)\in S_0}P_{A,B|X,Y}(a,b|x,y)
\een

Now the Hardy paradox consists of two ingredients. First, the set $S_0$ is chosen in such a way, 
that for classical boxes $\zhardy=0$ implies $\phardy=0$. 
Then one finds a quantum state, and measurements, such that $\zhardy=0$, but $\phardy>0$. Clearly, if the two ingredients are met, then the so-obtained quantum box cannot be classical, and it violates local realism. This alone is not enough 
for randomness amplification against general no-signaling adversaries though. In the previous section we had to show, that the Hardy output is random, i.e. bounded away from 
zero and one for all no-signaling boxes exhibiting the paradox. We have actually shown more using the no-signaling constraints, but just having the lower and upper bound 
for Hardy output is enough to amplify randomness of an $\epsilon$-SV source  for arbitrary $\epsilon$. 

We are now interested in the question, whether every Hardy paradox, existing or to-be discovered in the future, can be used 
for randomness amplification. This gives rise to the following basic question:  

{\it Consider a set $S_0$ of input-output combinations. Suppose that setting the probabilities $P_{A,B|X,Y}(a,b|x,y)$ to be zero for these combinations imposes that the probability $P_{A,B|X,Y}(a_H,b_H|x_H,y_H)$ for some distinguished $s_H=(a_H,b_H,x_H,y_H)$ is also zero for all classical boxes. Do the no-signaling constraints then imply that $P_{A,B|X,Y}(a_H,b_H|x_H,y_H)$ is bounded away from $1$ whenever $P_{A,B|X,Y}(a_H,b_H|x_H,y_H) > 0$ ?}
%

We may also consider the following noisy version of this question:

{\it Consider a set $S_0$ of input-output combinations. Suppose that setting the probabilities $P_{A,B|X,Y}(a,b|x,y)$ to be zero for these combinations imposes that the probability $\phardy = P_{A,B|X,Y}(a_H,b_H|x_H,y_H)$ for some distinguished $s_H=(a_H,b_H,x_H,y_H)$ is also zero for all classical boxes. Consider the quantity 
\be
\bhardy= \phardy - \zhardy
\ee
with 
\be
\zhardy=\sum_{(a,b,x,y)\in S_0} P_{A,B|X,Y}(a,b,|x,y)   
\ee
Do the no-signaling constraints imply that  $P_{A,B|X,Y}(a_H,b_H|x_H,y_H)$ is bounded away from  0 and 1 whenever $\bhardy>0$ ?} 

Note, that if for any given $S_0$ and $s_H$ the answer to the above question is positive, then we may construct a randomness amplification scheme
for arbitrary $\epsilon$ using this paradox. Indeed, this follows from Lemma \ref{lem:pzBsv}, which shows that for arbitrary $\epsilon>0$,
when the $\bhardy>0$ then also the observed value $\bsv>0$. 
We can then use Protocol II (Fig. \ref{protocolsingle-2}) to obtain certified randomness. 

\subsection{Hardy frames.}
Before we approach our question we introduce the notion of Hardy frames, which encode the set $S_0$ and the entry $\phardy$.
Consider a table of conditional probabilities $\{P_{A,B|X,Y}(a,b|x,y)\}$. Suppose that it is only 
partially filled: with some entries being set to zero (Hardy zeros),
and with the left upper corner being $\phardy$ - the probability of the Hardy output. The zeros must be in positions such
that any classical box which satisfies these zeros must obey $\phardy=0$.  Then we call such a table a {\it Hardy frame}.

To exhibit a {\it Hardy paradox}, a box must be compatible with some Hardy frame (i.e. has zeros in the places where 
the frame has zeros),  with  nonzero $\phardy$. To exhibit   a {\it noisy version} of Hardy paradox,  
the box must satisfy
$\phardy-\zhardy>0$ where $\zhardy$ is the sum of the entries in the box, 
for which the Hardy frame prescribes zeros.  One can see, that  $\phardy - \zhardy>0$ implies that the box must be non-classical.



Let us note that sometimes the zero constraints do not even allow for the existence of a classical box that satisfies these constraints.
For instance, in the CHSH Bell scenario, if we ask for zeros in the positions satisfying $a\oplus b\not =x \cdot y$, the only box compatible with such zeros is the Popescu Rochrlich box \cite{PR94}:
\ben
\left[
\begin{array}{c|c}
	\bea{cc}  \frac12 & 0  \\ 0   & \frac12  \\ \eea   
	& 
	\bea{cc}  \frac12 &  0    \\ 0   & \frac12  \\ \eea  
	\\
	\hline
	\bea{cc}  \frac12  & 0  \\ 0 & \frac12 \\ \eea     
	&
	\bea{cc}  0 & \frac12  \\ \frac12   & 0  \\ \eea  \\
\end{array}
\right]
\een
This then is not a very useful frame because the only box compatible with it cannot be realized quantum mechanically, so that the frame gives rise to a Hardy paradox that cannot be implemented in a lab. 

Also, a Hardy frame can enforce $p_H=0$ for arbitrary no-signaling boxes (not only the classical ones). 
Then we call it a {\it trivial} Hardy frame, as it cannot produce a Hardy paradox, since the latter requires $\phardy>0$.
The following is an example of a trivial Hardy frame:
\ben
\left[
\begin{array}{c|c}
	\bea{cc}  p_H & \emp  \\ \emp   & \emp  \\ \eea   
	& 
	\bea{cc}  0 &  0    \\ \emp   & \emp  \\ \eea  
	\\
	\hline
	\bea{cc}  \emp  & \emp  \\ \emp & \emp \\ \eea     
	&
	\bea{cc}  \emp & \emp  \\ \emp   & \emp  \\ \eea  \\
\end{array}
\right]
\een
It is trivial, since it enforces $p_H=0$ for all boxes, including quantum and no-signaling ones. 
The Hardy frame corresponding to the original Hardy paradox is the following
\ben
\left[
\begin{array}{c|c}
	\bea{cc}  p_H & \emp  \\ \emp   & \emp  \\ \eea   
	& 
	\bea{cc} \emp & 0  \\ \emp   & \emp  \\ \eea     
	\\
	\hline
	\bea{cc}  \emp  & \emp  \\ 0   & \emp \\ \eea     
	&
	\bea{cc}  0 & \emp  \\ \emp   & \emp  \\ \eea  \\
\end{array}
\right]
\een
Some quantum boxes allow $p_H>0$, while being compatible with the above frame.
One may also create other similar Hardy frames, note that in order to enforce $\phardy = 0$ for every classical box, 
we need to set at least one $x$ to be zero in each of the rectangles below:
\ben
&&\left[
\begin{array}{c|c}
	\bea{cc}  p_H & \emp  \\ \emp   & \emp  \\ \eea   
	& 
	\bea{cc}  x & \emp  \\ \emp   & \emp  \\ \eea  
	\\
	\hline
	\bea{cc}  x & \emp  \\ \emp   & \emp \\ \eea     
	&
	\bea{cc}  x & \emp  \\ \emp   & \emp  \\ \eea  \\
\end{array}
\right]
\quad
\left[
\begin{array}{c|c}
	\bea{cc}  p_H & \emp  \\ \emp   & \emp  \\ \eea   
	& 
	\bea{cc}  \emp & x  \\ \emp   & \emp  \\ \eea  
	\\
	\hline
	\bea{cc}  x & \emp  \\ \emp   & \emp \\ \eea     
	&
	\bea{cc}  \emp & x  \\ \emp   & \emp  \\ \eea  \\
\end{array}
\right]
\nonumber \\
&&\left[
\begin{array}{c|c}
	\bea{cc}  p_H & \emp  \\ \emp   & \emp  \\ \eea   
	& 
	\bea{cc}  x & \emp  \\ \emp   & \emp \\ \eea  
	\\
	\hline
	\bea{cc}  \emp & \emp  \\ x   & \emp \\ \eea     
	&
	\bea{cc}  \emp &  \emp  \\ x   &  \emp  \\ \eea  \\
\end{array}
\right]
\quad
\left[
\begin{array}{c|c}
	\bea{cc}  p_H & \emp  \\ \emp   & \emp  \\ \eea   
	& 
	\bea{cc}  \emp & x  \\ \emp   & \emp  \\ \eea  
	\\
	\hline
	\bea{cc}  \emp & \emp  \\ x   & \emp \\ \eea     
	&
	\bea{cc}  \emp & \emp  \\ \emp   & x  \\ \eea  \\
\end{array}
\right]
\een

\subsection{Randomness in Hardy paradoxes with binary observables}
\label{subsec:rand-2outs}

We begin answering the question by showing that when the two parties measure binary observables, irrespective of the input sizes, the answer to the question is positive, i.e., $\phardy < 1$ if $\phardy > 0$ for all no-signaling boxes. 
\begin{prop}
	\label{prop:bin-out-Hardy}
	Let $G$ denote a two-party Hardy paradox, $a,b \in \mathcal{A}, \mathcal{B}$ the outputs of two parties and $x,y \in \mathcal{X}, \mathcal{Y}$ the corresponding inputs.  
	Let $\phardy = P_{A,B|X,Y}(a_H,b_H|x_H,y_H)$ denote the maximum value of the non-zero Hardy probability in general no-signaling theories. Then, if $|\mathcal{A}| = |\mathcal{B}| = 2$, we have that $P_{A,B|X,Y}(a_H,b_H|x_H,y_H) < 1$.  
\end{prop}

\begin{proof}
	Suppose to the contrary that there exist two-party Hardy paradoxes with binary outputs such that classically $P_{A,B|X,Y}(a_H,b_H|x_H,y_H) = 0$ and $P_{A,B|X,Y}(a_H,b_H|x_H,y_H) = 1$ in general no-signaling theories. Let $G$ denote any Hardy paradox with this property, with $|\mathcal{A}|, |\mathcal{B}|$ inputs for the two parties, i.e., $G$ belongs to the Bell scenario (($|\mathcal{A}|,2$),($|\mathcal{B}|,2$)). 
	Let $C_{\text{G}}$ denote the set of events $(a,b,x,y)$ which form the Hardy constraints $S_0$ in $G$, i.e., for which $G$ imposes the Hardy constraint $P_{A,B|X,Y}(a,b|x,y) = 0$.  We will  construct from $G$ another Hardy paradox $G'$ that shares the same property, but with one less input for either Alice or Bob. The fact that there do not exist Hardy paradoxes with this property in the CHSH Bell scenario with $|\mathcal{A}| = |\mathcal{B}| = 2$ then completes the proof. Notice that the latter can be readily seen by the fact that only one non-local no-signaling vertex, namely the PR-box exists in the CHSH Bell scenario, and this has only entries $0$ and $\frac{1}{2}$.
	
	Let $V_{\text{ns}}(G)$ denote the set of vertices of the no-signaling polytope for the Bell scenario (($|\mathcal{A}|,2$),($|\mathcal{B}|,2$))  that satisfy all the Hardy constraints $S_0$ and obey $P_{A,B|X,Y}(a_H,b_H|x_H,y_H) = 1$. Similarly, let $V_{\text{c}}(G)$ denote the set of vertices of the classical polytope for this Bell scenario that satisfy $S_0$ and obey $P_{A,B|X,Y}(a_H,b_H|x_H,y_H) = 0$. We pick one non-local no-signaling vertex $Q_{\text{ns}}$ from $V_{\text{ns}}(G)$ and an adjacent local vertex $Q_{\text{c}}$ from $V_{\text{c}}(G)$. We construct a new Hardy paradox $G'$ with part of the new Hardy constraints $S'_0$ given by the set of events $(a,b,x,y)$ with $x \neq x_H, y \neq y_H$ for which $Q_{\text{c}}(a,b|x,y) = 0$ and $Q_{\text{ns}}(a,b|x,y) = 0$.
	Note that these constraints partly define the edge between the two vertices $Q_{\text{c}}$ and $Q_{\text{ns}}$ and contain as a subset the events from $S_0$ for the corresponding inputs. In other words, none of the other boxes from $V_{\text{ns}}$ and $V_{\text{c}}$ satisfy the constraints above. Adding a single event from $S_0$ for which $x = x_H$ or $y = y_H$ then gives the set of Hardy constraints $S'_0$ for $G'$. More precisely, we consider the case that $x = x_H$, and choose an event $(a_H,\tilde{b},x_H,\tilde{y})$ such that $Q_{\text{c}}(b|y) = 1$. We now obtain the set of events which define the Hardy constraints $S'_0$ for $G'$, i.e., 
	\begin{eqnarray}
	S'_0 := &&\{(a,b|x,y) : x \neq x_H, y \neq y_H, \nonumber \\ && Q_{\text{c}}(a,b|x,y) = 0 \vee Q_{\text{ns}}(a,b|x,y) = 0\} \bigcup \nonumber \\ &&\{ (a_H,\tilde{b},x_H,\tilde{y}) : (a_H,\tilde{b},x_H,\tilde{y}) \in C_{\text{G}} \}.
	\end{eqnarray} 
	Evidently, $Q_{\text{ns}}$ satisfies the constraints in $S'_0$ with $Q_{\text{ns}}(a_H,b_H|x_H,y_H) = 1$ and there exists a classical box $Q'_{\text{c}}$ that satisfies the constraints with $Q'_{\text{c}}(a_H,b_H|x_H,y_H) = 0$. However, this also implies $Q_{\text{ns}}(a_H, \tilde{b} \oplus 1 | x_H, \tilde{y}) = 1$ and $Q'_{\text{c}}(a_H,\tilde{b} \oplus 1 | x_H, \tilde{y}) = 0$. 
	
	We now have the Hardy paradox consisting of the constraints $P(a,b|x,y) = 0$ for $(a,b,x,y) \in S'_0$ and with inputs $x \in \mathcal{X}$, $y \in \mathcal{Y} \setminus y_H$. And all classical boxes that satisfy the constraints in $S'_0$ obey $Q'_{\text{c}}(a_H,\tilde{b} \oplus 1 | x_H, \tilde{y}) = 0$ while there exists a no-signaling box $Q_{\text{ns}}(a_H, \tilde{b} \oplus 1 | x_H, \tilde{y}) = 1$, where in these instances we consider the boxes $Q'_{\text{c}}$ and $Q_{\text{ns}}$ with input $y_H$ truncated. 
\end{proof}

\subsection{Randomness from $2\times n$ Hardy paradoxes.}
\label{subsec:rand-2xn}
Now we shall show that in the scenario with $2$ observables for Alice and $n$ for Bob (with arbitrary output sizes), a Hardy paradox always implies randomness. 

\subsubsection{Upper bounds for $\phardy$: noiseless case.}
\begin{prop}
\label{prop:upperbound1xn}
For arbitrary Hardy frame, scenario with $2$ observables for Alice and $n$ for Bob, any no-signaling box compatible with the frame 
has the probability $p_H$ of the Hardy output bounded as follows  $p_H\leq \frac{n-1}{n}$. The bound is tight, i.e. 
there exists explicit no-signaling box, which saturates the bound. 
\end{prop}

\begin{proof}
Let Alice and Bob observables be $A_0,A_1$ and $B_0,B_1,\ldots, B_n$ respectively, where $A_0,B_0$ is Hardy input. 
Consider all zeros of the first row of the matrix. 
In each block column labeled by $B_i$, $i\geq 1$, we permute columns in such a way, 
that the zeros are pushed most to the right. This divides each block column labeled by $B_i$
into two block columns: $B_i^{(1)}$ - gathering columns with zero in first row, and  
$B_i^{(0)}$ consisting of the rest columns from $B_i$. Note, that notation at the moment may seem a bit awkward - we 
named  $B_i^{(0)}$ those block columns that do {\it not} have zero in first row.

Consider now the first column of the matrix. Permute all rows in block row labeled by $A_1$, 
to push all zeros from the first column down. The block row, consisting of rows with zero in first column, 
we call $A_1^{(0)}$. 

The remaining entries of the block row  will be divided as follows. Consider one of the  rows. 
Clearly, its cross section with at least one block column $B_i^{(0)}$ must consist entirely of zeros. 
Otherwise, we could find a "rectangle" without zeros, and this would allow for a classical box to have nonzero Hardy output. 
We can now gather all rows, whose cross-section with $B_i^{(0)}$ is filled with zeros 
by permuting rows. By suitable permutation we thus obtain, that first group of rows has cross-section with $B_1^{(0)} $ 
filled with zeros, the second one with  $B_1^{(0)}$, and so on. These groups are blocks of rows, and we will denote them 
by $A_1^{(i)}$ subsequently. In other words, the block row  $A_1^{(i)}$ is the one whose cross-section with  $B_i^{(0)}$ is filled with zeros. 

Finally, we also divide the block row $A_0$ into $A_0^{(0)}$ being just the first row and  $A_0^{(1)}$ consisting of the rest of the rows.
Similarly $B_0$ is divided into $B_0^{(0)}$ being just the first column and $B_0^{(1)}$ consisting of the rest of the columns.
The resulting division into blocks of the total matrix is shown in Fig. \ref{fig:blockmatrix} (version with $\delta_i=\tilde \delta_i=0$
for all $i=1,\ldots,n-1$). 

\begin{figure}[bth]

\centering
\includegraphics[width=0.5\textwidth]{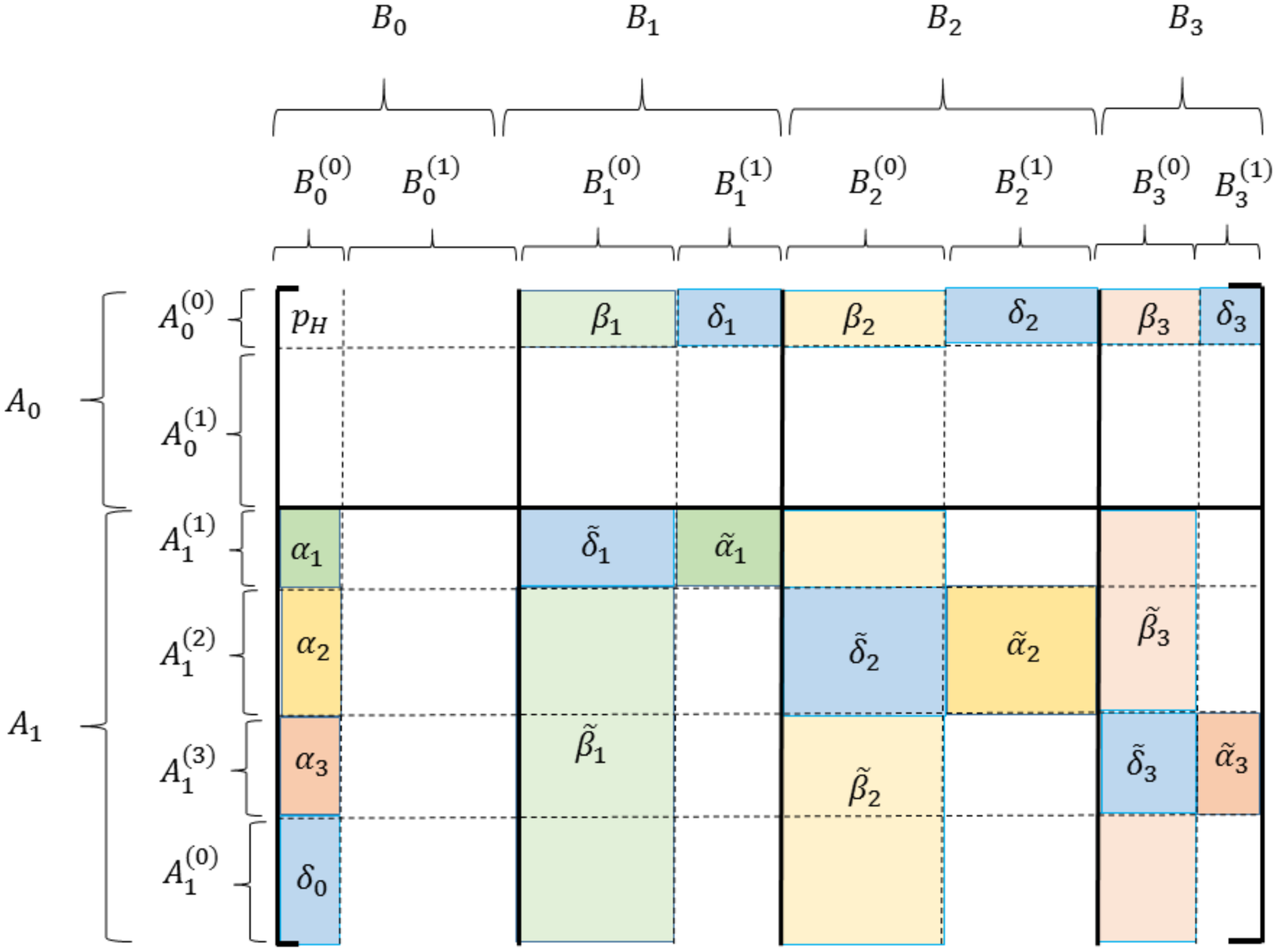}
\caption{\label{fig:blockmatrix} Hardy frame for $2$ Alice settings and $n$ Bob's settings for $n=4	$. 
All the Greek symbols, except of $\tilde\beta_i$ denote the sum of elements of the block they sit;
the $\tilde \beta_i's$ denote the sum of elements of the whole light green, light yellow and light orange areas, respectively, {\it excluding} the first row (i.e. excluding the blocks denoted by $\beta_i$'s).
For noiseless version, the blue blocks are filled with zeros, i.e. $\delta_i=0$ and $\tilde\delta_i=0$.}
\end{figure}
Here we list the relevant blocks of the matrix, together with sums of their elements. 
\bei
\item  $A_0^{(0)}\times B_0^{(0)}$  single element block $\sum=p_H$
\item  $A_0^{(0)}\times B_i^{(1)}$  matrix with one row $\sum=\beta_i$
\item  $\left((A_1^{(i)})^c/A_0\right)\times B_i^{(1)}$  rectangular matrix $\sum=\tilde\beta_i$
\item $A_1^{(i)}\times B_0^{(0)}$ matrix with one column $\sum=\alpha_i$ 
\item $A_1^{(i)}\times B_i^{(1)}$ rectangular matrix $\sum=\tilde\alpha_i$ 
\eei
where $i=1,\ldots,n-1$. Let us denote sum of the elements in the block $Y$ by $|Y|$.  We then have 
\ben
&& |A_0^{(0)}\times B_0^{(0)}|=p_H \nonumber \\
&& |A_0^{(0)}\times B_i^{(1)}|= \beta_i \nonumber \\
&& \left|\left((A_1^{(i)})^c/A_0\right)\times B_i^{(1)}\right|= \tilde\beta_i \nonumber \\
&& |A_1^{(i)}\times B_0^{(0)}|=\alpha_i \nonumber \\
&& |A_1^{(i)}\times B_i^{(1)}|= \tilde\alpha_i \nonumber \\
\een
where $i=1,\ldots,n-1$. 
We shall now consider inequalities obtained from nosignaling and normalization, in analogy to the case of 
$2\times2$. The inequalities implied by no-signaling will be determined by arrows in Fig. \ref{fig:2n-propagation},

\begin{figure}[bth]
\includegraphics[width=0.5\textwidth]{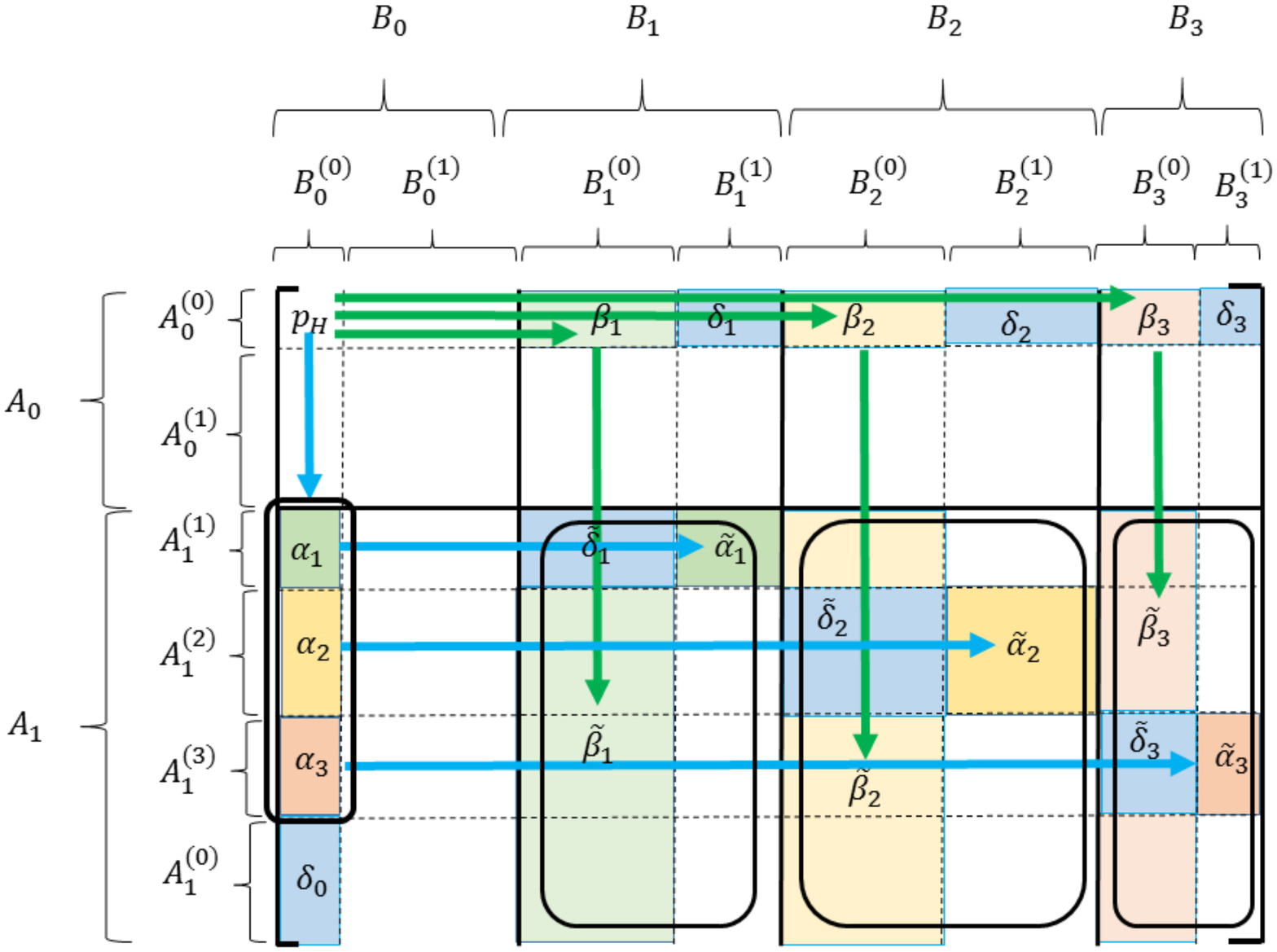}
\caption{\label{fig:2n-propagation} "Propagation" leading to inequalities used in proofs of Props. \ref{prop:upperbound1xn} and  \ref{prop:upperboundnoisy}. Arrow go from smaller to larger (up to $\delta$ quantities), the round rectangles mean normalization (within the rectangles the sum is equal to $1$.}
\end{figure}
where the rules are analogous to those in Fig. \ref{fig:phns}.  We get
\be
\label{eq:nshardy}
\tilde\beta_i\geq \beta_i\geq p_H, \quad \tilde\alpha_i\geq \alpha_i\quad \sum_{i=1}^{n-1}\alpha_i\geq p_H
\ee
On the other hand normalization gives
\ben
\label{eq:normhardy}
n-1=\sum_{i=1}^{n-1} |A_1\times B_i|\geq \sum_{i=1}^{n-1} |(A_i^{(1)})^c\times B_i^{(1)}|+ \nonumber \\
\sum_{i=1}^{n-1} \left|\left(A_1^{(i)})^c/A_0\right)\times B_i^{(0)}\right|=  \sum_{i=1}^{n-1} \tilde \alpha_i + \sum_{i=1}^{n-1} \tilde \beta_i
\een
Combining these we obtain
\be
n-1\geq  \sum_{i=1}^{n-1} \alpha_i + \sum_{i=1}^{n-1} \beta_i \geq p_H +(n-1) x = n p_H
\ee
hence $p_H\leq \frac{n-1}{n}$.
We shall now see that the bound is tight, in the Bell scenario $(2,n,n,n)$, i.e., with $2$ observables for Alice, $n$ for Bob, with $n$ outputs for each input. 
We first set the Hardy probability to be $\phardy = P_{X,Y|A,B}(0,0|0,0) = \frac{n-1}{n}$, together with 
$P_{X,Y|A,B}(1,n-1|0,0) = \frac{1}{n}$. The non-zero entries for the remaining setting pairs are as follows:
\begin{itemize}
	\item (i) $(A,B) = (0,k)$ for $k =1, \dots, n-1$. $P_{X,Y|A,B}(0,j|0,k) = \frac{1}{n}$ for $j = 0, \dots, n-2$, and $P_{X,Y|A,B}(1,n-1|0,k) = \frac{1}{n}$.	
	\item (ii) $(A,B) = (1,0)$. $P_{X,Y|A,B}(j,0|1,0) = \frac{1}{n}$ for $j = 0, \dots, n-2$, and $P_{X,Y|A,B}(n-1,n-1|1,0) = \frac{1}{n}$. 
	\item (iii) $(A,B) = (1,k)$ for $k=1, \dots, n-1$. $P_{X,Y|A,B}(j,n-k+j|1,k) = \frac{1}{n}$ for $j = 0, \dots, n-1$, where the addition is performed modulo $n$.  
\end{itemize}
Below, we show explicitly the box for $n=3$
\ben
\left[
\begin{array}{c|c|c}
	\bea{ccc}  \frac{2}{3} & 0 & 0  \\  0  & 0 &\frac{1}{3} \\ 0 & 0 & 0 \\ \eea   
	& 
	\bea{ccc} \frac{1}{3} & \frac{1}{3} & 0  \\ 0   & 0 & \frac{1}{3} \\ 0 & 0 & 0 \\ \eea & 	
	\bea{ccc} \frac{1}{3} & \frac{1}{3} & 0  \\ 0   & 0 & \frac{1}{3} \\ 0 & 0 & 0 \\ \eea  
	\\
	\hline
	\bea{ccc}  \frac{1}{3}  & 0 & 0 \\ \frac{1}{3}   & 0 & 0 \\ 0 & 0 & \frac{1}{3} \\ \eea     
	&
	\bea{ccc}  0  & 0 & \frac{1}{3} \\ \frac{1}{3}   & 0 & 0 \\ 0 & \frac{1}{3} & 0 \\ \eea  &
	\bea{ccc}  0 & \frac{1}{3} & 0 \\ 0  & 0  & \frac{1}{3} \\ \frac{1}{3} & 0 & 0 \eea  \\
\end{array}
\right]
\een
By inspection, the box is seen to be no-signaling with $P_{X|A}(0|0) = \frac{2}{3}, P_{X|A}(1|0) = \frac{1}{3}, P_{X|A}(0|1) = P_{X|A}(1|1) = P_{X|A}(2|1) = \frac{1}{3}$ and $P_{Y|B}(0|0) = \frac{2}{3}, P_{Y|B}(2|0) = \frac{1}{3}, P_{Y|B}(0|1) = P_{Y|B}(1|1) = P_{Y|B}(2|1) = \frac{1}{3}$, $P_{Y|B}(0|2) = P_{Y|B}(1|2) = P_{Y|B}(2|2) = \frac{1}{3}$.

\end{proof}

Thus, for $2$ inputs of Alice, arbitrary no-signaling box that is compatible with Hardy frame gives rise to no-signaling box, which has probability of Hardy output bounded away from 1.

\subsubsection{Upper bounds for $\phardy$: noisy case}

Let us now prove a noisy version of Proposition \ref{prop:upperbound1xn}. Namely, we now assume that the no-signaling box has,
instead of zeros from the Hardy frame, some nonzero elements, that sum up to $\zhardy$. 

\begin{prop}
\label{prop:upperboundnoisy}
Consider arbitrary Hardy frame, with scenario with $2$ observables for Alice and $n$ for Bob.
Consider any no-signaling box for which the entries in places where the frame has zeros, sum up to $z_H$. 
Then  we have $\phardy\leq \frac{n-1}{n}+ \frac1n z_H$. 
Moreover, the bound is tight, i.e. there exists explicit no-signaling box, which saturates the bound.
\end{prop}

\begin{proof}
Consider no-signaling box described in the Proposition. 
Such box is depicted in Fig. \ref{fig:blockmatrix}. In addition to the blocks listed in the proof of the previous lemma,
we have the following blocks:
\bei
\item $A_0^{(0)}\times B_1^{(i)}$ matrix with one row $\sum = \delta_i$
\item  $A_1^{(i)}\times B_i^{(0)}$ rectangular matrix $\sum = \tilde\delta_i$
\item $A_1^{(0)}\times B_0^{(0)}$ matrix with one column $\sum = \delta_0$.
\eei
Now, no-signaling gives rise to a modified version of \eqref{eq:nshardy}
\ben
&&\beta_i+\delta_i\geq p_H, \quad \tilde\beta_i + \tilde\delta_i \geq \beta_i \nonumber \\
&&\tilde\alpha_i+\tilde\delta_i\geq \alpha_i,\quad \sum_{i=1}^{n-1}\alpha_i+\delta_0\geq p_H.
\een
Normalization gives a version of \eqref{eq:normhardy}
\ben
n-1=\sum_{i=1}^{n-1} |A_1\times B_i|\geq \sum_{i=1}^{n-1} \tilde \alpha_i + \sum_{i=1}^{n-1} \tilde \beta_i +
\sum_{i=1}^{n-1} \tilde \delta_i.
\een
Combining the above inequalities, we obtain
\be
n-1  + \delta_0 +\sum_{i=1}^{n-1}(\tilde\delta_i 	+\delta_i) \geq (n-1) p_H+p_H.
\ee
Since by definition $z_H\geq  \delta_0 +\sum_{i=1}^{n-1}(\tilde\delta_i 	+\delta_i)$ we obtain
\be
p_H \leq \frac{n-1}{n} + \frac1n z_H. 
\ee
This ends the proof of the bound. To show tightness, consider the following construction.
We first set the Hardy probability $\phardy$ to be 
$p_{H} = P_{X,Y|A,B}(0,0|0,0) = \frac{n-1}{n} + \frac{1}{n} z_H$, 
together with $P_{X,Y|A,B}(1,n-1|0,0) = \frac{1}{n} - \frac{1}{n} z_H$. 
The non-zero entries for the remaining setting pairs are as follows:
\begin{itemize}
	\item (i) $(A,B) = (0,k)$ for $k =1, \dots, n-1$. $P_{X,Y|A,B}(0,j|0,k) = \frac{1}{n}$ for $j = 0, \dots, n-2$, $P_{X,Y|A,B}(1,n-1|0,k) = \frac{1}{n} - \frac{1}{n} \zhardy$, and $P_{X,Y|A,B}(0,n-1|0,k) = \frac{z_H}{n}$.

	\item (ii) $(A,B) = (1,0)$. $P_{X,Y|A,B}(j,0|1,0) = \frac{1}{n}$ for $j = 0, \dots, n-2$, $P_{X,Y|A,B}(n-1,n-1|1,0) = \frac{1-z_H}{n}$ and $P_{X,Y|A,B}(n-1,0|1,0) = \frac{\zhardy}{n} $.

	\item (iii) $(A,B) = (1,k)$ for $k=1, \dots, n-1$. $P_{X,Y|A,B}(j,n-k+j|1,k) = \frac{1}{n}$ for $j = 0, \dots, n-1$, where the addition is performed modulo $n$.  
	\end{itemize}
	\end{proof}
Below, we show explicitly the box for $n=3$
\ben
\left[
\begin{array}{c|c|c}
	\bea{ccc}  \frac{2+z_H}{3} & 0 & 0  \\  0  & 0 & \frac{1-z_H}{3} \\ 0 & 0 & 0 \\ \eea   
	& 
	\bea{ccc} \frac{1}{3} & \frac{1}{3} & \frac{z_H}{3}  \\ 0   & 0 & \frac{1-z_H}{3} \\ 0 & 0 & 0 \\ \eea & 	
	\bea{ccc} \frac{1}{3} & \frac{1}{3} & \frac{z_H}{3}  \\ 0   & 0 & \frac{1-z_H}{3} \\ 0 & 0 & 0 \\ \eea  
	\\
	\hline
	\bea{ccc}  \frac{1}{3}  & 0 & 0 \\ \frac{1}{3}   & 0 & 0 \\ \frac{z_H}{3} & 0 & \frac{1-z_H}{3} \\ \eea     
	&
	\bea{ccc}  0  & 0 & \frac{1}{3} \\ \frac{1}{3}   & 0 & 0 \\ 0 & \frac{1}{3} & 0 \\ \eea  &
	\bea{ccc}  0 & \frac{1}{3} & 0 \\ 0  & 0  & \frac{1}{3} \\ \frac{1}{3} & 0 & 0 \eea  \\
\end{array}
\right]
\een
By inspection, the box is seen to be no-signaling with $P_{X|A}(0|0) = \frac{2+z_H}{3}, P_{X|A}(1|0) = \frac{1-z_H}{3}, P_{X|A}(0|1) = P_{X|A}(1|1) = P_{X|A}(2|1) = \frac{1}{3}$ and $P_{Y|B}(0|0) = \frac{2+z_H}{3}, P_{Y|B}(2|0) = \frac{1-z_H}{3}, P_{Y|B}(0|1) = P_{Y|B}(1|1) = P_{Y|B}(2|1) = \frac{1}{3}$, $P_{Y|B}(0|2) = P_{Y|B}(1|2) = P_{Y|B}(2|2) = \frac{1}{3}$.

\subsubsection{Randomness}
For arbitrary Hardy paradox, we consider the same Bell inequality $\bsv$ as for the simplest one \eqref{eq:B-SV-Hardy}
but this does not change anything in the proof.  
Then lemma \ref{lem:pzBsv} still holds (proof almost the same)  with 
\be
\pmin=(\frac12-\epsilon)^{\log(2n)} , \quad \pmax= (\frac12+\epsilon)^{\log(2n)}.
\ee 
Thus when from estimation with a multi-testing procedure we know that $\bsv>0$ then also $\bhardy>0$. Thus we have  to show that 
$\bhardy>0$ guarantees randomness. 
Below we show that  for arbitrary $2\times n$ paradox, $\bhardy>0$  implies that $\phardy$ is bounded away from 0 and 1.

\begin{lemma}
\label{lem:boundpHsimple2xn}
Let $\phardy=p(a_H,b_H|x_H,y_H)$, and $\zhardy=\sum_{(x,y,a,b)\in S_0} p(a,b|x,y)$ where $S_0$ are the entries with zeros in Hardy frame.
Then, assuming no-signaling, 
for the case of $n$ Alice's observables and $2$ Bob's observables we have 
\be
\label{eq:boundpH}
\bhardy \leq \phardy \leq 1-\frac{1}{n-1} \bhardy
\ee
and 
\be
 \zhardy\leq 1 - \frac{n}{n-1}\bhardy
\ee
\end{lemma}
\begin{proof}
The first inequality of  Eq. \eqref{eq:boundpH} is obvious. 
Regarding the second one, we have from Proposition \ref{prop:upperboundnoisy}
\be
\phardy\leq \frac{n-1}{n}+ \frac{1}{n} \zhardy
\ee
By assumption $\zhardy\leq \bhardy -\phardy$. Inserting it into the above equation we obtain 
\be
\label{eq:phbh}
\phardy \leq 1 - \frac{1}{n-1}\bhardy.
\ee
The third inequality we get as  follows:
\be
\zhardy=\phardy - \bhardy \leq  1 - \frac{1}{n-1}\bhardy - \bhardy=1 - \frac{n}{n-1}\bhardy.
\ee
where the inequality comes from  \eqref{eq:phbh}.
\end{proof}

\subsubsection{Randomness from other inputs}
We have seen above that estimating $B_H \geq \delta > 0$ guarantees $\delta \leq p_H \leq 1 - \frac{\delta}{n-1}$ certifying randomness in the outputs for the input $(A_0,B_0)$. In particular, assigning bit value $0$ to the output pair $(0,0)$ and the bit value $1$ to the rest of the output pairs, one obtains a partially random bit $S$ with the property that $0 < P_{S|X,Y}(s=0|A_0,B_0) < 1$. Is it also possible to certify randomness in the other inputs? While this is not necessarily the case for all inputs, in what follows, we answer this question in the affirmative for all input pairs $(A_0,B_i)$ with $i = 1, \dots, n-1$. 

Specifically, we consider the noiseless case, and show that for each of the input pairs $A_0, B_i$, the value $\beta_i$ is strictly bounded below $1$. Let us consider first $(A_0, B_1)$ and show that $\beta_1 < 1$, the proof for the remaining $B_i$ will follow along analogous lines. The proof is by contradiction. Suppose that $\beta_1 = 1$, so that measuring $A_0$ returns a deterministic output $0$. This will contradict the property that the set of Hardy constraints constitute a proof of non-locality, due to the following fact.

\textbf{Fact:} \textit{Any no-signaling box with two observables for Alice and $n \geq 2$ observables for Bob, in which one of Alice's observables returns a deterministic output, must of necessity be local, i.e., admit a decomposition into a convex mixture of local deterministic boxes.} 

The above rather straightforward fact is just an extension of the observation that any no-signaling box with a single input for Alice and $n \geq 2$ inputs for Bob admits a convex decomposition into local deterministic boxes. To simulate such a box locally, Alice and Bob share a random variable $R$ with the same size and distribution as Alice's output alphabet for input $A_1$, i.e.,  $Q_R(r) = P_{A|X}(a|A_1)$. On input $A_1$, Alice outputs $a = r$ and on input $A_0$, Alice deterministically outputs $a= 0$, i.e., $P_{A|X,R}(a|x,r) = \delta_{a,r}$ when $x=1$ and $P_{A|X,R}(a|x,r) = \delta_{a,0}$ when $x=0$. Bob in turn knowing $R, Y$ outputs $P_{B|R,Y}(b|r,y)$ as per the distribution $\frac{P_{A,B|X,Y}(r,b|x=1,y)}{Q_R(r)}$. In both cases, we obtain the required local box $P_{A,B|X,Y}$ as
\begin{eqnarray}
P_{A,B|X,Y}(a,b|x,y) = \sum_r Q_R(r) P_{A|X,R}(a|x,r) P_{B|Y,R}(b|y,r). \nonumber
\end{eqnarray}

We thus obtain by contradiction that $\beta_1 < 1$ for any no-signaling box that satisfies the Hardy constraints (and is therefore non-local), and similarly $\beta_i < 1$ for all $i = 2, \dots, n$. Now, noting that $\beta_i > 0$ (which follows from $p_H > 0$ when $B_H > 0$), we may obtain a partially random bit. Specifically, assigning bit value $0$ to the outputs in $(A_0^{(0)}, B_i^{(0)})$, and bit value $1$ to the remaining outputs, we obtain partially random bits $S_i$ with the property that $0 < P_{S_i|X,Y}(s_i=0|A_0,B_i) < 1$. This allows for randomness to be obtained in about half the runs of the experiment, namely all the runs where Alice measures $A_0$, rather than restricting ourselves to the specific runs where the input pair $A_0, B_0$ was measured. \\

\subsection{Randomness in Hardy paradoxes beyond $2 \times  n$ scenarios}
\label{subsec:rand-beyond2xn}

We now proceed to Bell scenarios beyond $2 \times n$, i.e., where both Alice and Bob measure at least three observables each, with at least one observable having more than two outputs. In this case, we find that the randomness is not generic, and a linear programming check is needed in each specific instance to ensure that the Hardy outputs certify randomness against a no-signaling adversary. 

Specifically, we show that there exist Hardy paradoxes for the number of outputs of each party exceeding two, that do not allow for randomness. In other words, for the number of outputs $m \geq 3$, there exist Hardy paradoxes where the non-zero probability is strictly below unity in quantum theory, while there exist no-signaling strategies that allow for this non-zero probability to be $1$, thus disallowing any randomness. 

\begin{figure}[t]
	\centerline{\includegraphics[scale=0.36]{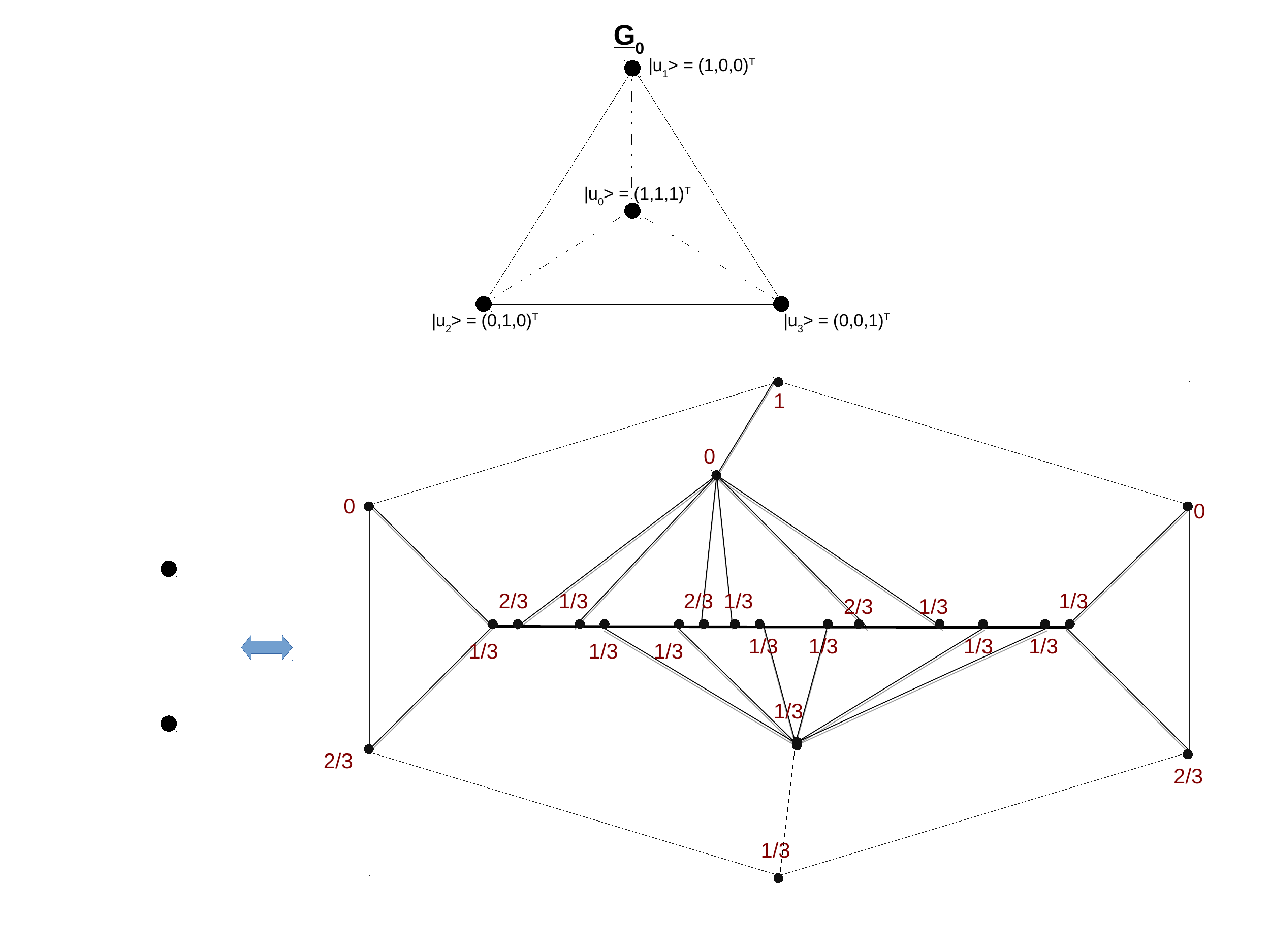}}
	\caption{\label{fig:gadg-hardy} The construction of the basic unit $G_0$ is shown in the top figure. The unit $G_0$ is constructed out of three virtual edges (shown by a dotted edge, that represents the graph below) connecting the vertices representing $(1,0,0)^T, (0,1,0)^T$ and $(0,0,1)^T$ to the central vertex representing $(1,1,1)^T$.}
\end{figure}

\begin{prop}
	\label{prop:3x3}
	Let $m_A = m_B = m$ denote the number of outputs of parties Alice and Bob in a two-party Hardy paradox and let $P_q(a^*,b^*|x^*, y^*)$, $P_{ns}(a^*,b^*|x^*,y^*)$ denote the maximum value of the non-zero probability in the Hardy paradox in quantum theory and general no-signaling theories, respectively. Then, for all $m \geq 3$, there exist two-party Hardy paradoxes such that $0 < P_q(a^*,b^*|x^*,y^*) < 1$ and $P_{ns}(a^*,b^*|x^*,y^*) = 1$. 
\end{prop}

\begin{proof}
	We construct Hardy paradoxes with the required property for all $m \geq 3$ by first constructing special sets of vectors exhibiting state-dependent contextuality using the $01$-gadgets proposed in \cite{RRHH+17, CPSS18}. We detail the construction of these vector sets for $m = 3$ (a contextuality test in dimension three), the construction for higher $m$ follows in an analogous manner. 
	
	As explained in \cite{RRHH+17}, a $01$-gadget is a graph $G$ that has an orthogonal representation by vectors \cite{CSW14, RH14} belonging to a Hilbert space of dimension $\omega(G)$ and with two distinguished non-adjacent vertices $v_1 \nsim v_2 \in V(G)$ (represented by distinct vectors $|v_1 \rangle, |v_2 \rangle$) such that in every $\{0,1\}$-coloring $f$ of $G$, either $f(v_1) = 0$ or $f(v_2) = 0$. 
	Moreover, given any two distinct non-orthogonal vectors $|v_1 \rangle, |v_2 \rangle \in \mathbb{C}^m$, it is shown in \cite{RRHH+17} that one can construct a finite set of vectors $S \subset \mathbb{C}^m$ with $|v_1 \rangle, |v_2 \rangle \in S$ such that the orthogonality graph $G_S$ of the set of vectors is a $01$-gadget, with the corresponding vertices $v_1, v_2$ being the distinguished vertices. 
	
	Given this background, the construction works via the basic unit $G_0$ shown in Fig. \ref{fig:gadg-hardy}. As shown in the Figure, the unit $G_0$ is obtained by connecting the vectors $|u_1 \rangle, |u_2 \rangle, |u_3 \rangle$ from a basis in $\mathbb{C}^3$ to a single vector $|u_0 \rangle$ via three appropriate $01$-gadgets. The exact set of vectors depends on the inner products $|\langle u_1 | u_0 \rangle|$, $| \langle u_2| u_0 \rangle|$ and $| \langle u_3 | u_0 \rangle|$ and is shown in Prop. 2 of \cite{RRHH+17}. In the Figure, we show the construction for the basis vectors $(1,0,0)^T, (0,1,0)^T, (0,0,1)^T$ each connected via a $01$-gadget (consisting of $22$ vectors which are explicitly constructed in Prop. 2 of \cite{RRHH+17}) to $(1,1,1)^T$. Now, assuming the usual Kochen-Specker rules, in any $\{0,1\}$-coloring $f$, one of the three vectors $|u_1 \rangle, |u_2 \rangle, |u_3 \rangle$ must be assigned value $1$, i.e.,  
	$f(u_1) = 1 \vee f(u_2) = 1 \vee f(u_3) = 1$. By the property of the $01$-gadgets connecting these three vertices to $u_0$, in any $\{0,1\}$-coloring of the graph $G_0$ we must have $f(u_0) = 0$. A similar construction can readily be obtained in any $\mathbb{C}^m$ for $m \geq 3$, using $01$-gadgets corresponding to the inner product $| \langle u_i | u_0 \rangle| = \frac{1}{\sqrt{m}}$ for $i=1, \dots, m$ between the vectors representing the distinguished vertices. 
	
	We obtain the Hardy paradoxes with the required property that $P_c(a^*,b^*|x^*,y^*) = 0$, $0 < P_q(a^*,b^*|x^*,y^*) < 1$ and $P_{ns}(a^*,b^*|x^*,y^*) = 1$ by following the construction of Hardy paradoxes from local contextuality sets given in the previous section. To elaborate, we first complete the maximum bases to obtain the set $Q_{\text{max}}$ of maximum cliques of $G_0$. In the corresponding Hardy paradox, Alice and Bob perform measurements in these bases and observe a set of conditional probability distributions $\{P_{A,B|X,Y}(a,b|x,y)\}$. Let $S_{G_0} = \{|u_{z,c}\rangle \}$ denote the faithful orthogonal representation and consider the set $\mathcal{S}_B$ defined as in Eq.(\ref{eq:SB-Hardy}) by $\mathcal{S}_B := \{((x,a),(y,b)) : \langle u_{(x,a)} |u_{(y,b)} \rangle = 0 \} \nonumber \\ \bigcup \; \; \{((x^*,a^*),(y^*,b^*))\}$. Here, the input-output pair $((x^*,a^*),(y^*,b^*))$ corresponds to the vector $|u_0 \rangle$ for both Alice and Bob. The indicator vector for the events that appear in the Hardy paradox are given as in Eq.(\ref{eq:bell-indicator}) by 
	\begin{equation}
	\textbf{B}(a,b,x,y) = \left\{
	\begin{array}{lr}
	1 & : ((x,a), (y,b)) \in \mathcal{S}_B \\
	0 & : \text{otherwise} 
	\end{array}
	\right.
	\end{equation} 
	The statement that $f(u_0) = 0$ in any $\{0,1\}$-coloring then translates to the statement that in any classical strategy $P_c(a^*,b^*|x^*,y^*) = 0$.  
	
	In the quantum strategy, the players share a maximally entangled state  $| \psi \rangle  = \frac{1}{\sqrt{m}} \sum_{i=0}^{m-1} | i, i \rangle$ (for the specific $G_0$ under consideration $m=3$). Upon receiving the input $x,y$ the players measure in the basis $\{\Pi_x\} = \{|u_{(x,a)} \rangle \langle u_{(x,a)} | \}_{a=0,\dots,m-1}$ and $\{\Sigma_y \} = \{|u_{(y,b)} \rangle \langle u_{(y,b)} | \}_{b=0,\dots,m-1}$  corresponding to the received clique. They return the outcomes $a,b$ of the measurement. This strategy satisfies the constraints of the Hardy paradox, i.e., $P_q(a,b|x,y) = 0$ for all $((x,a),(y,b)) \in \mathcal{S}_B \setminus ((x^*,a^*),(y^*,b^*))$ and 
	\begin{eqnarray}
	P_q(a^*,b^*|x^*,y^*) &=& | \langle \psi | (|u_{(x^*,a^*)} \rangle \otimes |u_{(y^*,b^*)} \rangle) |^2 \nonumber \\
	&=& \frac{1}{m} | \langle u_0 | u_0 \rangle|^2 = \frac{1}{m}.
	\end{eqnarray}

	On the other hand, for the graph $G_0$ an assignment $f: V(G_0) \rightarrow [0,1]$ exists and is explicitly given in Fig. \ref{fig:gadg-hardy} for which $f(u_0) = 1$ and $f(u_1) = f(u_2) = f(u_3) = \frac{1}{3}$. This assignment can be used to construct the marginal strategies of each party, i.e., $P_{ns}(a|x) = f(u_{(x,a)})$ and $P_{ns}(b|y) = f(u_{(y,b)})$.
	We then construct the joint distributions as $P_{ns}(a,b|x,y) = P_{ns}(b|y,(x,a)) P_{ns}(a|x)$ with
	\begin{eqnarray}
	P_{ns}(b|y,(x,a)) = \left\{
	\begin{array}{lr}
	1 & : u_{(x,a)} \cong u_{(y,b)}\\
	0 & : \text{otherwise} 
	\end{array}
	\right.
	\end{eqnarray} 
	In other words, the strategy allows for perfect correlations between the corresponding vertices on the two copies of $G_0$ for Alice and Bob. This gives a no-signaling box which satisfies all the constraints in the Hardy paradox for which $P_{ns}(a^*,b^*|x^*,y^*) = 1$. Analogously, the fact that the $01$-gadgets allow for an assignment $f(u_0) = 1$ with the other distinguished vertex taking value $f(u_i) = \frac{1}{m}$ for $i = 1, \dots, m$ in the construction for general $m \geq 3$ then gives the required no-signaling strategy for these cases.  
\end{proof}

\subsection{Increasing the probability of the Hardy output: Construction of Hardy paradoxes using local contextuality sets}
\label{subsec:Hardy-gadgets}
 
One way to boost the noise-tolerance of the protocol as seen from the calculations in the previous section is to increase the Hardy probability $p_Q^*$ while ensuring $p_{NS}^* < 1$. This is a question that has been studied previously in the literature and some solutions have been proposed \cite{BBMH97, MV14, CCXS+13}. A "ladder paradox" was introduced in \cite{BBMH97} in the $(2,k,2)$ scenario of two parties, each choosing from $k$ binary inputs. In the limit of a large number of inputs ($k \rightarrow \infty$), the non-zero probability was shown to approach $0.5$, with the corresponding state getting close to the maximally entangled state. An extension to two qudit systems for $d > 2$ considered in \cite{CCXS+13} makes use of the following conditions in the $(2,2,d)$ scenario: $P(A_1 < B_0) = 0$, $P(B_0 < A_0) = 0$, $P(A_0 < B_1) = 0$ and $P(A_1 < B_1) > 0$. While classically it is easy to see that the above conditions cannot be satisfied simultaneously, it was verified numerically that the maximum value of the non-zero probability in this case is $\approx 0.417$ for large $d$. In \cite{MV14}, Man\v{c}inska and Vidick proposed a generalization of Hardy's paradox, increasing the probability of the Hardy output to the entire interval $(0,1]$, but at the expense of a larger number of outputs and increasing dimensionality of the shared quantum state. More precisely they showed that in the $(2,2,2^d)$ scenario, the probability of the Hardy output can be boosted to $1 - (1-p_Q^*)^d$ where $p_Q^* = \frac{5 \sqrt{5} - 11}{2}$, so that infinite dimensional states are required to achieve the entire interval $(0,1]$. 

While the paradoxes in \cite{BBMH97, CCXS+13} also satisfy $p_{NS}^* < 1$, in the Man\v{c}inska-Vidick game, an appropriate PR-type box achieves the value $1$ for the Hardy probability so that this game does not allow for randomness amplification against a no-signaling adversary. For the ladder paradox, by the characterization of the vertices of the no-signaling polytope \cite{JM05, BP95} in the $(2,k,2)$ scenario, we have that $p_{NS}^* = 1/2$. For the qudit Hardy paradoxes by Chen et al. \cite{CCXS+13}, we verified using linear programming that $p_{NS}^* = 1 - \frac{1}{d-1}$. As such, both of these constructions can be used in our device-independent protocol for amplification. The price to pay in the first case is an increase in the number of inputs since $p_{Q}^* \rightarrow 1/2$ only as the number of inputs $k$ tends to infinity. In the second case, the price to pay is an increase in the dimensionality of the shared quantum state, since the Hardy probability here in quantum theory was verified numerically to approach a maximum of $\approx 0.417$ as the dimensionality of the shared state tends to infinity. To remedy these deficiencies, we introduce a fundamental procedure allowing to construct Hardy paradoxes with arbitrary value of the non-zero probability in $(0,1]$ in the next section.

It is well-known that single-party non-contextuality inequalities cannot directly be applied for device-independent information processing. The reason for this lack of applicability is that any experimental test of a non-contextuality inequality tends to make the assumption that the same projector is being measured in multiple distinct contexts, which is contrary to the device-independent requirement. One way to overcome this is to construct a two-party inequality based on the local contextuality, so that the spatial correlations between two parties ensures that the local projectors remain the same in different contexts \cite{RBHH+15}. Indeed, it is known that every Kochen-Specker system gives rise to a two-party Bell inequality that is algebraically violated by quantum mechanical correlations (as such this is also called a pseudo-telepathy game) \cite{BBT05,RW04}. However, the known Kochen-Specker systems involve measuring a large number of projectors, which translates to multple inputs for each of the measuring parties. To reduce the number of projectors and inputs, we instead use local (state-dependent) contextuality sets called $01$-gadgets that we recently studied in \cite{RRHH+17}. As stated in the previous subsection, a $01$-gadget is a graph $G$ that has an orthogonal representation by vectors \cite{CSW14} belonging to a Hilbert space of dimension $\omega(G)$ and with two distinguished non-adjacent vertices $v_1 \nsim v_2 \in V(G)$ (represented by distinct vectors $|v_1 \rangle, |v_2 \rangle$) such that in every $\{0,1\}$-coloring $f$ of $G$, either $f(v_1) = 0$ or $f(v_2) = 0$. 

The $01$-gadgets admit a $\{0,1\}$ assignment, and as such do not give rise to a two-party pseudo-telepathy game; nevertheless, we show that one can construct a Hardy paradox based on the local $01$-gadget. This gives rise to a remarkable device-independent application of local contextuality where the randomness essentially arises from the contextuality exhibited by the $01$-gadget.

\begin{figure}[t]
	\centerline{\includegraphics[scale=0.33]{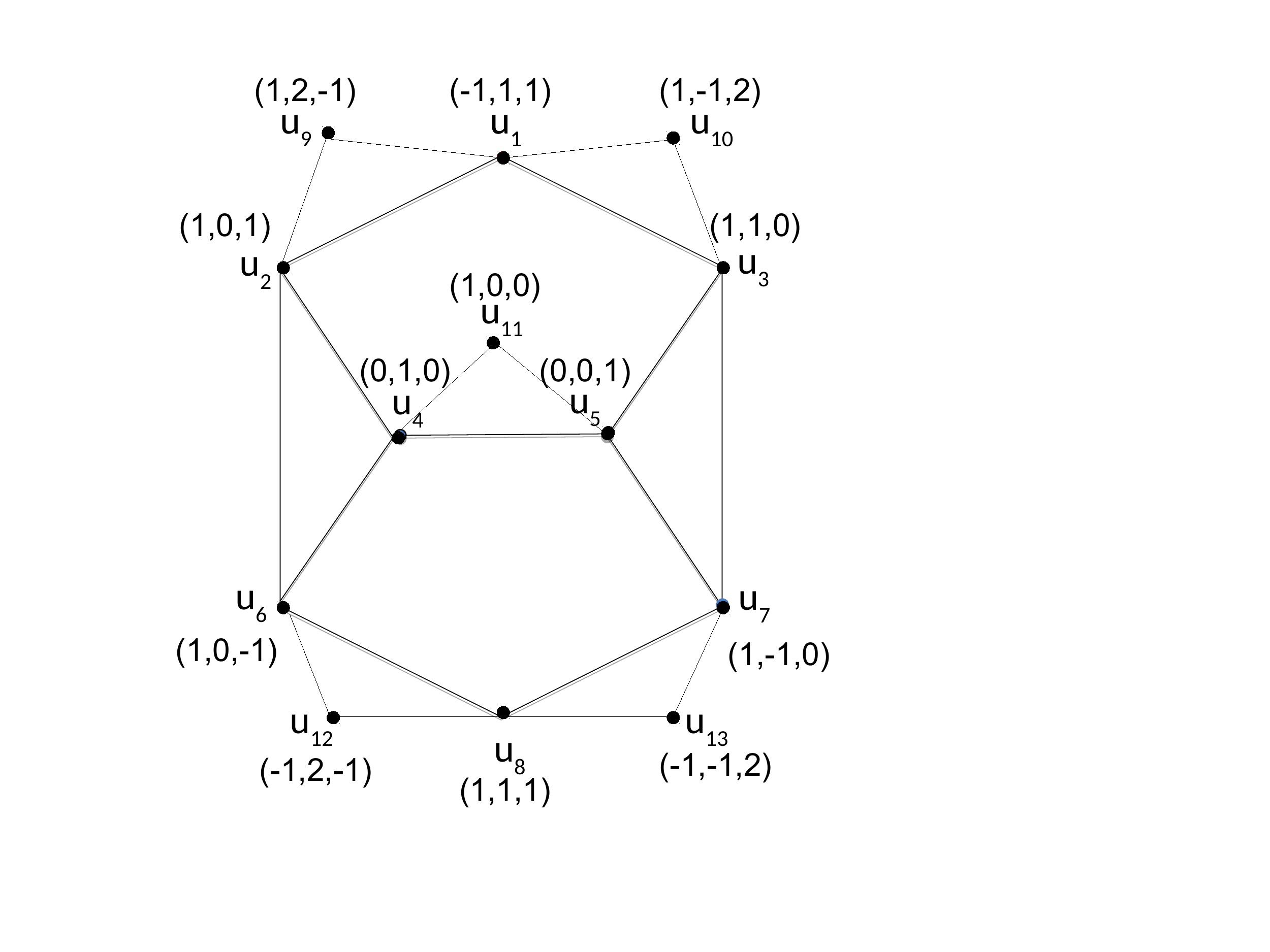}}
	\caption{The $8$-vertex ``Clifton" graph that was used by Kochen and Specker in their construction of the $117$ vector KS set, shown with all bases completed in dimension three These bases serve as the inputs in the corresponding Hardy paradox as explained in the text.}
	\label{fig:Clifton2}
\end{figure}

The simplest (with fewest projectors) $01$-gadget is the $8$-projector Clifton gadget $G_{\text{Clif}}$ \cite{Clifton93} with the corresponding vectors given by 
	\begin{eqnarray}
\label{eq:Clif-orth-rep}
&&	| u_1 \rangle = \frac{1}{\sqrt{3}}(-1,1,1)^T, \; \; |u_2 \rangle = \frac{1}{\sqrt{2}}(1,1,0)^T, \nonumber \\ 
&& |u_3 \rangle = \frac{1}{\sqrt{2}} (0,1,-1)^T, |u_4 \rangle = (0,0,1)^T, \nonumber \\
&&	|u_5 \rangle = (1,0,0)^T, \; \; |u_6 \rangle = \frac{1}{\sqrt{2}}(1,-1,0)^T, \nonumber \\
&& |u_7 \rangle = \frac{1}{\sqrt{2}}(0,1,1)^T, \; \; |u_8 \rangle = \frac{1}{\sqrt{3}}(1,1,1)^T. 
\end{eqnarray}
This gadget is shown in Fig. \ref{fig:Clifton2} with all bases completed in dimension three. 

The Hardy paradox corresponding to the Clifton gadget belongs to the class $(2, 7, 3)$ indicating that it involves two parties Alice and Bob, each making one of seven possible measurements and obtaining one of three possible outcomes. We label the measurement settings of Alice $x$ and those of Bob $y$ with $x, y = 1, \dots, 7$. These measurement settings are obtained as follows. We first complete the possible measurement bases in the graph to obtain the set $Q_{\text{max}}$ of bases (maximum cliques in the corresponding graph) as shown in the Fig.\ref{fig:Clifton2}. For the specific Clifton gadget under consideration this set is given as
\begin{eqnarray}
Q_{\text{max}} = \{(1,2,9), (1,3,10), (2,4,6), (3,5,7), \nonumber \\
(4,5,11), (6,8,12), (7,8,13)\}.
\end{eqnarray}
The measurement settings $x,y$ of the two parties respectively, correspond to the maximum cliques in $Q_{\text{max}}$, so $x, y  \in \{1, \dots, 7\}$. The outcomes (corresponding to the respective projectors) of Alice are labeled $a$ and those of Bob $b$ with $a, b \in \{1, 2, 3\}$. The two parties observe a set of conditional probability distributions $\{P_{A,B|X,Y}(a, b|x, y)\}$. Let $\{|u_{z,c} \rangle\}$ denote the vectors (a faithful orthogonal representation of the graph $G$ \cite{CSW14}) such as in Eq.(\ref{eq:Clif-orth-rep}). Consider the set $\mathcal{S}_B$ be defined as follows
\begin{eqnarray}
\label{eq:SB-Hardy}
\mathcal{S}_B := \{((x,a),(y,b)) : \langle u_{(x,a)} |u_{(y,b)} \rangle = 0 \} \nonumber \\ \bigcup \; \; \{((x^*,a^*),(y^*,b^*))\}.
\end{eqnarray}
Here the input-output pair $((x^*,a^*),(y^*,b^*))$ corresponds to the distinguished vertices $u_{(x^*,a^*)}, u_{(y^*,b^*)}$ of the gadget and will give rise to the non-zero probability in the Hardy paradox.  For instance, for the specific gadget under consideration this refers to the input-output pairs $((x^*,a^*),(y^*,b^*)) = ((1,1),(7,2))$ corresponding to the input basis $(1,2,9)$ for Alice and the basis $(7,8,13)$ for Bob with the respective outcomes corresponding to projector $|u_1 \rangle \langle u_1|$ for Alice and projector $|u_8 \rangle \langle u_8|$ for Bob. 
The correlation expression is then given by
\begin{equation}
\label{bip-Bell-ineq}
\textbf{B} \cdot \{P(a, b| x, y)\} = \sum_{a,b, x, y} \textbf{B}(a,b, x, y) P(a, b | x, y),
\end{equation}
where $\textbf{B}$ is the indicator vector with entries 
\begin{equation} \label{eq:bell-indicator}
\textbf{B}(a,b,x,y) = \left\{
\begin{array}{lr}
1 & : ((x,a), (y,b)) \in \mathcal{S}_B \\
0 & : \text{otherwise} 
\end{array}
\right.
\end{equation} 
Since a $\{0,1\}$-coloring exists for the gadget and is the assignment used in any local deterministic box, it follows that the classical value of the correlation expression in Eq.(\ref{bip-Bell-ineq}) is exactly $0$. Crucially, the probability $P_{A,B|X,Y}(a^*,b^*|x^*,y^*)$ for the distinguished vertices $u_{(x^*,a^*)}, u_{(y^*,b^*)}$ is also zero due to the property of the gadget that in any $\{0,1\}$ assignment both these vertices cannot be assigned value $1$. The local deterministic box achieving this value is precisely given for each of the parties by the $\{0,1\}$-coloring of the gadget, i.e.,  both the parties, given their measurement setting, return the outcome corresponding to value $1$ in the $\{0,1\}$-coloring.

The quantum strategy is given as follows. 
\begin{enumerate}
	\item The players share a maximally entangled state  $| \psi \rangle  = \frac{1}{\sqrt{d}} \sum_{i=0}^{d-1} | i, i \rangle$ (for the specific gadget under consideration $d=3$). 
	
	\item Upon receiving the input $x,y$ the players measure in the basis $\{\Pi_x\} = \{|u_{(x,a)} \rangle \langle u_{(x,a)} | \}_{a=1,\dots,d}$ and $\{\Sigma_y \} = \{|u_{(y,b)} \rangle \langle u_{(y,b)} | \}_{b=1,\dots,d}$  corresponding to the received clique. They return the outcomes $a,b$ of the measurement. 
\end{enumerate}

%
Now for all $((x,a),(y,b)) \in \mathcal{S}_B \setminus \{((x^*,a^*),(y^*,b^*))\}$, we have that  
\begin{eqnarray}
\langle \psi | \left(|u_{(x,a)} \rangle \otimes | u_{(y,b)} \rangle \right) &=& \frac{1}{\sqrt{d}} \sum_{i=0}^{d-1} \langle i | u_{(x,a)} \rangle \langle i | u_{(y,b)} \rangle \nonumber \\
&=& \frac{1}{\sqrt{d}} \overline{\langle u_{(x,a)} |  \overline{u}_{(y,b)} \rangle} = 0, 
\end{eqnarray}
since the corresponding vectors are orthogonal. These correspond to the zero constraints in the obtained Hardy paradox. On the other hand, for the distinguished vertices, from the fact that a faithful representation of the gadget exists, we have that the corresponding probability is non-zero. For the specific gadget under consideration we have from the orthogonal representation in Eq.(\ref{eq:Clif-orth-rep}) that  $P_{A,B|X,Y}(a^*,b^*|x^*,y^*) = \frac{1}{27}$. This gives the non-zero Hardy probability in the obtained Hardy paradox.  

The above Hardy paradox generalizes in a straightforward manner to any $01$-gadget. 
For a large class of Hardy paradoxes obtained from $01$-gadgets, we show that the Hardy probability is also strictly bounded below unity for general no-signaling strategies so that we may employ these paradoxes in randomness amplification using the Protocol II of Fig. \ref{protocolsingle-2}. 


It has been an open question as to how large the non-zero probability in the Hardy paradox can get and whether Hardy paradoxes can be constructed for maximally entangled states \cite{ACY16}. From the construction of the Hardy paradox from $01$-gadgets outlined above, 
and the construction of a $01$-gadget for any two distinct vectors in $\mathbb{C}^d$ for $d \geq 3$ given in Proposition \ref{prop:fin-gadg-const} (Theorem 2 of \cite{RRHH+17}) we obtain Hardy paradoxes with the non-zero probability taking on the spectrum of values in $(0,1/3]$ at the expense of complexity of the gadget. 

\begin{prop}[Theorem 2 of \cite{RRHH+17}]
	\label{prop:fin-gadg-const}
	Let $|v_1 \rangle$ and $|v_2 \rangle$ be any two non-orthogonal vectors in $\mathbb{C}^d$ with $d \geq 3$. Then there exists a 01-gadget in dimension $d$ with $|v_1 \rangle$ and $|v_2 \rangle$ being the two distinguished vertices.
\end{prop}   

In fact, it is possible to obtain Hardy paradoxes for the entire spectrum $(0,1]$ as we now show. To do this, we work in $\mathbb{R}^4$, i.e., we augment the gadget in $\mathbb{R}^3$ with the additional vertex $(0,0,0,1)^T$ to obtain a gadget in dimension four. As shown in Prop. \ref{prop:fin-gadg-const} it is possible to obtain a gadget with any two vectors as distinguished vertices $u_1, u_2$ in $\mathbb{R}^4$ by embedding the gadget in $\mathbb{R}^3$ in this manner. We now form four copies of the gadget $G^{(1)}, G^{(2)}, G^{(3)}, G^{(4)}$ in $\mathbb{C}^4$, with the corresponding vectors in each copy being mutually orthogonal, i.e., 
$|u_k^{(1)} \rangle, |u_k^{(2)} \rangle, |u_k^{(3)} \rangle, |u_k^{(4)} \rangle$ form a complete basis (the vertices form a maximum clique). That such a construction is always possible is based on the fact that in dimensions $4$ and $8$ there exist division algebras \cite{Cohn03} (the quaternions and octonions) so that one can rotate each vector $(a,b,c,d) \in \mathbb{R}^4$ in a set of vectors, to the orthogonal vectors $(b,-a,-d,c), (c,d,-a,-b), (d,-c,b,-a)$ by multiplication by the orthogonal units $i,j$ or $k$ of the algebra, i.e., to every vector $(a,b,c,d) \in \mathbb{R}^4$, one can associate the real orthogonal matrix
\begin{eqnarray} 
\label{eq:real-ort-mat}
\begin{pmatrix} a & b & c & d\\ b & -a & d & -c \\ c & -d & -a & b \\ d & c & -b & -a \end{pmatrix}
\end{eqnarray}
A similar construction also exists in dimension eight by means of the octonions. 
Let $G$ denote the newly formed orthogonality graph with $V(G) = V(G^{(1)}) \bigcup V(G^{(2)}) \bigcup V(G^{(3)}) \bigcup V(G^{(4)})$ and the edge set formed by the orthogonality constraints defined by the three faithful representations of the copies as before. The construction is illustrated by means of the $8$-vertex Clifton gadget in Fig. \ref{fig:hardy-(0,1]}.

\begin{figure}[t]
	\centerline{\includegraphics[scale=0.39]{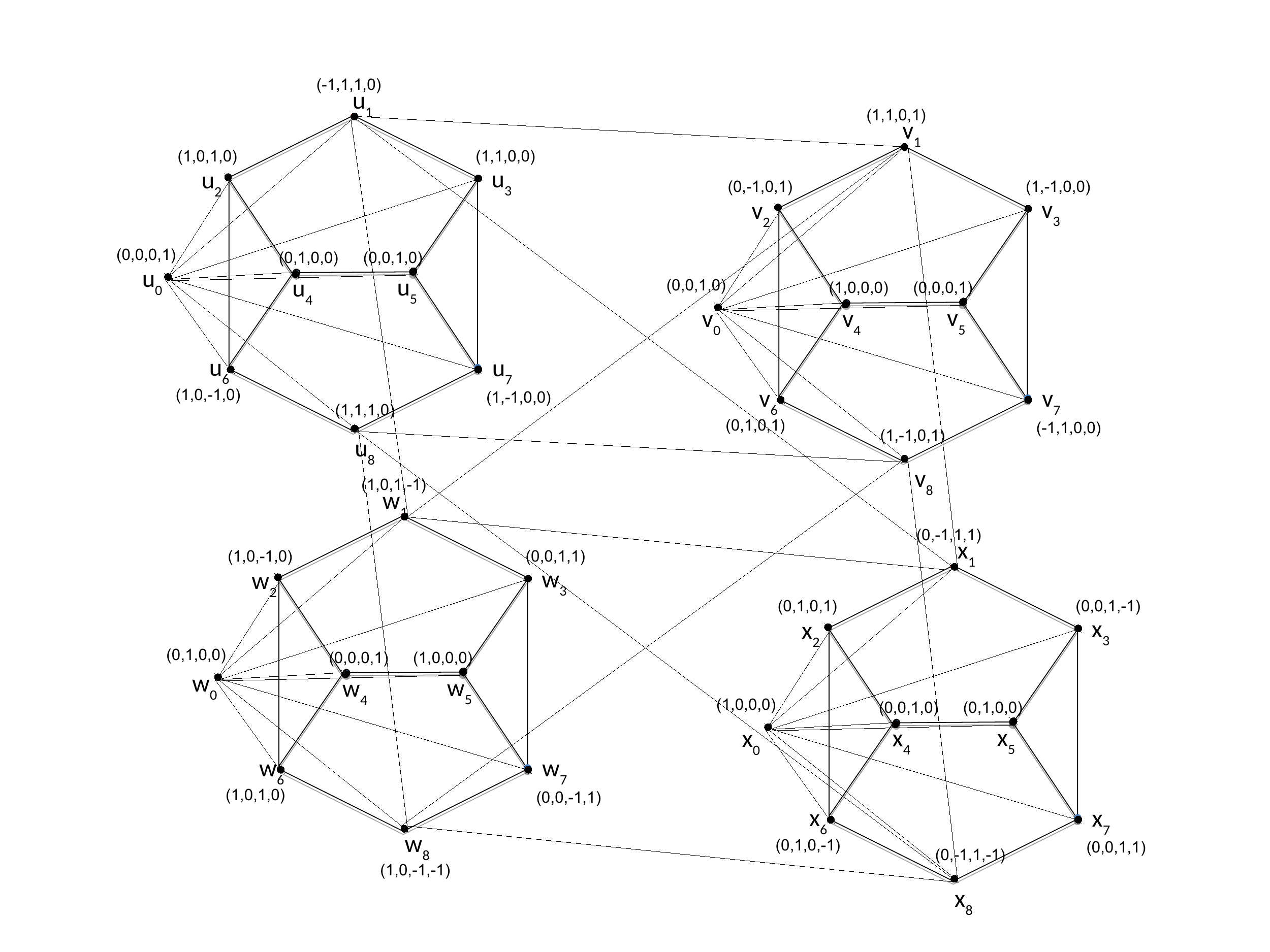}}
	\caption{\label{fig:hardy-(0,1]} An illustration of the Hardy paradox construction in dimension four by means of the real orthogonal matrix in Eq.(\ref{eq:real-ort-mat}). We embed the Clifton graph in dimension four by adding the extra vertex $u_0$ corresponding to the vector $(0,0,0,1)^T$. We then obtain four copies of the resulting gadget in dimension four by multiplication by the orthogonal unit as in Eq.(\ref{eq:real-ort-mat}) (not all resulting edges are shown for clarity). The two distinguished vertices $u_1$ and $u_8$ give rise in this manner to distinguished measurement bases given as $\{|u_{1/8} \rangle, |v_{1/8} \rangle, |w_{1/8} \rangle, |x_{1/8} \rangle\}$. In the scenario of the gadget constructed in Prop. \ref{prop:fin-gadg-const} with $|v_1 \rangle$ and $|v_2 \rangle$ identical, measurements on a maximally entangled ququart state by Alice and Bob in these respective bases give rise to a Hardy paradox with non-zero probability in $(0,1]$ as explained in the text.}
	\end{figure} 

Now, we form the set $Q_{\text{max}}(G)$ of bases (maximum cliques in $G$) as before, with the measurement settings $x,y$ of the two parties corresponding to these maximum cliques. In particular, we now denote $x^*,y^*$ as the maximum cliques formed by the four orthgonal vectors corresponding to the distinguished vertices $u_1$ and $u_2$, i.e., $u_1^{(1)}, \dots, u_1^{(4)}$ and $u_2^{(1)}, \dots, u_2^{(4)}$. From a faithful representation of the gadget $\{|u_{z,c} \rangle\}$, we form the constraint set $\mathcal{S}_B$
\begin{eqnarray}
\mathcal{S}_B & := &\{((x,a),(y,b)) : \langle u_{(x,a)} | u_{(y,b)} \rangle = 0\} \nonumber \\ &&\bigcup  \cup_{k=1}^{4} \{((x^*,u_1^{(k)}),(y^*,u_2^{(k)}))\}.
\end{eqnarray}
Now, the input-output pairs $((x^*,u_1^{(k)}),(y^*,u_2^{(k)}))$ correspond to the four pairs of distinguished vertices of the $01$-gadget in $\mathbb{R}^4$ and will give rise to the non-zero probabilities in the Hardy paradox. In the quantum strategy, the players share a two-ququart maximally entangled state $\frac{1}{2} \sum_{i=0}^3 | i, i  \rangle$ and measure in the basis $\{\Pi_x\} = \{| u_{(x,a)} \rangle \langle u_{(x,a)} |\}_{a=1,\dots, 4}$ and $\{\Sigma_y \} = \{|u_{(y,b)} \rangle \langle u_{(y,b)} | \}_{b=1,\dots,d}$  corresponding to the received input clique. They return the outcomes $a,b$ of the measurement. We have for all $((x,a),(y,b)) \in \mathcal{S}_B \setminus \cup_{k=1}^{4} \{((x^*,u_1^{(k)}),(y^*,u_2^{(k)}))\}$ that 
\begin{eqnarray}
\langle \psi | \left(|u_{(x,a)} \rangle \otimes | u_{(y,b)} \rangle \right) &=& \frac{1}{2} \overline{\langle u_{(x,a)} |  \overline{u}_{(y,b)} \rangle} = 0.
\end{eqnarray}
As before, these form the zero constraints of the corresponding Hardy paradox. 
By the properties of the $01$-gadgets, in any classical theory we have that for each of the distinguished pairs, the probability $P(u_1^{(k)}, u_2^{(k)}|x^*, y^*) = 0$ for all $k =1, \dots, 4$. In the quantum strategy for the measurements $x^*, y^*$ however, we have that 
\begin{eqnarray}
\label{eq:quant-prob}
P(u_1^{(k)}, u_2^{(k)}|x^*, y^*) = \frac{1}{4} | \langle u_1^{(k)} |  \overline{u}_{2}^{(k)} \rangle|^2 
\end{eqnarray} 
This gives the non-zero Hardy probability in the corresponding Hardy paradox. 
Choosing a gadget with the distinguished vectors identical in Prop. \ref{prop:fin-gadg-const}, we see that the probability $P(u_1^{(k)}, u_2^{(k)}|x^*, y^*)  = \frac{1}{4}$ for all $k = 1, \dots, 4$. We thus obtain a Hardy paradox with the maximum possible contradiction, i.e., a set of events which have probability $0$ in any classical theory, which however are certain to occur in quantum theory. Thus, from Eq.(\ref{eq:quant-prob}) and the gadget construction in Prop. \ref{prop:fin-gadg-const}, we see that one can obtain Hardy paradoxes with the non-zero probability in the entire spectrum $(0,1]$ and obtain the following proposition.

\begin{prop}
	\label{prop:Hardy-gadg}
	There exist Hardy paradoxes for the maximally entangled state $\frac{1}{\sqrt{d}} \sum_{i=0}^{d-1} |i, i \rangle$ for all $d \geq 3$ with the non-zero probability taking any value in $(0,\frac{1}{d}]$. In dimensions four and eight, there exist Hardy paradoxes for the maximally entangled state with the non-zero probability taking any value in $(0,1]$.
\end{prop}

\section{Appendix: Experimental Implementation of the simplest Hardy paradox}
\label{sec:Appex-exp}
We now describe the implementation in a quantum optical setup of the Hardy paradox in the simplest Bell scenario.  
In our experiment, the physical qubits are single-photon polarization states and the computational basis corresponds to the horizontal (H) and vertical (V) polarization, i.e., $|H \rangle \equiv | 0 \rangle$  and $|V \rangle \equiv| 1 \rangle$. 
Experimentally, to perform the Hardy test, a suitable two-qubit photon polarisation non-maximally entangled state 
\begin{eqnarray}
\label{eq:hardy-stexp}
|\psi_{\theta} \rangle &=& \frac{\cos{(\theta)}}{\sqrt{1+\cos{(\theta)}^2}}  \left( |HV \rangle + |VH\rangle \right)\\ \nonumber &+&
\frac{\sin{(\theta)}}{\sqrt{1+\cos{(\theta)}^2}}| VV \rangle 
\end{eqnarray}
is prepared, where ideally the parameter $\theta$ takes the optimal value for the Hardy test given by $\theta = \arccos{\left(\sqrt{\frac{\sqrt{5}-1}{2}}\right)}$. 

\begin{figure}
	\includegraphics[width=1\columnwidth]{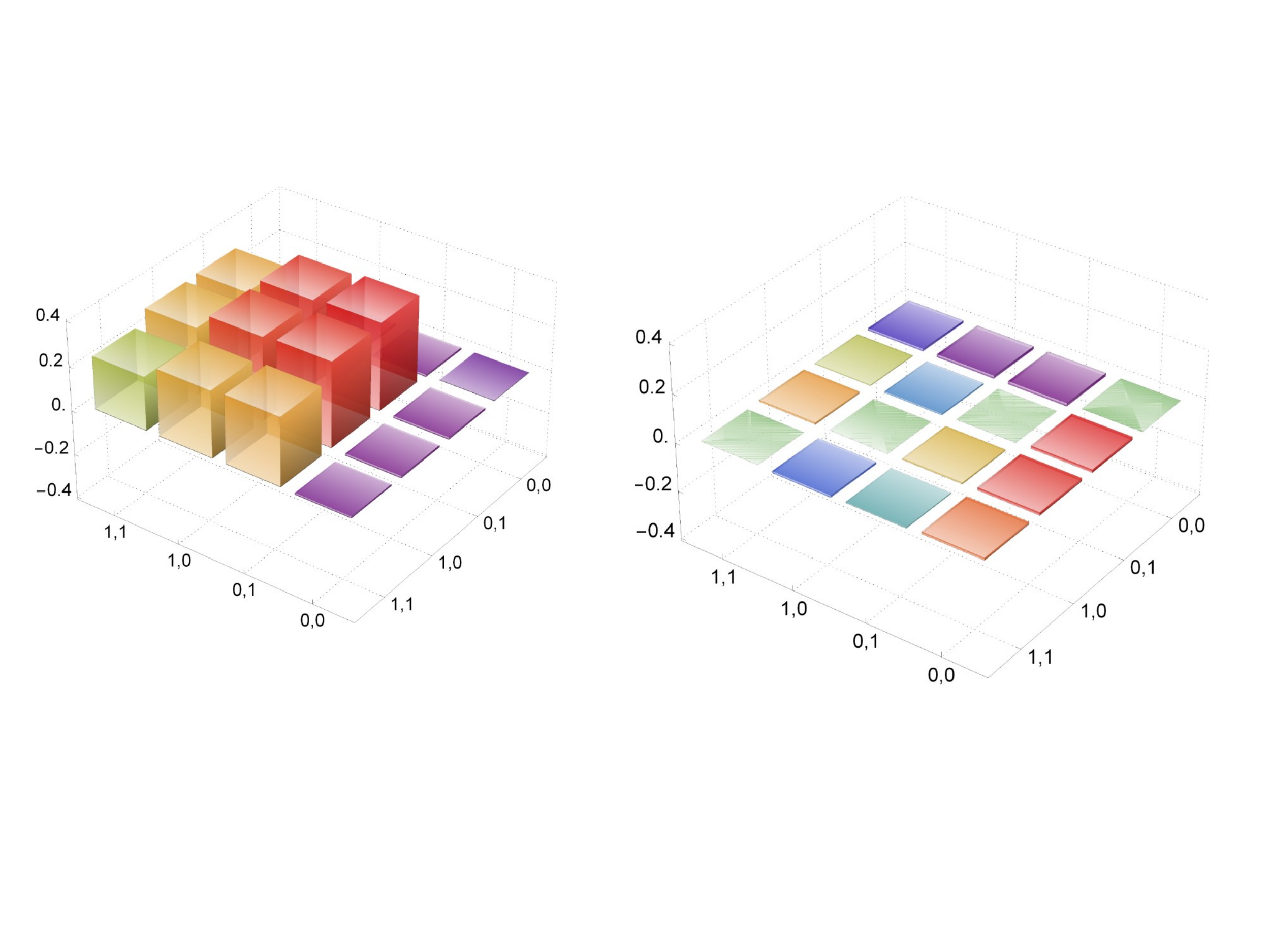}
	\caption{\label{rhoexpdensity}(Color online) Experimental two qubit density matrix obtained using quantum state tomography, real(left) and the imaginary(right) parts. In the axes, the labels $0$ and $1$ refer to the horizontal ($H$) AND vertical ($V$) polarizations, respectively}
\end{figure}

\textit{State Preparation.-} A UV pump laser at $390 nm$ was focused onto two beta barium borate (BBO) crystals places in cross-configuration to produce photon pairs emitted into two spatial modes "a" and "b" through type-I SPDC process. 
Two crystals were used for compensation of longitudinal and transversal walk-offs. The emitted photons were coupled into single-mode optical fibers and passed through a narrow-bandwidth interference filter  to secure well-defined spatial and spectral emission modes. To observe the desired state, we used a half wave plate (HWP) oriented at $27.86$ degrees and placed after the output fiber coupler in each of the two modes. 
We have performed the full state tomography, where we measured photon polarisation of each of the two photons in three mutually unbiased basis (Horizontal-Vertical, Diagonal-Anti-diagonal, Left-circular-Right-circular). These polarization measurements
were performed by using HWPs, quarter wave plates (QWP) and polarizing beam splitters (PBS) followed by single photon detectors (actively quenched Si-avalanche
photodiodes (Si-APD)). An FPGA based counting system
was used to record the number of multi-fold coincidence events with
a detection time window of 1.7 ns.  From experimentally collected data, we reconstructed the density matrix $\rho_{\exp}$ corresponding to the two-photon state by considering maximum likelihood estimation. The Real and Imaginary parts of the density matrix are shown in Fig. \ref{rhoexpdensity}.
We have estimated the fidelity $F = \langle \psi_{\theta}| 	\rho_{\exp}|\psi_{\theta} \rangle$ of the experimentally prepared state $\rho_{\exp}$ with respect to the optimal $\psi_{\theta}$, to be $F = 0.997 \pm 0.001$.

\textit{Hardy test.-} For the Hardy test, the optimal settings of Alice and Bob (for $x,y=1$) and (for $x,y=0$) are given by
\begin{eqnarray}
\label{eq:hardy-measexp}
&&\{|H \rangle, |V \rangle\} \; \; \;  \; \text{for} \; x,y=1, \nonumber \\
&&\{\sin{\theta} | H \rangle - \cos{\theta} |V \rangle, \cos{\theta} | H \rangle + \sin{\theta} | V \rangle\} \; \; \text{for} \; x,y = 0. \nonumber \\
\end{eqnarray}
In the experimental Hardy test, the rate of two photon coincidence count was $10^5$ per second and the measurement time for each setting was $5$ hours. The measured coincidences counts, experimentally obtained joint probability of Alice and Bob $P_{A,B|X,Y}(a,b|x,y)$ and the corresponding errors (standard deviations in the estimated probabilities) are listed in the table \ref{Table1-expn}. During the measurement, the total number of coincidence counts for Alice's and Bob's settings  $(x = 0, y = 0)$, $(x = 0, y = 1)$, $(x = 1, y = 0)$, and $(x = 1, y = 1)$ were $1.90\times 10^9$, $1.91\times 10^9$, $1.91 \times 10^9$, and $1.93 \times 10^9$ respectively. The errors are due to cross-talks originating from the
PBS extinction and absorption wave-plate setting errors, wave-plates offset error, wave-plates retardance tolerance and error due to Poissonian statistics of the
incoming photons. As can be seen from the Table, the high fidelity of state preparation and measurements allowed us to achieve the high-quality Hardy test required for the DI application to randomness amplification.

\begin{table}[t]
	\centering
	\caption{\label{Table1-expn}{The theoretical and experimentally obtained conditional joint probability $P_{AB|XY}(ab|xy)$ and corresponding errors (standard deviations in the estimated probabilities) and coincidence counts for $a, b = \{0,1\}$  the outputs of Alice and Bob's measurements respectively and $x, y = \{0,1\}$ Alice and Bob's settings.}}
	{\begin{tabular}{ |l|l|l|l| }
			\hline
			{\bf $(ab| xy)$ } & {\bf Theory}& {\bf Experiment}& {\bf Error}  \\ \hline \hline
			{\bf $(00|00)$ }& 0.0902& 0.0911& 0.0025  \\
			{\bf $(01|00)$ }& 0.1459& 0.1496& 0.0036  \\
			{\bf $(10|00)$ }& 0.1459& 0.1485& 0.0043  \\
			{\bf $(11|00)$ }& 0.6180& 0.6008& 0.0052  \\
  
        			{\bf $(11|00)$ }& 0.2360& 0.2405& 0.0019  \\
			{\bf $(01|01)$ }&      0& 0.0015& 0.0005  \\
			{\bf $(10|01)$ }& 0.1459& 0.1345& 0.0043  \\
			{\bf $(11|01)$ }& 0.6180& 0.6234& 0.0034  \\
		
			{\bf $(00|10)$ }& 0.2360& 0.2504& 0.0019  \\
			{\bf $(01|10)$ }& 0.1459& 0.1413& 0.0046  \\
			{\bf $(10|10)$ }&      0& 0.0015& 0.0002  \\
			{\bf $(11|10)$ }& 0.6180& 0.6069& 0.0039  \\
		
			{\bf $(00|11)$ }&      0& 0.0008& 0.0002  \\
			{\bf $(01|11)$ }& 0.3819& 0.3715& 0.0037  \\
			{\bf $(10|11)$ }& 0.3819& 0.3916& 0.0039  \\
			{\bf $(11|11)$ }& 0.2361& 0.2361& 0.0064  \\
		
			\hline
	\end{tabular} }
\end{table}

We remark that while experimentally we have achieved the high fidelities required for a DI implementation, this is still under the caveat of the fair sampling assumption and the locality loophole. While it is possible by a shielding argument to circumvent the locality loophole for DI randomness generation, in future it would be desirable to repeat the experiment in a manner suited to handle the detection loophole. It is noteworthy that novel experimental techniques in the form of high-efficiency superconducting detectors are available currently, so that such an experimental implementation to close the freedom-of-choice loophole as studied in this paper, may be expected shortly.

\section{Appendix: Randomness from  experimental data}

\subsection{Convenient form of the extractor parameters}

Let us first rewrite the general results of Ref. \cite{BRGH+16} to a form more convenient for the experimental data.
Recall that the result was a scheme for a randomness amplifying device which transforms the weak private randomness
into a strong one with the data from the Bell inequality experiment. One of the parameters of the device is 
the probability of acceptance $q(\text{ACC})$ \cite{BRGH+16}. A potential consumer would want that even for a small $q(\text{ACC})$, the min-entropy $h_{\text{min}}$ is guaranteed - of course, for an extremely small $q(\text{ACC})$ there is no guarantee of good $h_{\text{min}}$ since the box may simply be classical. Therefore, we make sure that for $q(\text{ACC})$ chosen above some threshold value there is a guarantee of good $h_{\text{min}}$, and this threshold range should be acceptable to the customer, as it goes down with the number of runs of the Hardy experiment $n$.


\textit{Bound on the min-entropy of the device output.-}
The main technical Proposition to derive the bound on the min-entropy of the device output is given as Prop. 19 of \cite{BRGH+16}. Using Prop. 19 of \cite{BRGH+16} and taking $q(\text{ACC}) \geq \sqrt{\gamma^{\mu n} + 2 \epsilon_{Az}}$ for real parameters $\gamma, \mu \in (0,1)$ we obtain that given 
acceptance, we have that the outputs constitute a min-entropy source with min-entropy 
\begin{equation}
h_{box}\geq -\frac14 \log_2 (\gamma^{\mu n} + 2 \epsilon_{Az})
\label{eq:hbox}
\end{equation}
with probability 
\begin{equation}
\label{eq:PrGood}
\text{Pr}({\rm GOOD})\geq 1-(\gamma^{\mu n} + 2 \epsilon_{Az})^{1/4} 
\end{equation}
(in other words, upon acceptance the satisfaction of the min-entropy condition in Eq. \eqref{eq:hbox} constitutes the GOOD event).
Here $\mu$ is the number of boxes for which the upper bound $\gamma$ holds on the output probabilities, and 
$\epsilon_{Az}=2 \exp(-n \delta_{Az}^2/4)$, where $\delta_{Az}$ is the parameter from Lemma
\ref{lemmaazuma} of Appendix I.

One can then estimate the min-entropy of the outputs $h_{box}$ as follows:
\begin{equation}
h_{box} \geq \min \frac{n}{4} \left\{ -\mu \log_2 \gamma,  \frac{ \delta_{Az}^2 }{4} \log_2 e \right\} -\frac{3}{4}
\label{eq:h-box-bound}
\end{equation}

Now we shall exploit two Lemmas to maximise the above quantity over some free parameters at fixed experimental values and given $\epsilon$. To this end, let us introduce new notation for the quantity in Eq. (\ref{eq:means}) 
in the Lemma \ref{lemmaazuma} of Appendix I namely $\delta_{exp}=L_n$ (the experimentally observed Bell value).
We also put $\delta_{box}$ as a lower bound on the true Bell value $0 < \delta_{box}\leq \overline{L}_n$ (cf. Eq.(\ref{eq:cond-means})) .
Note that $\delta_{Az}$ is a free parameter here. In what follows 
we choose $\delta_{Az}=
\delta_{exp} - \delta_{box}$ and let $\delta_{Az}$
 vary over the  range $0<\delta_{Az}<\delta_{exp}$.
Then, if we substitute in the Lemma  \ref{lem:num-good-runs} of Appendix I the value
$\frac{\delta}{2}=\delta_{box}$
and apply the Lemma together with the concluding formula
(\ref{eq:box-upper-bound}) we find that the  
parameters $\mu$ and $\gamma$ satisfy
\begin{equation}
\mu\geq\frac{\delta_{exp}- \delta_{Az}- \kappa}{\frac{1}{16}-\kappa},\quad \gamma= 1- \frac{\kappa}{2(\frac14-\epsilon^2)^2}
\end{equation}
where $0<\kappa<\delta_{exp}-\delta_{Az}$  
can be chosen arbitrarily. 

Now let us take the minimum value of $\mu$ satisfying the above condition, namely $\mu = \frac{\delta_{exp}- \delta_{Az}- \kappa}{\frac{1}{16}-\kappa}$, substitute it into the RHS of (\ref{eq:h-box-bound}) and write the lower bound $h_{bound}(\delta_{exp},\epsilon,n)$ on $h_{box}$ explicitly as the maximum of the following function
\begin{eqnarray}
&&F(\delta_{exp},\delta_{Az}, \kappa, \epsilon, n) \equiv \nonumber \\
&& \frac{n}{4} \min \bigg\{-\left(\frac{\delta_{exp}- \delta_{Az}- \kappa}{\frac{1}{16}-\kappa} \right)\log_2\left( 1 - \frac{\kappa}{2(\frac{1}{4}-\epsilon^2)^2} \right), \nonumber \\
&&  \qquad \qquad \log_2(e) \frac{\delta_{Az}^2}{4}  \bigg\}-\frac{3}{4} 
\end{eqnarray}
over the free parameters $\delta_{Az}, \kappa$.
In this way we have the following estimate useful for further analysis of experimental data:
\begin{equation}
h_{box}\geq h_{bound}(\delta_{exp},\epsilon,n)\equiv \max_{\delta_{Az},\kappa} F(\delta_{exp},\delta_{Az}, \kappa, \epsilon, n)
\label{eq:entropy-bound}
\end{equation}
where maximum is taken over the ranges $0<\delta_{Az}<\delta_{exp}$ and  $0<\kappa<\delta_{exp}-\delta_{Az}$. 
Note that the final result, as intended, depends only on the 
experimental values of the observed Bell value and the number of runs $\delta_{exp}$, $n$ and the parameter 
$\epsilon$ characterising the weakness of the SV source. 

\textit{Quality of the final output bits from the Randomness Amplification protocol.-}
Let us now estimate the quality of the final output bits from our randomness amplifier. Recall that we apply Raz's extractor \cite{Raz2005} to the $SV$-source as well as the min-entropy source $h_{min}$ obtained from the device output above,.
When the event denoted as GOOD occurs, and provided that the min-entropies and the number of runs satisfy Raz's constraints, we obtain $m_{\rm Raz}$ bits (to be determined later) with distribution $\{p_i\}, i = 1, \ldots, 2^{m_{\rm Raz}}$ satisfying 
\begin{equation}
\text{dist}_{extr}(\{p_i\}, \{1/2^{m_{\rm Raz}}\})\leq 2^{-1.5 m_{\rm Raz}},
\end{equation}
where $\text{dist}$ denotes the distance between probability distributions given by $\text{dist}(P, Q) = \sum_a \bigg| P(a) - Q(a) \bigg|$.
 
However this comfortable situation occurs only when the GOOD event happens. From Eq.(\ref{eq:PrGood}) with probability $\text{Pr}(\text{BAD})\leq  (\gamma^{\mu n} + 2 \epsilon_{Az})^{1/4}$ there is an unwanted BAD event. 
Also in this case we apply the extractor (since we do not know whether the BAD or GOOD event occurred). 
In this case, we have $\text{dist} \leq 2$. Overall, the $m_{\rm Raz}$ bits obtained from the extractor satisfy
\begin{eqnarray}
\label{eq:pi-dist}
&& \text{dist}(\{p_i\}, \{1/2^{m_{\rm Raz}}\}) \nonumber \\ 
&&\leq \text{Pr}(\text{GOOD}) 2^{-1.5 m_{Raz}} + 2 \text{Pr}(\text{BAD})  \nonumber \\ 
&&\leq  2^{-1.5 m_{Raz}}  + 2  
(\gamma^{\mu n} + 2 \epsilon_{Az})^{1/4} 
\label{eq:dist}
\end{eqnarray}
The main figure of merit, i.e., the {\it composable distance} from uniform of the final $k$ bits output by the protocol with distribution $\{q_i\}, i = 1, \ldots, 2^k$ ($\{q_i\}$ is the marginal of $\{p_i \}$ on the $k$-bit output), is then bounded as follows
\begin{equation}
\label{eq:comp-d}
d_{\text{comp}}\leq 2^k \text{dist}(\{q_i\}, \{1/2^k\}). 
\end{equation}
Note that above we chose to obtain a less number of bits $k$ rather than $m_{Raz}$, because as seen in Eq. \eqref{eq:comp-d}, the composable distance is bounded  by the non-composable distance guaranteed by the extractor of \eqref{eq:dist} multiplied by the factor $2^k$. 
Choosing  $k$ smaller than $m_{Raz}$ allows us to then make the composable distance small. 

Now from Eq.(\ref{eq:pi-dist}, the probability distribution of the final $k$ bits satisfies
\begin{equation}
q_i\leq \frac{1}{2^k}(1 + \Delta)
\end{equation}
where 
\begin{equation}
\Delta=\left(2^{-1.5 m_{Raz}}  + 2 (\gamma^{\mu n} + 2 \epsilon_{Az})^{1/4} \right) 2^{k}
\label{eq:Delta}
\end{equation}
Finally, the quality of our output can then be determined by the security parameter $t$, i.e. we will demand 
that $\Delta\leq 2^{-t}$.

In this manner, we have obtained the following conclusion. 
Provided that the number ofbits taken from the SV source to be fed into the extractor, the number of runs, as well as 
the  min-entropy of the SV source and the min-entropy of the box (bounded in Eq. \eqref{eq:hbox}) satisfy Raz's constraints,
we obtain $k$ bits satisfying 
\begin{equation}
q_i\leq \frac{1}{2^k} \left(1 + 2^{-t} \right)
\end{equation}
provided 
\begin{equation}
\left(2^{-1.5 m_{Raz}}  + 2 (\gamma^{\mu n} + 2 \epsilon_{Az})^{1/4} \right) 2^{k}  \leq 2^{-t},
\label{eq:condkt}
\end{equation}
for security parameter $t > 0$.

\subsection{ Application to the present experiment}
For our protocol we have 
\begin{eqnarray}
&&\delta_{exp}=f(00,00)(1/2-\epsilon)^2 - \bigg( f(01,01) + \nonumber \\
&&\qquad \quad +f(10,10)+f(00,11)\bigg) (1/2+\epsilon)^2
\end{eqnarray}
where the frequencies are $f(00,00)=n(00,00)/n$, $f(00,01)=n(00,01)/n$ etc. 
The numbers $n(ab,xy)$ denote the number of events where settings $x,y$ were chosen and the outputs $a,b$ were observed, and are obtained directly from the experiment (these are displayed explicitly in Table I of the main text, and the corresponding frequencies $f(ab, xy)$ are also displayed as the experimentally obtained joint probabilities $P_{AB, XY}(ab, xy)$ in the Table I).

In what follows we shall choose the number of runs (boxes) $n$ and the number of  bits taken from the SV-source $n_{SV}$ to be identical, i.e.,
\begin{equation}
n_{SV}=n.
\end{equation}
However, 
as per the original Raz randomness extractor constraints 
quoted as (1R-4R) below, we shall choose to keep different 
symbols for consistency with Raz's paper \cite{Raz2005}.

{\it Checking Raz's constraints.}  
The Raz randomness extractor is a function \cite{Raz2005} that maps the pair of
sequences of bits of length $n_{SV}$ and $n$ 
with min-entropies $h_{SV}$ and $h_{box}$ respectively into $m_{Raz}$ final bits, whose distance from 
the maximally mixed distribution is bounded from above by $2^{-1.5 m_{Raz}}$.
In our case the min-entropy from the SV source is given by $h_{SV} \geq -n\log(1/2 + \epsilon)$, 
while in place of the min-entropy from the devices $h_{box}$ we shall eventually substitute 
the lower bound $h_{bound}$ derived in Eq.(\ref{eq:entropy-bound}).

(i) \textit{Constraint (1R):}  
\begin{equation}
n_{SV} \geq 6  \log n_{SV}+ 2 \log n
\end{equation}
Since we set $n_{SV} = n$, this condition reads $n\geq 8 \log n$ which holds for $n\geq 64$ and is the case in our experiment.

(ii) \textit{Constraint (2R):}
\begin{equation}
h_{SV} \geq (1/2+\delta_{Raz}) n_{SV}+3 \log n_{SV} + \log n
\end{equation} 
where $\delta_{Raz}$ is a parameter to be fixed later.
Since $h_{SV} \geq -n\log(1/2 + \epsilon)$ from above, this constraint reads as 
\begin{equation}
\delta_{Raz} \leq -\log(1/2+\epsilon) -1/2 - 4 \log(n) /n
\label{eq:cons-delta-Raz}
\end{equation}
In our experiment, we have $4\log(n)/n \leq 2 \times 10^{-8}$. 
We shall just take 
\begin{equation}
\delta_{Raz} = -\log(1/2+\epsilon) -1/2 -2 \times 10^{-8}
\end{equation}
Thus $\delta_{Raz}>0$ for $\epsilon\leq 0.0207$. Interestingly, this is just a bit below the ultimate $\epsilon$ that Raz's extractor can tackle, which is coincidentally equal to the epsilon value for the CHSH inequality $\epsilon_{CHSH}= 1/\sqrt{2}-1/2$.

(iii) \textit{Constraint (3R):} 
\begin{equation}
h_{box}\geq 5 \log(n_{SV}-h_{SV}).
\end{equation}
This is definitely satisfied if 
\begin{equation}
h_{box} \geq 5 \log n \approx 165
\end{equation}
Now for the whole range $\epsilon\in\{0,\epsilon_{CHSH} \}$ we have $h_{box}\geq 371$, so that the constraint 
is satsified for all $\epsilon$ for which Raz's extractor works.

(vi) \textit{Constraint (4R):} 
\begin{equation}
m_{Raz}\leq \delta_{Raz} \min\{n_{SV}/8,h_{box}/40 \} -1
\end{equation}
Since we chose $n_{exp}=n_{SV}$, clearly $n_{SV}/8 \geq h_{box}/40$. 
Thus we obtain that from Raz's extractor one can get 
\begin{equation}
m_{Raz}=\lfloor \delta_{Raz} h_{box}/40 \rfloor -1
\label{eq:mraz}
\end{equation}
bits  whose distance from the maximally mixed probability distribution is $2^{-1.5 m_{Raz}}$. Of course, due to the monotonicity of 
the distance, also a subset of those bits 
satisfies the same condition for the distance from the uniform distribution.

{\it Amplifying randomness.}
Using Eqs. \eqref{eq:dist} and \eqref{eq:mraz} we get 
\begin{eqnarray}
&& {\rm dist}(\{p_i\}, \{1/2^{m_{Raz}} \})\leq  2^{-1.5 m_{Raz}}  + 2 \times
2^{-4 h_{box}}  \nonumber \\
&&\qquad \qquad \leq 2^{-3/80 (\delta_{Raz} h_{box})+3} +2^{- h_{box}} \nonumber \\
&&\qquad \qquad \leq 2^{-3/80 (\delta_{Raz} h_{box}) +4} 
\end{eqnarray}
Thus Eq. \eqref{eq:condkt} now
reads  
\begin{equation}
2k \leq \frac{3}{80} \delta_{Raz} h_{box}-3  - t 
\label{eq:2k}
\end{equation}
Let us now apply the bound in Eq.\eqref{eq:entropy-bound} and  Eq.\eqref{eq:cons-delta-Raz} 
Then, since  $4 \log(n)/n\leq 2 \times 10^{-8}$  in our experiment, we obtain 
\begin{eqnarray}
\delta_{Raz} h_{box} \geq h_{bound}(\delta_{exp},\epsilon, n) \left(-\log(\frac12 +\epsilon)-\frac12 - 2 \times 10^{-8} \right) 
\nonumber \\
\end{eqnarray}
Plugging it into \eqref{eq:2k} we obtain the final
experimental result.

{\it Experimental randomness amplification: final conclusion.-}
 Our experimental data allows us to obtain 
	\begin{equation}
	k(\epsilon, t)=\frac{g(\delta_{exp},\epsilon,n) n - (t+3)}{2}
	\end{equation}
	bits satisfying
	\begin{equation}
	q_i \leq \frac{1}{2^k}\left(1+2^{-t}\right),
	\end{equation}
	where 
	\begin{eqnarray}
    &&	g(\delta_{exp},\epsilon, n)= 
	\frac{3}{80} h_{bound}(\delta_{exp},\epsilon, n) \bigg(-\log(\frac12 +\epsilon)-\frac12 \nonumber \\
	&& \qquad \qquad \qquad  \qquad - 2 \times 10^{-8} \bigg) 
	\end{eqnarray}
	with the function $h_{bound}$ defined by the formula Eq.(\ref{eq:entropy-bound}).
	
	Note that here $f(ab,xy)=n(abxy)/n$ 
	where $n$ is the number of runs of experiment, and $n(abxy)$ is the 
	number of runs, in which settings $xy$ were chosen, and outcomes $ab$ are obtained. 
	Specifically in our experiment 
	\begin{eqnarray}
	&&n=7655734250 \\ \nonumber
	&&f(00,00)=0.022667675540\\ \nonumber
	&&f(01,01)=0.000384051993\\ \nonumber
	&&f(10,10)=0.000363028536\\ \nonumber
	&&f(00,11)=0.000203651270
	\end{eqnarray}

\begin{figure}[bth]
	\includegraphics[width=0.5\textwidth]{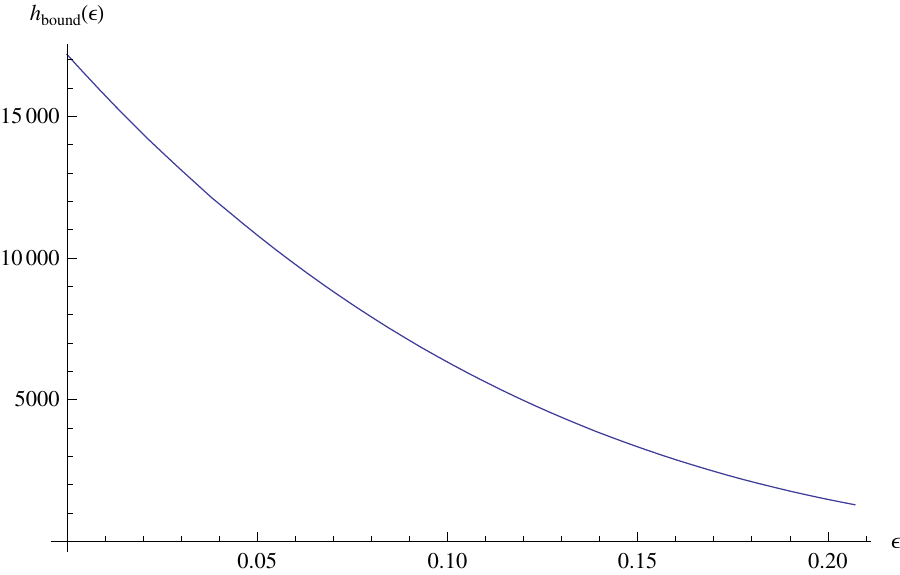}
	\caption{\label{fig:hboxrate} The min-entropy produced by outputs in the Hardy test versus the parameter $\epsilon$ of the SV source.}
\end{figure}

The experimentally obtained (lower bound on) min-entropy and the number of output bits  $k$  are depicted in Fig.  \eqref{fig:hboxrate} and Fig. \eqref{fig:krate}. The latter is also presented in the main text. 

\begin{figure}[bth]
	\includegraphics[width=0.5\textwidth]
	{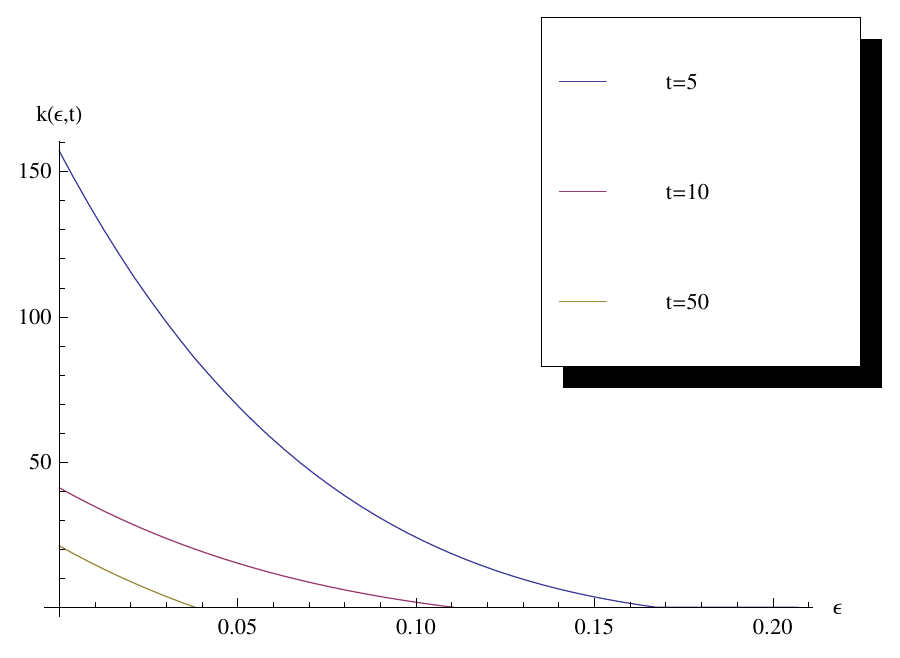}
	\caption{\label{fig:krate} The number of output bits $k(\epsilon, t)$ versus $\epsilon$ for three values of security parameter $t$ (recall that the final $k$ bits deviate from uniform by $2^{-t-1}$).}
\end{figure}

\end{document}